\documentclass[aps,11pt,showkeys,nofootinbib]{revtex4-2}

\usepackage{amssymb,amsmath,amsfonts,amsthm,wasysym}
\usepackage{amsbsy} 
\usepackage{epsfig}
\usepackage{latexsym}
\usepackage{comment}
\usepackage{tensor}
\usepackage{adjustbox}
\usepackage{graphicx}
\usepackage[]{cancel}
\usepackage{xcolor}
\usepackage{subcaption}
\usepackage{hyperref}

\makeatletter
\def\l@subsubsection#1#2{}
\makeatother

\def\be{\begin{equation}}
\def\ee{\end{equation}}
\def\dd{{\rm d}}
\def\bes{\begin{eqnarray}}
\def\ees{\end{eqnarray}}

\DeclareMathOperator{\sgn}{sgn} 

\begin{document}

\title{Towards anisotropic cosmology in group field theory}
\author{Andrea Calcinari}
\email{acalcinari1@sheffield.ac.uk}
\author{Steffen Gielen}
\email{s.c.gielen@sheffield.ac.uk}
\affiliation{School of Mathematics and Statistics, University of Sheffield, Hicks Building, Hounsfield Road, Sheffield S3 7RH, United Kingdom}
\date{March 17, 2023}

\begin{abstract}
In cosmological group field theory (GFT) models for quantum gravity coupled to a massless scalar field the total volume, seen as a function of the scalar field, follows the classical Friedmann dynamics of a flat Friedmann--Lema\^itre--Robertson--Walker (FLRW) Universe at low energies while resolving the Big Bang singularity at high energies. An open question is how to generalise these results to other homogeneous cosmologies. Here we take the first steps towards studying anisotropic Bianchi models in GFT, based on the introduction of a new anisotropy observable analogous to the $\beta$ variables in Misner's parametrisation. In a classical Bianchi I spacetime, $\beta$ behaves as a massless scalar field and can be used as a (gravitational) relational clock. We construct a GFT model for which in an expanding Universe $\beta$ initially behaves like its classical analogue before ``decaying'' showing a previously studied isotropisation. We support numerical results in GFT by analytical approximations in a toy model. One possible outcome of our work is a definition of relational dynamics in GFT that does not require matter.
\end{abstract}

\maketitle

\tableofcontents
\newpage
\section{Introduction}

A major challenge for discrete approaches to quantum gravity is the derivation of an effective (emergent) continuum description which can be compared with classical general relativity or more general gravitational theories. The challenge arises on many levels, for instance in recovering the usual notions of a spacetime manifold from combinatorial structures \cite{manifoldGraph}; recovering an effective description in terms of coordinates and restoring the continuum notion of diffeomorphisms or coordinate changes \cite{BiancaDiffeo}; and understanding the intricate interplay between a continuum and semiclassical limit. Deriving such a description, however, is crucial for understanding the phenomenology of such quantum gravity theories and ensuring their compatibility with observation given that, e.g., new fifth-force degrees of freedom at low energies would have to be compatible with tight experimental bounds \cite{ExpConstraints}. A common approach in this situation is to restrict to situations of high symmetry, in particular spatially homogeneous cosmology or spherically symmetric black holes. While by assumption they no longer include all degrees of freedom, symmetry-restricted models would be expected to capture at least some phenomena of the underlying theory (as they do in classical general relativity), while also connecting directly to phenomenology given the obvious relevance of cosmological and black hole spacetimes. A prominent example is loop quantum gravity (LQG), whose cosmological sector has been studied in loop quantum cosmology \cite{LQCBojo, *Ashtekar+Singh_LQC, *Banerjee_2012} while there are also a number of effective black hole models including LQG discretisation effects \cite{LQBH1,*LQBH2,*LQBH3}.

Effective cosmological models have recently been constructed in the GFT approach to quantum gravity \cite{FreidelGFT,*Oriti_GFTandLQG}, itself closely related to the spin foam definition of LQG dynamics \cite{PerezSFQG} and to matrix and tensor models \cite{matrix,*ColourTensor}. Just as LQG, GFT models are fundamentally defined in terms of discrete, combinatorial structures, interpreted loosely as ``quanta of spacetime''. One can associate geometric notions such as areas, volumes and angles to these degrees of freedom by incorporating concepts from LQG, but there is no simple way of giving these an effective continuum interpretation, in particular given that the conceptual status of a given discretisation in GFT is similar to a Feynman graph in quantum field theory, i.e., only one term in an infinite expansion. What can be done relatively straightforwardly, however, is to derive dynamical equations for {\em global} observables such as the total (spatial) volume of a certain geometry, which can then be contrasted with globally homogeneous cosmological models. These equations are usually derived for certain GFT states whose properties make them good candidates for spatially homogeneous geometries. After some prior groundwork \cite{GFTcosmoLONGpaper,*Gielen_2016} a breakthrough in this line of research came when, in GFT models for quantum gravity coupled to a massless matter scalar field, a ``relational'' volume observable (corresponding to the volume of space for a given value of the scalar field) was shown to satisfy the Friedmann dynamics of general relativity at low energies while also replacing the classical Big Bang singularity by a bounce \cite{Oriti_2016,*BOriti_2017}. Similar results have been obtained using different methods and from different starting points \cite{toy,relham_Wilson_Ewing_2019,*relhamadd,Gielen_2020,Marchetti2021}, emphasising their robustness: one can understand the main properties of the resulting cosmological dynamics from classical solutions for a single field mode. While important for the phenomenology of GFT and for connecting to approaches such as loop quantum cosmology, these results have so far been restricted to the case of a flat homogeneous, isotropic Universe.\footnote{Inhomogeneities can be included perturbatively as in \cite{GFTsepUniv,*LucaGFTpert}, and match physical expectations at least in a long-wavelength limit.} Some studies have included anisotropies perturbatively \cite{de_Cesare_2017,Mairi_2017} showing that they decay leading to isotropisation, but there is so far no characterisation of, e.g., an anisotropic Bianchi cosmology.

Here we take the first steps towards the study of anisotropic cosmologies in group field theory, focussing on the simplest possible case of Bianchi I cosmology with local rotational symmetry so that two out of the three directional scale factors are taken to be equal. There are at least two, initially quite separate, challenges involved in this extension of past work. The first is to find a characterisation of anisotropies in group field theory, i.e., to define an observable that can distinguish isotropic and anisotropic geometries and quantify the amount of anisotropy. Here the key idea we use is the Misner parametrisation of Bianchi models (see, e.g., \cite{BojoBook}; we will also review this below) in terms of a volume degree of freedom and two relative anisotropy variables, the Misner variables $\beta_\pm$. In a classical Bianchi I model $\beta_\pm$ behave as free, massless scalar fields in a flat FLRW geometry\footnote{This property makes the Misner parametrisation particularly natural in classical general relativity. For comparison we should mention that in loop quantum cosmology the situation is different, as one quantises an LQG-corrected Hamiltonian constraint and the particular type of corrections makes Misner variables less convenient \cite{AshtekarEdBianchiI,*BojowaldMisner}.}, which have already been studied in GFT. On the other hand, the discreteness of geometry in GFT means that we cannot simply take over a continuum definition of anisotropy, so the construction of an analogue $\beta_\pm$ variable requires careful thought. The second challenge is to understand which simplifying approximations used in past work need to be relaxed in order to allow for anisotropies in the effective description. For instance, while the work of \cite{Oriti_2016,*BOriti_2017} only used ``isotropic'' states interpreted as describing simplicial building blocks for which all faces have equal area, it is known (see, e.g., \cite{Gielen_2020}) that this microscopic restriction to isotropy is neither necessary nor sufficient to obtain the correct (flat FLRW) Friedmann dynamics: the more relevant assumption is to restrict to a single field mode in the Peter--Weyl expansion in representation data. Hence, in order to describe anisotropic geometries, multiple Peter--Weyl modes must be taken into account, but it is not clear how many (and which) modes are needed to capture physical anisotropies. 

In this paper we show how to tackle the first challenge; we define an $\beta_\pm$ analogue with a clear geometric interpretation, quantum ambiguities that disappear for large quanta, and correct physical properties -- constant velocity and hence linear evolution -- at least for a certain cosmological period of time, before the isotropisation observed in \cite{Mairi_2017,de_Cesare_2017} sets in and the anisotropy disappears. The second challenge is partially addressed, given that the $\beta_\pm$ dynamics partially match expectations from classical relativity, but the observed isotropisation does not correspond to a classically expected behaviour and, more importantly, anisotropies do not backreact on the effective Friedmann equation as expected. This suggests that while our constructions will be useful for future work, our model needs further refinement to reproduce the correct physics of a classical Bianchi Universe.

Any monotonically evolving quantity in a cosmological model can be used as a relational clock: all other dynamical variables can be written, at least in principle, as functions of this ``clock''. In a vacuum Bianchi I model, the anisotropy variables $\beta_\pm$ have this property and hence, in contrast to what is often done in quantum cosmology and in particular in GFT, no coupling to matter would be needed to be able to express the dynamics in relational terms. The fact that we have defined a new quantity with monotonic evolution in GFT cosmological models hence raises the possibility of defining relational evolution in GFT without adding matter fields, which might help in understanding the ``problem of time'', or possible dependence of dynamics on the choice of clock, in GFT (see \cite{Axel_Steffen_scalars} for some work on this issue in models with multiple possible clocks).

The remaining parts of the paper are structured as follows. In section \ref{reviewSec} we review basic ideas of the GFT formalism, possible definitions of a canonical Hilbert space quantisation, and application to cosmology: we show how one can derive effective Friedmann equations by restricting to a single field mode and neglecting interactions. Readers familiar with GFT cosmology may skip this review section. Similarly, section \ref{relcosmology} is a review of classical FLRW and Bianchi cosmologies written in relational terms, using a scalar field clock; this is the classical theory that any effective description of GFT can be compared to. Section \ref{anisoGFT} includes the main new results: we motivate the introduction of an effective $\beta_\pm$ variable used to characterise anisotropies in GFT. We then propose models based on a few Peter--Weyl modes and study the effective dynamics of both the anisotropies and the spatial volume, comparing both with the dynamics of general relativity.  We also propose a simplified ``toy model'' in which some of our main results, in particular the linear growth in anisotropy which matches classical expectations, can be derived analytically rather than numerically as in the main part. We conclude in section \ref{conclusions}. An appendix gives details on the classical and quantum geometry of tetrahedra as used in LQG and GFT.

\section{Short review of GFT cosmology}
\label{reviewSec}

In this section we briefly summarise past work on deriving effective cosmological dynamics from group field theory. In this past work, effective Friedmann equations were obtained after truncating the full dynamics and choosing simple GFT states, following two different approaches. The two approaches, which we will call {\em algebraic} and {\em deparametrised}, will be introduced in section \ref{canonicalGFT}.

\subsection{Basics of group field theory}

We are interested in GFT models for simplicial gravity coupled to a (free, massless) scalar field $\chi$. In such models one defines a group field $\varphi$ whose arguments are $d$ elements of a Lie group $G$ (hence the name ``group field'') and a real variable corresponding to the scalar matter field $\chi\in\mathbb{R}$,
\begin{equation}
	\varphi: G^d \times \mathbb{R} \rightarrow \mathbb{K}\,, 
\end{equation}
where $\mathbb{K}$ can be $\mathbb{R}$ or $\mathbb{C}$. When applied to four-dimensional quantum gravity, $d=4$ and one usually takes $G$ to be the local gauge group of general relativity: $G$ is typically $SO(3,1)$ or $SL(2,\mathbb{C})$ in the Lorentzian case, $SO(4)$ or $Spin(4)$ in the Euclidean case, or their rotation subgroup $SU(2)$ which is the gauge group of loop gravity (or the Ashtekar--Barbero formulation of classical general relativity \cite{barbero}). The last choice is the one we will use here. To implement a notion of discrete gauge invariance in the resulting simplicial gravity description, one usually requires invariance of the field under the right diagonal group action,
\begin{equation}\label{rightdiagonalaction}
	\varphi(g_I,\chi) \equiv \varphi(g_1,\dots, g_d , \chi) = \varphi(g_1 h,\dots ,g_d h, \chi)\,, \qquad \qquad \forall h \in G \, .
\end{equation}
The action has the general form
\begin{equation}\label{SGFT}
	S[\varphi,\bar{\varphi}] = \int {\rm d}g\; {\rm d}g'\; {\rm d}\chi\;\bar{\varphi}(g_I,\chi) K(g_I,g'_I)\varphi(g'_I,\chi) + V[\varphi,\bar{\varphi}]\,,
\end{equation}
where for a real field $\bar{\varphi} = \varphi$. Here $\int {\rm d}g$ stands for an integration over $d$ copies of the group, using the Haar measure normalised to unity. The action is therefore split into a quadratic part and an interaction part $V$. The kernel $K$ is assumed to respect the symmetries associated to a minimally coupled massless scalar field on a curved background, namely shift ($\chi\rightarrow\chi+c$) and sign reversal ($\chi\rightarrow-\chi$) symmetries. For this reason, $K$ cannot depend explicitly on $\chi$, but it is in general a differential operator in $\chi$, which does not involve odd powers \cite{Oriti_2016,*BOriti_2017,Li_2017}. $\chi$ usually plays the role of a relational time variable, so that other observables are defined relative to $\chi$, as is common in many approaches to quantum cosmology, in particular in loop quantum cosmology. The kinetic term can be written as an expansion in derivatives with respect to $\chi$,
\begin{equation}\label{genericK}
	K(g_I,g'_I) = \sum_{n=0}^\infty K^{(2n)} (g_I,g'_I)\frac{\partial^{2n}}{\partial \chi^{2n}} = K^{(0)}(g_I,g'_I) + K^{(2)}(g_I,g'_I)\partial_\chi^2 + \dots\,,
\end{equation}
which is usually truncated after the second term: starting from the simplest $K$ that only includes a constant ``mass term'' suggested by the relation to spin foam models \cite{Reisenberger_2000}, a Laplacian term is generated by radiative corrections \cite{Geloun_2013}. One could stop here, given that no higher derivative terms are required, or make the weaker assumption that higher derivatives are present but suppressed, i.e., $|K^{(2n)}/K^{(0)}| \ll |K^{(2)}/K^{(0)}|^n$ for $n>1$. For concreteness, we will follow previous work (see, e.g., \cite{GFTcosmoLONGpaper,Gielen_lowspin}) and assume that $K$ has the minimal form
\begin{equation}\label{kineticterm}
	K = m^2-\partial^2_\chi + M^2\sum_{I=1}^d \Delta_{g_I}	\, ,
\end{equation}
where $m$ and $M$ are coupling constants and $\Delta_{g_I}$ is the Laplace--Beltrami operator acting on the $I$-th group argument. Evidently, this corresponds to $K^{(0)}=m^2+M^2\sum_I \Delta_{g_I}$, $K^{(2)}=-1$ in (\ref{genericK}). 

\begin{figure}[ht]
	\begin{center}
			\includegraphics[width=.22\textwidth]{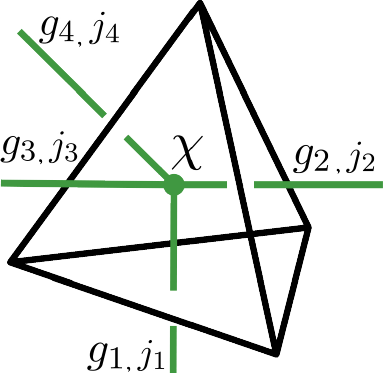}
			\hspace{2.5 cm}
		\includegraphics[width=.23\textwidth]{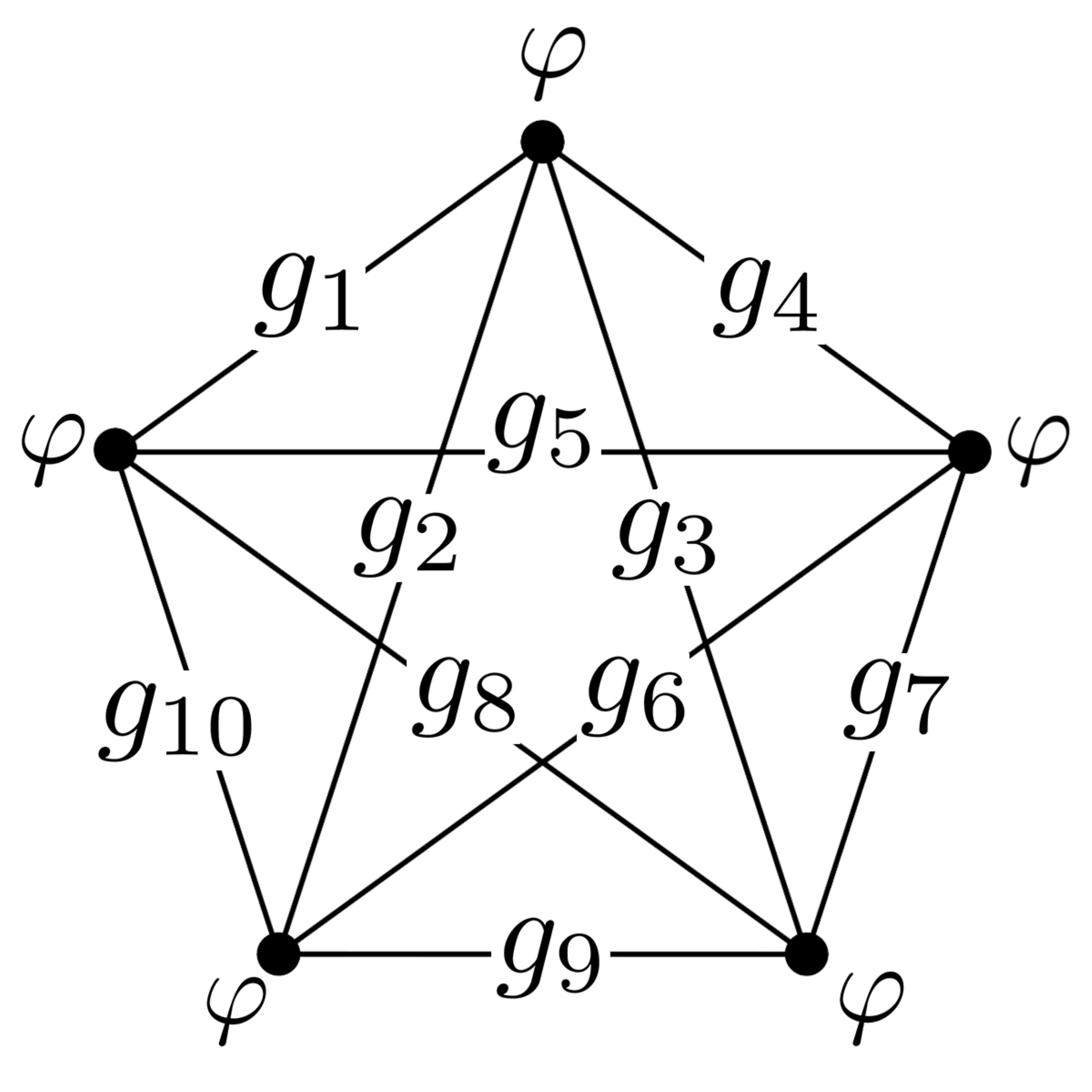}
		\end{center}
	\caption{\small Left: GFT quantum as open spin network vertex (green) or quantum tetrahedron (black). Right: Graph of five vertices, representing a four-simplex as gluing of five tetrahedra (three-simplices). Here each field is represented by a node with four legs, and a shared link corresponds to matching group arguments.}
	\label{tetra+5graph}
\end{figure}

To construct an interaction term in simplicial gravity models, one can think of the group field $\varphi$ as representing a ($d-1$)-simplex. The interaction term then describes the gluing of such simplices to form $d$-dimensional structures, here a $d$-simplex, similarly to what happens in tensor models \cite{ReggeGFT,ColourTensor}. This means that an appropriate interaction term consists of products of fields that are paired according to a pattern which encodes the combinatorics of a $d$-simplex. In four dimensions, we could for example choose (here for a real field)
\begin{equation}\label{5interaction}
	V[\varphi] = \frac{\lambda}{5}\int ({\rm d}g)^5\; V(g^1_I,...,g^5_I)\prod_{a=1}^5\varphi(g^a_I,\chi) \,,
\end{equation}
where $\lambda$ is a coupling constant, $\int({\rm d}g)^5$ means integration over $G^{20}$, and $V(g^1_I,...,g^5_I)$ is a product of ten Dirac delta distributions on the group, imposing appropriate matching between group elements appearing as arguments of the fields $\varphi(g^j_I,\chi)$, in order to encode the pattern of gluings needed to form a four-simplex out of five tetrahedra (see right panel of figure \ref{tetra+5graph}). Such an interaction should allow the structure of four-dimensional spacetime to emerge from the dynamics of the theory.

We will now fix $G^d=SU(2)^4$. Then the field $\varphi(g_I,\chi)$ can also be represented as a four-valent spin network node. The $g_I$ are associated to the links dual to the faces of a tetrahedron (see left panel of figure \ref{tetra+5graph}), and the matter field $\chi$ ``sits" on the node dual to the tetrahedron. 

It is often convenient to use the Peter--Weyl theorem to express the group field as
\begin{equation}\label{decomposition}
	\varphi (g_I, \chi) = \sum_{\vec{j},\vec{m},\vec{n},\imath} \varphi^{\vec{j},\imath}_{\vec{m}}(\chi)\, \mathcal{I}_{\vec{n}}^{\vec{j},\imath}\prod_{I=1}^4 \sqrt{2j_I+1}\,D^{(j_I)}_{m_I,n_I}(g_I)\,,
\end{equation} 
where the $D^{(j_I)}_{m_I,n_I}(g_I)$ are Wigner matrices and $ \varphi^{\vec{j},\imath}_{\vec{m}}(\chi) \equiv \varphi_{m_1,\dots,m_4}^{j_1,\dots,j_4,\imath}(\chi)$ are complex functions\footnote{If the group field is real, the complex Peter-Weyl coefficients satisfy the reality condition \cite{Ooguri,relhamadd}\begin{equation}
		\overline{\varphi^{\vec{j},\imath}_{\vec{m}}(\chi)} = (-1)^{\sum_I(j_I-m_I)}\varphi^{\vec{j},\imath}_{-\vec{m}}(\chi)\,.
\end{equation}}. The intertwiners $\mathcal{I}_{\vec{n}}^{\vec{j},\imath}\equiv \mathcal{I}_{n_1,\dots,n_4}^{j_1,\dots,j_4,\imath}$ arise because of the invariance under group multiplication from the right \eqref{rightdiagonalaction}. Recall that intertwiners are elements of the Hilbert space of the tetrahedron (or four-valent node) $\mathcal{H}_4 = \text{Inv} \left[\bigotimes_{I=1}^4\mathcal{H}_{j_I}\right]$, where each $\mathcal{H}_{j_I}$ corresponds to the Hilbert space of an irreducible unitary representation of $SU(2)$ (see appendix \ref{AppA}). The sums are over representations $j_I\in \mathbb{N}_0/2$ and magnetic indices $m_I,n_I\, \in [-j_I\,,\,j_I]$, while $\imath$ labels the possible intertwiners for each set of given spins. One can picture each pair $(j_I,m_I)$ as living on a link emerging from the node, while $\imath$ lives on the node itself. 

The main motivation behind these definitions is the connection with spin foam models; one can understand GFTs as a more fundamental quantum field theory-like framework (or quantum gravity theory) into which spin foam models coming from LQG can be embedded. Namely, the Feynman expansion of the GFT partition function generates a sum over graphs $\Gamma$ that can be seen as dual to simplicial complexes (here again for a real field),
\begin{equation}\label{partitionfunction}
	Z_{\text{GFT}} = \int \mathcal{D}\varphi\; e^{-S[\varphi]} = \sum_\Gamma  \lambda^{n_V(\Gamma)}\mathcal{A}_\Gamma\,,
\end{equation}
where $n_V(\Gamma)$ is the number of interaction vertices and $\mathcal{A}_\Gamma$ is the Feynman amplitude. The Feynman integrals over momenta are  discrete sums (as the space on which the field theory is defined is compact) over group representations and intertwiners, associated with faces and edges of the two-complex dual to $\Gamma$. The GFT Feynman amplitudes are exactly what is usually called {\em spin foam amplitudes} \cite{Reisenberger_2000,BCmodelfromGFT}. In the LQG spin foam approach, one often makes a choice of a fixed two-complex, but a sum over two-complexes would be necessary if one hopes to capture the continuum dynamics of quantum gravity. The Feynman expansion \eqref{partitionfunction} provides precisely such a sum over two-complexes, with relative weights determined by $\lambda$, and hence a generating functional for the covariant quantisation of LQG. By choosing $K$ and $V$ in the GFT action (\ref{SGFT}) one can generate models which are related in a precise way to different spin foam models. 

We should mention that just as for the kinetic term $K$, the choice of interaction term(s) $V$ may also be guided by considerations beyond the correspondence with simplicial gravity. For instance, additional interaction terms could be generated by renormalisation group flow (see, e.g., \cite{Lahoche:2018hou} for an analysis of models with $G=U(1)$), or included in a general effective field theory description.
 
\subsection{Canonical quantisation of GFT}
\label{canonicalGFT}

The traditional way of thinking of quantum GFT is through the path integral as given in (\ref{partitionfunction}), and its expansion into Feynman graphs and amplitudes. As we discussed, this path integral directly connects to the covariant (spin foam) setting of LQG. More recently however, a lot of work has focused on the {\em canonical quantisation} of GFT, with the main goals of connecting to the canonical setting for LQG and -- most importantly for us here -- extracting effective cosmological dynamics which are more easily defined in a canonical setting, as they are in LQG where loop quantum cosmology has so far only been derived through canonical methods.

Two main approaches to canonical quantisation have been established in the GFT literature: one based on a kinematical Fock space of nondynamical spin network-like states on which dynamical equations are imposed in a suitable (usually mean-field) approximation \cite{Oriti_GFT2ndLQG}, and one where a time variable is selected before quantisation and used to directly obtain a physical Fock space and physical (relational) Hamiltonian \cite{relham_Wilson_Ewing_2019,*relhamadd}. These two types of quantisation are to some extent analogous to two approaches in usual canonical LQG: on the one hand, in LQG one can define a kinematical Hilbert space whose states -- though $SU(2)$ invariant and satisfying the (spatial) diffeomorphism constraint -- do not yet satisfy any dynamics. Physical states would only emerge after one has also implemented the Hamiltonian constraint (see, e.g., \cite{ThiemannBook}). Since this can usually not be done exactly, one can try to implement dynamics approximately, for example using coherent states to study a semiclassical effective theory \cite{GieselThiemann,Alesci1,*Alesci2,*Alesci3,Dapor_effective_2017}. On the other hand, one can introduce matter (often pressureless ``dust'' scalar fields) to define a gauge-fixed classical theory, in which the matter is used as a preferred standard of time. This deparametrised approach leads to a physical Hilbert space since the Hamiltonian constraint can be solved for the momentum canonically conjugate to ``dust time" \cite{LQG_dust1,AQG_thiemanngiesel,*gravityquantised_Lewandowski,*LQG_de-parametrised,*Giesel_scalarmaterialreference}. While this reduced Hamiltonian approach breaks time reparametrisation invariance by requiring a preferred time coordinate before quantisation, it automatically describes physical quantum states which satisfy dynamics, and allows bypassing the unresolved problem of implementing the Hamiltonian constraint on the kinematical Hilbert space.

In GFT, there is no Hamiltonian constraint with the same status as in canonical quantum gravity, i.e., related to diffeomorphisms in time. The two approaches we will review nevertheless share the same two different starting points of either imposing dynamical equations on an abstract Hilbert space, or directly defining a physical Hilbert space by choosing a matter clock. While in general these two types of quantisation cannot be equivalent, in the application to cosmology they lead to very similar effective dynamics, essentially since one requires semiclassical states with small quantum corrections to classical GFT solutions.

\subsubsection{Algebraic approach}

This approach requires a complex group field. Starting from the Peter--Weyl decomposition (\ref{decomposition}) and its complex conjugate, one defines a quantum theory by promoting the field modes to operators $\hat{\varphi}_J$ and $\hat{\varphi}^\dagger_J$ (we adopt the compact notation of \cite{Isha_thermalGFT} and use the labels $J=(\vec{j},\vec{m},\imath)$ and $-J = (\vec{j},-\vec{m},\imath)$) and postulating their commutation relations according to bosonic statistics,
\begin{equation}
\left[ \hat{\varphi}_J(\chi), \hat{\varphi}^\dagger_{J'}(\chi') \right] = \delta_{JJ'}\delta(\chi-\chi') \,.
\end{equation}

 One can then construct an abstract (unphysical) Fock space in the usual way, starting from a vacuum $|0\rangle$ satisfying $\hat\varphi_J|0\rangle=0$. The Fock space contains fundamental quanta (pictorially ``atoms of space" \cite{ORITI2017235}) created and annihilated by field operators, which are interpreted as equivalent to LQG spin-network vertices. As in the kinematical Hilbert space of LQG, there is no notion of dynamics so far: this will only be implemented ``on average" later on, similarly to what happens in condensed matter physics. The Fock vacuum $|0\rangle$ is a state with no topological nor geometrical meaning and information. Then the one-quantum Hilbert space has states
 
\begin{equation}
\biggr|\; \adjincludegraphics[valign=c,width=.09\textwidth]{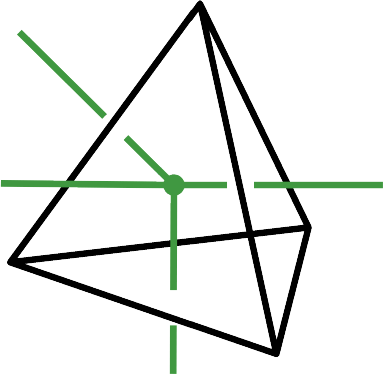} \biggr>=	|J,\chi \rangle = \hat{\varphi}^\dagger_J(\chi) |0\rangle \,, 
\end{equation}
interpreted as (open) spin network vertices decorated with four Peter--Weyl labels (or group elements) and a scalar $\chi$.

At this stage one usually defines general one-body operators (see, e.g., \cite{Oriti_GFT2ndLQG})
\begin{equation}\label{OBop}
	\hat{O} (\chi) = \sum_J O^J \, \hat{\varphi}^{\dagger}_J(\chi)\hat{\varphi}_J(\chi)\,,
\end{equation}
where $O^J$ is the LQG matrix element of the desired operator evaluated on a single spin network node, e.g., the volume\footnote{Concretely, after choosing a basis of intertwiners $\imath$ that diagonalises the LQG volume operator on a spin network node, one associates a volume $V_{\vec{j},\imath}$, given by the eigenvalue of this operator, to the quanta generated by $\varphi^\dagger_J(\chi)$. See appendix \ref{AppA} for more details on the spectrum and eigenstates of this LQG volume operator. These properties can be derived from a more general setting of quantisation of tetrahedra in terms of $SU(2)$ recoupling theory \cite{Barbieri_1997,PolyhedraIta,Baez_1999}, without working necessarily within LQG.} $V_{\vec{j},\imath}$ which plays a particularly important role in cosmology. To give explicit formul\ae, the number and volume operators read
\begin{equation}\label{NVspincondensate}
	\hat{N}(\chi) = \sum_J \hat{\varphi}^{\dagger}_J(\chi)\hat{\varphi}_J(\chi)\,,\qquad \text{and} \qquad \hat{V}(\chi) = \sum_J V_{\vec{j},\imath} \hat{\varphi}^{\dagger}_J(\chi)\hat{\varphi}_J(\chi)\, .
\end{equation}
Notice that these operators are defined ``relationally", in the sense that they are given as functions of the matter field $\chi$, associated to the creation and annihilation operators. At the kinematical level, this $\chi$-dependence does not represent time evolution: operators defined for different $\chi$ are independent. One then usually assumes that equations of motion are satisfied on average,
\begin{equation}\label{quantumaction}
	\left< \frac{\delta S[\hat{\varphi},\hat{\varphi}^\dagger]}{\delta \hat{\varphi}^\dagger}\right> =0\,,
\end{equation}
where $S$ is the action of our GFT model. Specifying suitable coherent states to compute expectation values will then give effective cosmological equations, as explained in the next section. 

\subsubsection{Deparametrised approach}\label{BDP}

If, on the other hand, one wants to follow a deparametrised approach, one can work with a real GFT field. Working in the spin representation, equation \eqref{SGFT} then reads \cite{relham_Wilson_Ewing_2019,*relhamadd}
\begin{equation}\label{realaction}
	S[\varphi] =  \frac{1}{2} \int {\rm d}\chi \sum_{J} \varphi_{-J}(\chi)K_{J}(\chi) \varphi_J(\chi)+ V[\varphi]\,,
\end{equation}
where $K_{J} = K^{(0)}_{J}+K^{(2)}_{J}\partial_\chi^2$ is defined as in \eqref{genericK}. In this paper we will assume \eqref{kineticterm} so that $K_{J} = m^2-\partial_\chi^2-M^2\sum_I j_I(j_I+1)$ (using the fact that the Laplace--Beltrami operator acts as a Casimir on Wigner matrices, $\Delta_g D^{(j)}_{mn} (g)= -j(j+1) D^{(j)}_{mn} (g)$), but we can keep $K^{(0)}_{J}$ and $K^{(2)}_{J}$ general for the time being.

Now one can define the conjugate momentum $\pi_{J}(\chi)$ to the group field $\varphi_{J}(\chi)$, and the Legendre transform of the Lagrangian $\mathcal{L}$ with respect to $\chi$ gives a relational Hamiltonian
\begin{equation}
	\mathcal{H} = -\frac{1}{2} \sum_{J} \left[\frac{\pi_{J}(\chi)\pi_{-J}(\chi)}{K^{(2)}_{J}}+K^{(0)}_{J}\varphi_{J}(\chi)\varphi_{-J}(\chi)\right]-V[\varphi] \, ,
\end{equation}
which determines the dynamics of any observable $\mathcal{O}$ through Poisson brackets ${\rm d}\mathcal{O}/{\rm d}\chi = \{\mathcal{O},{\mathcal{H}}\}$. The field and its momentum are promoted to operators with the usual canonical {\em equal-time} commutation relation
\be
\left[ \hat{\varphi}_J(\chi), \hat{\pi}_{J'}(\chi) \right] = {\rm i}\,\delta_{JJ'} \,.
\ee

The key difference with the previous approach is that these operators already satisfy dynamical equations, implemented through the Heisenberg equations of motion. This has a cost: we needed to specify our time variable once and for all from the very beginning to define the conjugate momentum of the field and the relational Hamiltonian.

One can now define creation and annihilation operators $\hat{a}^{\dagger}_J$ and $\hat{a}_J$ as in any bosonic quantum field theory (not to be confused with $\hat{\varphi}^{\dagger}_J$ and $\hat{\varphi}_J$) with their own (equal-time) commutation relations,
\begin{equation}
	\left[ \hat{a}_J(\chi), \hat{a}^\dagger_{J'}(\chi) \right] = \delta_{JJ'} \,,
\end{equation}
 and use these to construct a physical Fock space. This space is ``smaller" than the one introduced in the earlier algebraic approach since the states are already interpreted as physical states, not subject to any constraints.\footnote{For complex group fields, one can obtain this physical Fock space from a group averaging construction in which one imposes a GFT constraint (exactly) on a larger kinematical Fock space as described above \cite{frozenf}.} Dynamics for any operator $\hat{\mathcal{O}}$ are given by relational Heisenberg equations
\begin{equation}\label{Heisenberg}
	{\rm i}\frac{{\rm d} \hat{\mathcal{O}}}{{\rm d}\chi} = [\hat{\mathcal{O}},\hat{\mathcal{H}}]\,.
\end{equation}
The relational Hamiltonian operator for the free theory (i.e., for $V[\varphi]=0$) can be expressed as a sum of single-mode Hamiltonians $\hat{\mathcal{H}} = \sum_{J} \hat{\mathcal{H}}_{J}$, and the $\hat{\mathcal{H}}_J$ can themselves be written in terms of creation and annihilation operators. For a mode such that $K^{(0)}_{J}$ and $K^{(2)}_{J}$ have different signs the single-mode Hamiltonian takes the form \cite{relham_Wilson_Ewing_2019,relhamadd}
\begin{equation}\label{relham}
	\hat{\mathcal{H}}_{J} =  \frac{1}{2} M_{J} \left(\hat{a}^{\dagger}_{J}\hat{a}^{\dagger}_{-J}+\hat{a}_{J}\hat{a}_{-J}\right)\,,
\end{equation}
where $M_{J} = -\sgn\left(K^{(0)}_{J}\right)\sqrt{\left|K^{(0)}_{J}/K^{(2)}_{J}\right|}$. For the case of interest in the rest of this paper, where the kinetic term is \eqref{kineticterm} so that $K_J$ is as given below \eqref{realaction},
\be
M_{J} =  m_{\vec{j}} := -\sqrt{m^2-M^2\sum_I j_I(j_I+1)}\,.
\ee
Only modes for which the argument of the square root is positive will correspond to such a Hamiltonian. The coupling $m_{\vec{j}}$ only depends on the spin labels, so magnetic and intertwiner indices can be dropped in many expressions.  The Hamiltonian \eqref{relham} is a squeezing operator which does not leave the Fock vacuum invariant, but creates pairs of excitations with opposite values for the magnetic indices. Expanding space arises from the instability of the Fock ``vacuum'', realising a type of ``geometrogenesis'' (the term was coined in \cite{Graphity} and then used in the GFT literature \cite{GFTquantumST_Oriti,GFTcosmoLONGpaper,DisappEmergenceSpaceTime}). The rate of squeezing or expansion is determined by $|M_J|$; for models in which $|M_J|$ takes a maximum for some $J$ (such as, in our case, for the modes of lowest $j_I$), these modes will always dominate asymptotically \cite{Gielen_lowspin}. This instability is to be contrasted with modes for which $K^{(0)}_{J}$ and $K^{(2)}_{J}$ have the same sign; these modes are stable with constant particle number and quickly become insignificant compared to the unstable modes, which is why they are usually ignored. Of course, this also means that to obtain a realistic cosmology we have to assume that at least one mode has a Hamiltonian of squeezing type \eqref{relham}. 

In order to extract the simplest cosmological interpretation, we are only interested in the total particle number and volume operators, here given by
\begin{equation}\label{NVunderlined}
	\hat{\underline{N}}(\chi) = \sum_J \hat{a}^{\dagger}_J(\chi)\hat{a}_J(\chi)\,, \qquad \text{and} \qquad \hat{\underline{V}}(\chi) = \sum_J V_{\vec{j},\imath} \hat{a}^{\dagger}_J(\chi)\hat{a}_J(\chi)\,, 
\end{equation}
where we distinguish the notation from (\ref{NVspincondensate}) by underlining them to keep in mind that these are different operators from the ones defined in the algebraic approach. We can however compare the two approaches at the level of expectation values, as illustrated below.

The operators we are interested in here are always a sum of single-mode operators. Such one-body operators are extensive and often simplify because one is only interested in certain modes.

\subsection{Emergent FLRW Universe from free theory and single mode}\label{SecC}

In order to link this theory with cosmology, some simplifying assumptions are needed. First of all, the cosmological sector of GFTs often deals with regimes in which interactions can be neglected, so we will only consider the kinetic term of the action (i.e., the free theory). This approximation is often interpreted \cite{GFTcosmoLONGpaper, relham_Wilson_Ewing_2019} as corresponding to homogeneity given that interactions between spin-network nodes (and correlations between the GFT quanta) are negligible. Given the instability of the theory, such an approximation can only hold for a finite amount of time before the number of quanta is too large \cite{Oriti_2016,*BOriti_2017}. 

Moreover, we now focus on a single Peter--Weyl mode. This restriction is motivated by computational simplicity and the fact, mentioned above, that often one mode quickly dominates dynamically so that this approximation becomes better and better with time. There is evidently a certain clash with the first approximation of negligible interactions, which gets worse over time. In any case, this second approximation means that a cosmological spacetime expands or contracts by modifications to the combinatorial structure of the spin network (i.e., by changing number of the GFT quanta), rather than by changing the spin labels on the network (i.e., transitioning between GFT quanta of different spin representations). Furthermore, one can use the insights gained from loop quantum cosmology, where all the spins are usually fixed to only one value \cite{LQCBojo,*Banerjee_2012,*Ashtekar+Singh_LQC}, to motivate this single-mode restriction in GFT. Such assumptions behind loop quantum cosmology models are often motivated by a suggestion that cosmological expansion or contraction is indeed realised by a changing graph structure in full LQG \cite{AshtekarEdBianchiI,*BojowaldMisner}.

One then also needs to specify a particular type of states, and here coherent states are used to implement a notion of semiclassicality similar to what is often done in quantum cosmology: in a macroscopic Universe quantum fluctuations over expectation values should be small. Different choices of coherent states are examined with respect to this criterion, e.g., in \cite{Gielen_2020}.

\subsubsection{Algebraic approach}

The next step in the algebraic approach is to define states in which \eqref{quantumaction} is to be evaluated. The simplest choice is given by field coherent or ``condensate'' states, whose key property is
\begin{equation}\label{coherentmean}
 \hat{\varphi}_{J} (\chi) |\sigma\rangle = \sigma_{J}(\chi) |\sigma\rangle\,.
\end{equation}
This property allows replacing (after normal ordering) all field operators by the collective variable $\sigma_{J}(\chi)$ (sometimes called condensate wavefunction), which explicitly depends on $\chi$; the quantum equation of motion (\ref{quantumaction}) reduces to the classical GFT equation of motion. This mean-field approximation can be interpreted as the idea that cosmology arises from the ``hydrodynamics of quantum gravity" \cite{ORITI2017235}. Since all quanta are characterised by a single quantum state, it has been conjectured that these states may correspond, when coarse-grained, to homogeneous cosmological spacetimes. In this mean-field approach one can then derive effective dynamics for expectation values of the operators of interest, such as $\hat{N}(\chi)$ and in particular $\hat{V}(\chi)$, basically ignoring fluctuations: starting from the definitions \eqref{NVspincondensate} and making use of the property \eqref{coherentmean}, one finds the expectation values (also assuming that only a single Peter--Weyl mode contributes)
\begin{equation}\label{EXPMF}
	N(\chi) = \langle \hat{N}(\chi) \rangle= |\sigma_{J}(\chi)|^2\,, \qquad \text{and} \qquad V(\chi) = \langle \hat{V}(\chi) \rangle = V_{\vec{j},\imath} |\sigma_{J}(\chi)|^2\, .
\end{equation} 
The expectation values \eqref{EXPMF} can now be evaluated for a solution of the free theory $K_{J}\,\sigma_{J}(\chi)=0$, given that we have decided to neglect interactions. In particular, we will again assume the single-mode kinetic kernel 
\begin{equation}\label{KKK}
	K_{J} =   -\partial_\chi^2 + m^2_{\vec{j}}    \,.
\end{equation}
Using the notation of \cite{Gielen_lowspin}, the general solution to this equation of motion then reads
\begin{equation}\label{solutionMF}
	\sigma_{J} (\chi) = \alpha^+_J e^{m_{\vec{j}}\chi}+\alpha^-_J e^{-m_{\vec{j}} \chi}			\,,
\end{equation}
where the coefficients generally depend on the mode $J$. Substituted into \eqref{EXPMF}, one finds the volume
\begin{equation}\label{volumeMF}
		V(\chi)  = V_{\vec{j},\imath}\left[\left(|\alpha^+_J|^2+|\alpha^-_J|^2\right) \cosh(2m_{\vec{j}} \chi)+\left(|\alpha^+_J|^2-|\alpha^-_J|^2\right)\sinh(2 m_{\vec{j}} \chi) + 2\Re(\alpha^+_J\overline{\alpha^-_J})\right]\, .	
\end{equation}
Crucially, this form for the volume corresponds to a cosmological solution that satisfies Friedmann-like dynamics while also containing a cosmological bounce. Even though the original work \cite{Oriti_2016} was based on equilateral tetrahedra which were supposed to encode isotropy, one can retain four different spins $\vec{j} = (j_1,j_2,j_3,j_4)$ for the faces of the building blocks; the key ingredient for this result lies in the single-mode restriction alone.

To make this explicit, one can introduce the ``GFT energy" $E_J := -4m_{\vec{j}}^2\Re(\alpha^+_J\overline{\alpha^-_J})$, which is a conserved quantity associated to $\chi$ translations (but whose interpretation was not specified in \cite{Oriti_2016,Gielen_lowspin}). One can easily show that \eqref{volumeMF} satisfies an effective Friedmann equation
\begin{equation}\label{EFEORITI}
	\left(\frac{1}{V}\frac{\dd V}{\dd\chi}\right)^2 = 4m_{\vec{j}}^2 \left(1- \frac{V_{\vec{j},\imath}^2 Q_{J}^2}{m_{\vec{j}}^2 V(\chi)^2}\right) + \frac{4E_{J}V_{\vec{j},\imath}}{V(\chi)} \,,
\end{equation}
where $Q_{J} :=  2m_{\vec{j}}\Im(\alpha^+_J\overline{\alpha^-_J})$ is another conserved quantity, associated with the $U(1)$ symmetry of the complex GFT. In \cite{Oriti_2016}, it was proposed that this same quantity also plays the role of conjugate momentum to the scalar field, so that by defining an effective energy density $\rho=Q_{J}^2/(2V^2)$ and $\rho_c = m_{\vec{j}}^2/(2V_{\vec{j},\imath}^2)$, one can  rewrite \eqref{EFEORITI} in a form similar to loop quantum cosmology with a term $-\rho/\rho_c$, plus an extra term proportional to the GFT energy. In these terms, $\rho_c$ would (for a fixed mode) represent a universal upper bound for the density $\rho$, which allows a bounce at a minimum nonsingular volume, just like in loop quantum cosmology. However, it is not clear why $Q_{J}$ would be interpreted as the conjugate momentum of the matter field, as it would make more sense to associate this role to $E_{J}$ (see also, e.g., \cite{Axel_Steffen_scalars} for more on this).

The large volume limit of \eqref{EFEORITI} needs to be consistent with the classical Friedmann equation $(V'/V)^2=12\pi G$, which we will review in the next section. In this case, large volumes (or low energy densities) are obtained when $|\chi|$ is big enough so that we are away from the bounce. Hence agreement with the classical theory requires the identification between the GFT coupling and Newton's constant as $m_{\vec{j}}^2=3\pi G$. This identification should be seen as the ``emergence" of Newton's constant from more fundamental GFT parameters.

Finally, we note that an effective Friedmann equation very similar to \eqref{EFEORITI}, with some additional corrections, emerges from the ``effective relational'' approach of \cite{Marchetti2021} which uses different types of coherent states and different ways of approximating the full quantum dynamics. This certainly supports the robustness of the GFT cosmology approach overall. 

\subsubsection{Deparametrised approach}

The idea of deparametrisation was introduced in GFT cosmology in \cite{toy} (see also \cite{IshaDaniele} for related ideas) where an effective Friedmann equation similar to \eqref{EFEORITI} was recovered in a simple toy model with a squeezing Hamiltonian, without requiring a mean-field approximation. In this model the Hamiltonian generates evolution with respect to a preferred matter clock. The general idea of deparametrisation, using some degrees of freedom as coordinates parametrising the others, was then used in full GFT in the Hamiltonian formalism reviewed in section \ref{BDP}\footnote{Here we are only interested in a preferred clock or time coordinate, but more generally one could couple GFT to four scalar fields and obtain relational coordinates for time and space \cite{Gielen_GFT_matterframes}.}. Initially, coherent states were used to obtain approximate solutions for expectation values of operators of interest, but this limitation was  overcome in \cite{Gielen_2020} where the dynamics were solved for operators in the Heisenberg picture, without requiring a specific state. One can then choose specific states to compare expectation values with previous results; effective Friedmann equations could be found for many types of coherent states, e.g., based on the $\mathfrak{su(1,1)}$ algebra generated by the volume and Hamiltonian operators \cite{Bojo_2019}. 

Since the Hamiltonian for the free GFT is quadratic, one can solve the Heisenberg equation \eqref{Heisenberg} to find the (relational) time evolution of the number and volume operators \eqref{NVunderlined}. Then, again for a single Peter--Weyl mode, the expectation value of the volume operator $V(\chi)\equiv V_{\vec{j},\imath}\langle \hat{\underline{N}}(\chi)\rangle$ satisfies
\begin{equation}\label{VDP}
	V(\chi) = - \frac{V_{\vec{j},\imath}}{2}+\left(V(0)+\frac{V_{\vec{j},\imath}}{2}\right)\cosh(2m_{\vec{j}} \chi) +K(0) V_{\vec{j},\imath}\sinh(2m_{\vec{j}} \chi) \, ,
\end{equation}
where $K(0) \equiv\frac{{\rm i}}{2}\langle\hat{a}_J^{\dagger2}(0)-\hat{a}_J^2(0)\rangle$ and $V(0) = V_{\vec{j},\imath}N(0) = V_{\vec{j},\imath} \langle \hat{a}_J^{\dagger}(0)\hat{a}_J(0) \rangle$ are initial conditions. The latter can clearly be interpreted as the volume at $\chi = 0$. On the other hand, a nonvanishing $K(0)$ would imply an asymmetry with respect to $\chi=0$, i.e., different pre- and post-bounce scenarios. 

While \eqref{VDP} holds for any state, requiring a semiclassical description at least at late times again motivates the use of coherent states. Rather than specifying a state at all times as in \eqref{coherentmean}, here one defines a coherent state at a given ``initial" time $\chi_0$ (usually $\chi_0=0$), but does not assume that it will remain an eigenstate of the annihilation operator at later times. Again one convenient type of states is given by Fock coherent states defined by
\begin{equation}\label{coherentHAM}
	\hat{a}_{J}(0) |\underline{\sigma}\rangle = \underline{\sigma}_{J}|\underline{\sigma}\rangle \,,
\end{equation}
where $|\underline{\sigma}\rangle$ is an eigenstate of the annihilation operator only at $\chi_0$. The expression for the expectation value \eqref{VDP} is of almost the same form as \eqref{volumeMF}, but the analogue of the ``GFT energy" is now fixed to $m_{\vec{j}}^2$. From \eqref{coherentHAM}, we can see that $|\underline{\sigma}_{J}|^2 \equiv {N} (0)$ represents the expectation value of the number operator at the initial value of our clock $\chi$; the volume at later times is obtained from \eqref{VDP}.

From \eqref{VDP}, one can find an effective Friedmann equation similar to \eqref{EFEORITI},
\begin{equation}\label{EFESTEFF}
	\left(\frac{1}{V}\frac{\dd V}{\dd\chi}\right)^2 = 4m_{\vec{j}}^2 \left( 1+\frac{V_{\vec{j},\imath}}{V(\chi)} -\frac{V_{\vec{j},\imath}^2}{V(\chi)^2}\left[N(0)^2 +N(0) - K^2(0)  \right]\right)\,.
\end{equation}
One can now again define a critical energy density $\rho_c = m_{\vec{j}}^2/(2V_{\vec{j},\imath}^2)$ and rewrite the last term inside the brackets in the suggestive form $-\rho_{\text{eff}}/\rho_c$. In general, the effective energy density $\rho_{\text{eff}}$ takes the form $\rho_{\text{eff}}=\rho_\chi(\chi)$ + corrections, where the additional terms depend on the initial conditions, but where the matter energy density $\rho_\chi(\chi)$ is now defined by identifying the Hamiltonian $\mathcal{H} \equiv\langle \hat{\mathcal{H}}_J\rangle$ (cf. \eqref{relham}) 
with the conjugate momentum of $\chi$ (as one would expect), i.e., $\rho_\chi(\chi) = \mathcal{H}^2/(2V(\chi)^2)$. 

For our purposes, the main result is that \eqref{EFESTEFF} also reduces to the classical Friedmann equation in the late-time (or large-volume) limit. Moreover, it has corrections $\sim 1/V$ and $\sim 1/V^2$, very similar to the ones of \eqref{EFEORITI} obtained by mean-field approximation in the algebraic approach.

\subsubsection{Towards anisotropic cosmologies}

We have seen that, using a restriction to a single mode, the GFT literature is rich of ways to derive effective cosmological dynamics. The results differ in numerical factors (e.g., the high energy corrections coming from the cross term $2V_{\vec{j},\imath}\Re(\alpha^+_J\overline{\alpha^-_J})=-V_{\vec{j},\imath}E_{J}/(2m_{\vec{j}}^2)$ in \eqref{volumeMF} versus $-V_{\vec{j},\imath}/2$ in \eqref{VDP}), but represent the same general scenario. Whether we follow the algebraic approach (which is how the GFT cosmology programme began) or use a deparametrised point of view, we obtain comparable effective Friedmann equations, which in particular have the same large volume limit. They are also all similar to the effective dynamics of loop quantum cosmology, and describe an effective repulsive behaviour at high energies. 

We now want to extend this discussion to anisotropic cosmologies. Classically, incorporating anisotropy means going from the FLRW Universe to the more general class of Bianchi models. We aim to define a GFT model that can generalise the results presented in this section at least to the simplest anisotropic cosmology (Bianchi I). To do this, we need to consider two new aspects: understanding how anisotropies modify the effective Friedmann equation, and understanding the dynamics of the anisotropies themselves.

An anisotropic GFT model naively requires quanta of geometry which are non-equilateral tetrahedra. Moreover, in order to obtain a nontrivial time evolution for the anisotropies we will need to lift the single-mode restriction. Heuristically, this is because if the shape of non-equilateral tetrahedra is fixed there is no room for a dynamical notion of anisotropy. Multiple modes are required, so that the relative contributions of different shapes can change and a macroscopically ``average'' anisotropy can become dynamical.

Some preliminary work on anisotropic GFT cosmology was done in \cite{de_Cesare_2017} where anisotropies were seen as perturbations. Furthermore, an anisotropic trirectangular tetrahedron was used as building block in \cite{Mairi_2017} where a notion of ``dynamical isotropisation" was found, but without details on the evolution of a (global) anisotropy parameter. This will in fact be the main novelty in our work: we define a measure for anisotropies given by a parameter (corresponding to a Misner-like variable $\beta_\pm$ in classical cosmology) emerging from fundamental quantum gravity arguments.

\section{Relational definition of cosmological dynamics}
\label{relcosmology}

In this section we briefly recall how one can obtain a relational description of cosmological models in general relativity, starting from the Hamiltonian (Arnowitt--Deser--Misner, or ADM) form of the Einstein--Hilbert action
\begin{equation}
	\label{ADMaction}
	S =\int \dd t \int \dd^3x\,\left(\pi^{ab}\dot{q}_{ab}-N \mathcal{C}-N^a \mathcal{C}_a\right)\,.
\end{equation}
In addition to the metric tensor of three-dimensional spatial slices $q_{ab}$ and its conjugate momentum $\pi^{ab}$, there are four Lagrange multipliers: the lapse function $N$ and the shift vector field $N^a$. They multiply the Hamiltonian and (spatial-)diffeomorphism constraints, defined as
\begin{equation}\label{constraints}
	\mathcal{C} =  \frac{16\pi G}{\sqrt{q}}\left(\pi_{ab}\pi^{ab}-\frac{1}{2}(\pi\indices{^a_a})^2\right)-\frac{\sqrt{q}}{16\pi G}\,\tensor[^{(3)}]{R}{}\, ,\qquad \qquad \mathcal{C}_a = -2D_b\pi\indices{^b_a} \,,
\end{equation}
where $q = \det(q)$, $\tensor[^{(3)}]{R}{}$ is the Ricci scalar of $q_{ab}$, $D_a$ is the spatial covariant derivative compatible with $q_{ab}$ and indices are raised and lowered with $q_{ab}$. The constraints \eqref{constraints} need to vanish for physical solutions, as can be seen varying the action with respect to $N$ and $N^a$ respectively.

 We will be interested in gravity coupled to a (free, massless) scalar field $\chi$ with conjugate momentum $p_\chi$, which will serve as relational clock. Moreover we will only deal with homogeneous settings, therefore we do not have an energy gradient term for $\chi$, and we can set $N^a=0$. The action then reads
\begin{equation}
	\label{generalS}
	S =\int \dd t \int \dd^3 x \,\left(\pi^{ab}\dot{q}_{ab}+p_\chi\dot{\chi}-NC\right)\,,
\end{equation}
where now the total Hamiltonian constraint has a gravitational and a matter part (denoted $\mathcal{C}_\chi$),
\begin{equation}\label{generalC}
	C = \mathcal{C}+\mathcal{C}_\chi=  \frac{16\pi G}{\sqrt{q}}\left(\pi_{ab}\pi^{ab}-\frac{1}{2}(\pi\indices{^a_a})^2\right)-\frac{\sqrt{q}}{16\pi G}\,\tensor[^{(3)}]{R}{} + \frac{p_\chi^2}{2\sqrt{q}} =0\,.
\end{equation}
If all fields are assumed to be spatially homogeneous, the integral over space $\int\dd^3 x$ just gives a constant, which we can set to unity. This does not play any physical role in homogeneous settings.

\subsection{FLRW}

We now specialise the above framework to a spatially flat, homogeneous and isotropic FLRW Universe. Given the metric 
\begin{equation}
	\dd s^2 = -N(t)^2 \dd t^2+V(t)^{2/3}\left(\dd x^2+\dd y^2+\dd z^2\right)\,,
\end{equation}
the action is simply expressed in terms of the volume $V$ (related to the familiar scale factor via $V=a^3$) and its momentum $p_V$, as
\begin{equation}\label{FLRWC}
	S =\int dt \,\left(p_V\dot{V}+p_\chi\dot{\chi}-NC\right)\,,\quad NC = -{6\pi G N Vp_V^2 } +\frac{N p_\chi^2}{2V} \,.
\end{equation}

We want to describe dynamics in relational terms, using $\chi$ as time variable. Recall that the scalar field can act as clock because it is monotonic, since $p_\chi$ is a constant of motion as $\{p_\chi,NC\}=0$. The equation of motion for $\chi$ allows to choose the scalar field to be the time variable, fixing $N$:
\begin{equation}
	\dot{\chi} = \{\chi,NC\}=\frac{Np_\chi}{V} \qquad \Leftrightarrow\qquad N\dd t = \frac{V}{p_\chi}\dd\chi\,.
\end{equation}
Now the equation of motion for the volume, $\dot{V} = \{V,NC\} = -12\pi G N V p_V$, can be reformulated in relation to $\chi$. In particular, using $\dot{\chi}$ and $\dot{V}$ one can express $p_V$ as
\begin{equation}\label{PV}
	p_V = -\frac{1}{12\pi G} \frac{p_\chi}{V^2}\frac{\dd V}{\dd\chi}\,.
\end{equation}
Substituting \eqref{PV} in the vanishing of the Hamiltonian constraint \eqref{FLRWC}, we find the relational Friedmann equation
\begin{equation}\label{FLRWEFE}
	\left(\frac{1}{3V}\frac{\dd V}{\dd\chi}\right)^2 = \frac{4\pi G}{3}\,.
\end{equation}
The large-volume limit of the isotropic GFT models introduced in the previous section is compared to \eqref{FLRWEFE}. Note that \eqref{FLRWEFE} can be obtained equivalently by finding the lapse $N$ setting $\dot{\chi}=1$, or by evaluating $\{V,\mathcal{H}\}$ where the relational Hamiltonian is $\mathcal{H} \equiv p_\chi = \pm\sqrt{12\pi G}\,Vp_V$.

\subsection{Bianchi I}

We now move our attention to a Bianchi I cosmology, with metric 
\begin{equation}\label{bianchimetric}
	\dd s^2 = -N(t)^2 \dd t^2+a_1(t)^2 \dd x^2+a_2(t)^2 \dd y^2+a_3(t)^2 \dd z^2 \,.
\end{equation}
This generalises the flat FLRW metric to the case with separate scale factor $a_i(t)$ in each Cartesian direction $i=1,2,3$. We introduce the Misner parametrisation (see, e.g., \cite{BojoBook}) with a volume variable $V=a_1 a_2 a_3$,
\begin{equation}\label{misnervar}
	\begin{aligned}
		a_1&= V^{1/3}e^{\beta_++\sqrt{3}\beta_-}\,,\\
		a_2 &= V^{1/3}e^{\beta_+-\sqrt{3}\beta_-}\,,\\
		a_3&= V^{1/3}e^{-2\beta_+}\,.
	\end{aligned}
\end{equation}
The variables $\beta_{\pm}$ represent anisotropy parameters and have their own momenta $p_{\pm}$. Using this parametrisation, equations \eqref{generalS} and \eqref{generalC} become
\begin{equation}\label{BianchiIS}
	S= \int \dd t\; \left(p_V\dot{V}+p_+\dot{\beta}_++p_-\dot{\beta}_-+p_\chi\dot{\chi}-NC\right)\,,
\end{equation}
where
\begin{equation}\label{ncBI}
		NC=-6\pi GN Vp_V^2+\frac{2\pi G}{3}\frac{Np_+^2}{V}+\frac{2\pi G}{3}\frac{Np_-^2}{V}+\frac{Np_\chi^2}{2V}\,.
\end{equation}
Notice the similarity between the anisotropy variables and the scalar field $\chi$. Their contribution to the Hamiltonian is basically the same, expect for numerical factors; at least classically, a Bianchi I Universe is not different from an FLRW Universe with free massless scalar fields. Even though we still make use of a matter clock in this paper, this equivalence suggests that the anisotropy variables $\beta_\pm$ could play the role of a clock in GFT cosmological (anisotropic) models.     

For now, as before, we use $\chi$ as a clock. Thus, from the equations of motion
\begin{equation}\label{eoms}
	\dot{V} =-12\pi G NVp_V\, , \qquad \dot{\chi} =\frac{Np_\chi}{V}\, , \qquad \dot{\beta}_\pm = \frac{4\pi G}{3}\frac{Np_\pm}{V}\,,
\end{equation}
we extract relational dynamics as follows. One can use the first two equations of motion to obtain 	$p_V$ as in \eqref{PV}, which can then be substituted into the vanishing of \eqref{ncBI}, to get
\begin{equation}\label{vchi}
	\left(\frac{1}{3V}\frac{\dd V}{\dd\chi}\right)^2 =\left(\frac{4\pi G}{3}\right)^2 \frac{p_+^2+p_-^2}{p_\chi^2}+\frac{4\pi G}{3}\,.
\end{equation}
This relational (generalised) Friedmann equation reduces to \eqref{FLRWEFE} in the isotropic case. 

Moreover, we now have anisotropy degrees of freedom, whose dynamics are obtained in a similar fashion. We use the last two equations of motion \eqref{eoms} to write the momenta
\begin{equation}
	p_\pm = \frac{3p_\chi}{4\pi G}\frac{\dd\beta_\pm}{\dd\chi}\,.
\end{equation}
Substituting these into the Hamiltonian constraint, we obtain
\begin{equation}\label{betachi}
	\left(\frac{\dd\beta_\pm}{\dd\chi}\right)^2= \left(\frac{4\pi G}{3}\right)^2 \left(\frac{9V^2p_V^2}{p_\chi^2}-\frac{p_\mp^2}{p_\chi^2}\right)-\frac{4\pi G}{3}\,.
\end{equation}
\eqref{vchi} and \eqref{betachi} can also be combined into the relational equation
\begin{equation}\label{BianchiIEFE}
	\left(\frac{1}{3V}\frac{\dd V}{\dd\chi}\right)^2 =\left(\frac{\dd\beta_+}{\dd\chi}\right)^2 + \left(\frac{\dd\beta_-}{\dd\chi}\right)^2+ \frac{4 \pi G}{3}\,.
\end{equation}
This form has the advantage that it only relies on derivatives with respect to $\chi$ rather than canonical momenta. This equation will be used as a classical comparison for our anisotropic GFT cosmological model in the large volume limit.

\subsection{Bianchi II}

For the sake of completeness, we extend the above formalism to the Bianchi II case to see how curvature affects relational (classical) dynamics. We follow the same steps as before, but now we also have a curvature term appearing in \eqref{generalC} \cite{BojoBook},
\begin{equation}\label{curva}
	{}^{(3)}R = -\frac{1}{2}V^{-2/3} e^{4(\beta_++\sqrt{3}\beta_-)}\, ,
\end{equation}
and the Hamiltonian constraint \eqref{generalC} reads 
\begin{equation}
	C = - 6\pi G V p_V^2 + \frac{2\pi G}{3}\frac{p_+^2}{V}+ \frac{2\pi G}{3}\frac{p_-^2}{V} + \frac{V^{1/3}}{32\pi G}e^{4(\beta_++\sqrt{3}\beta_-)} + \frac{p_\chi^2}{2V} =0\,.
\end{equation}
Without repeating all the steps seen in the Bianchi I case, we recall that we use the equations of motion to obtain the momenta $p_V$ and $p_\pm$. Plugging them into $C=0$ one obtains
\begin{equation}
	\left(\frac{1}{3V}\frac{\dd V}{\dd \chi}\right)^2 =\left(\frac{4\pi G}{3}\right)^2 \frac{p_+^2+p_-^2}{p_\chi^2}+\frac{4\pi G}{3}+\frac{V^{4/3}}{12p_\chi^2}e^{4(\beta_++\sqrt{3}\beta_-)} \,,
\end{equation}
and
\begin{equation}
	\left(\frac{\dd\beta_\pm}{\dd\chi}\right)^2 = \left(\frac{4\pi G}{3}\right)^2 \left(\frac{9V^2p_V^2}{p_\chi^2}-\frac{p_\mp^2}{p_\chi^2}\right)-\frac{4\pi G}{3}-\frac{V^{4/3}}{12p_\chi^2}e^{4(\beta_++\sqrt{3}\beta_-)}\,.
\end{equation}
Finally, putting everything together one can write a Bianchi II generalisation of the (relational) Friedmann equation,
\begin{equation}\label{BianchiIIEFE}
	\left(\frac{1}{3V}\frac{\dd V}{\dd\chi}\right)^2 = \left(\frac{\dd\beta_+}{\dd\chi}\right)^2+\left(\frac{\dd\beta_-}{\dd\chi}\right)^2 +\frac{4\pi G}{3}+\frac{V^{4/3}}{12 p_\chi^2}e^{4(\beta_++\sqrt{3}\beta_-)}\,,
\end{equation}
which reduces to \eqref{BianchiIEFE} when the last term (which comes from $^{(3)}R$) is zero. Notice that in an expanding Universe this term becomes dominant at later times: the dynamics are initially close to Bianchi I and deviate only when the exponential expansion of the volume takes over. Because of this, and given that the Bianchi II scenario is the simplest Bianchi model involving spatial curvature, we will also compare our anisotropic GFT cosmology to these equations later.

\section{Anisotropic GFT model}
\label{anisoGFT}

In order to study anisotropic cosmologies in GFT, we need to define a notion of ``anisotropy variables'', analogous to the Misner variables $\beta_\pm$. We will study the (relational) dynamics of these variables and observe how they affect the effective Friedmann equation for the volume $V(\chi)$. We will compare these effective dynamical equations with those of Bianchi models, in particular with the simplest Bianchi I cosmology which would be the natural extension of the spatially flat FLRW Universe previously studied in GFT. As in this isotropic case, we will follow both the algebraic and deparametrised approaches. We will once again see that they give slightly different behaviours close to the bounce, but basically match otherwise. 

We will study two different observables, representing the degrees of freedom of a classical Bianchi I cosmology: the volume $V(\chi)$  and ``average anisotropy parameters'' $\beta_\pm(\chi)$, defined as
\begin{equation}\label{Vandbetaofchi}
V(\chi) = \sum_{J} V_{\vec{j},\imath} N_{J} (\chi)\,, \qquad \qquad	\beta_\pm(\chi) = \frac{1}{N(\chi)}\sum_{J} \beta_\pm^{\vec{j},\imath} N_{J} (\chi)\,,
\end{equation}
where $N_J(\chi)$ is the expectation value for the particle number in the mode $J$. $V(\chi)$ is defined as in previous work, namely as the expectation value of \eqref{NVspincondensate} or \eqref{NVunderlined}. On the other hand, the expression for $\beta_\pm(\chi)$ is different from the usual structure of operators in GFT (cf.~\eqref{OBop}). We think of anisotropies as determined by the shape of our geometric building blocks; these variables should be ``intensive'' and not simply grow with the number of quanta, therefore we divide by the total number $N(\chi) = \sum_{J} N_{J} (\chi)$ (see, e.g., \cite{LastAxel} for a similar concern related to a possible scalar matter quantum operator). $\beta_\pm(\chi)$ is not an expectation value, and we do not propose any definition of operators $\widehat{\beta}_\pm$ representing anisotropy; the overall $1/N(\chi)$ could not arise from taking an expectation value. Instead, the definition \eqref{Vandbetaofchi} introduces a semiclassical notion of anisotropy only.

At the end of section \ref{reviewSec}, we already argued that this setup requires including multiple Peter--Weyl modes. Indeed, for only a single $J$ in \eqref{Vandbetaofchi} $\beta_\pm(\chi)$ would clearly be constant, consistent with the idea that we would not incorporate any dynamical ``shape'' degrees of freedom. Moreover, we already know that the volume emerging from a single mode would give rise to an effective Friedmann equation of the form of \eqref{EFEORITI} or \eqref{EFESTEFF}, which corresponds to an isotropic Universe. These two statements agree with the classical observation that the Bianchi I relational dynamics \eqref{BianchiIEFE} do not differ from FLRW \eqref{FLRWEFE} if the anisotropy parameters are constant, as they only appear through derivatives. Another way of seeing this is to observe that a Bianchi I metric for which the Misner variables are constant can be brought to FLRW form by rescaling coordinates. Hence, to make $\beta_\pm(\chi)$ dynamical one needs to allow for contributions coming from multiple shapes, i.e., modes with different values for $\beta_\pm^{\vec{j},\imath}$.

In order to compute the sums \eqref{Vandbetaofchi}, we need to specify the single-mode expectation values $N_{J}(\chi)$ and the coefficients $V_{\vec{j},\imath}$ and $\beta_\pm^{\vec{j},\imath}$. The volume eigenvalues $V_{\vec{j},\imath}$, as introduced in \eqref{NVspincondensate} and \eqref{NVunderlined}, are imported from LQG; a procedure to (numerically\footnote{Given that these eigenvalues can only be computed numerically there is little or no hope to be able to solve the sums \eqref{Vandbetaofchi} in a closed form. We will have to truncate them and make some suitable choices for the modes.}) compute them is well-known \cite{Bianchi_Haggard_vol,*Bianchi_Haggard_letter}, and will be recalled in appendix \ref{AppA}. 
The expectation values $N_J(\chi)$ were discussed in section \ref{reviewSec}, and we have explicit expressions for the two approaches given in \eqref{volumeMF} and \eqref{VDP}, using that for a single mode $V(\chi)=V_{\vec{j},\imath}N_{J}(\chi)$. The definitions of $\beta_\pm^{\vec{j},\imath}$, on the other hand, deserve some further elucidation.

\subsection{Defining $\beta_\pm^{\vec{j},\imath}$: the trisohedral tetrahedron}\label{definingbeta}

Unlike for quantities such as areas and volumes, there is no fundamental operator in LQG corresponding to Misner variables $\beta_\pm$. Our task is therefore to give a proposal for $\beta_\pm^{\vec{j},\imath}$ as functions of such more fundamental geometrical quantities, defined at the level of each tetrahedron. These variables (areas and volume), in turn, are determined by the spins $\vec{j}$ and intertwiner $\imath$. We think of the fundamental tetrahedra as embedded into a manifold with Bianchi I metric, and reconstruct parts of that metric from geometric quantities of a tetrahedron, as originally proposed for GFT in \cite{GFTcosmoLONGpaper}. There is clearly some ambiguity in any such procedure, given that the required embedding is not part of the GFT formalism but additional input. Our proposal is a relatively direct extension of previous work in GFT cosmology: we follow the idea \cite{Oriti_2016,*BOriti_2017} that equal spins for a four-valent node (i.e., $\vec{j} = (j,j,j,j)$)  capture a discrete notion of isotropy; hence such a configuration should correspond to $\beta_\pm=0$. Departures from this microscopic notion of isotropy are, in a sense, assumed to add up coherently to a macroscopic definition of anisotropy.

For any excitation associated to $J=(\vec{j},\vec{m},\imath)$, the spins $\vec{j}$ determine the areas of faces of a tetrahedron (see figure \ref{tetra+5graph})\footnote{This is again a statement imported from LQG, where the eigenvalues of the area operator $\widehat{A}_I$ associated to the $I$-th face of a quantum tetrahedron are given by $l_0^2\sqrt{j_I(j_I+1)}$, $l_0$ being a fundamental length scale parameter.\label{LQGareaFootnote}}. However, these areas are not sufficient to determine the shape of a tetrahedron; there is a two-sphere's worth of different tetrahedra with given face areas \cite{Bianchi_Haggard_vol}, so additional assumptions are needed to identify a given quantum state with a configuration of a classical tetrahedron. The idea of \cite{Oriti_2016,*BOriti_2017} is that for equal spins we should choose a \textit{regular} tetrahedron, whose edges are all of equal length\footnote{Such a platonic solid is specified by a single number (e.g., edge length, height, distance between opposite edges) with fixed dihedral angles between faces, \textit{et cetera}.}. To justify this assumption one fixes the intertwiner $\imath$ to maximise the volume eigenvalue, as this would represent a situation which goes as close as possible to the classical picture. More generally, we decide to focus on \textit{orthocentric} \cite{Gerber_Ortho} simplices (discussed in some detail in appendix \ref{AppA}). Orthocentric tetrahedra maximise the volume for given face areas. The regular tetrahedron is a particular example of orthocentric tetrahedra. This motivates us to always choose the largest allowed volume eigenvalue for given face areas, and hence spins $\vec{j}$; we will then interpret our states as orthocentric tetrahedra. For a few specific shapes, we explicitly show in appendix \ref{AppA} that the largest volume eigenvalue for a given mode is indeed the closest one to the classical volume of an orthocentric tetrahedron for the same face areas. In this sense, a choice for $\imath$ is dictated by the comparison that we make between GFT models and classical geometry.  

The sum over $J$ then reduces to spins $\vec{j}$ and magnetic indices $\vec{m}$ only, since for each choice of spins the intertwiner $\imath$ is already fixed in all the sums from now on. Symbolically, we are left with
\begin{equation}\label{sum}
	\sum_J = \sum_{\vec{j}}\sum_{\vec{m}}\sum_{\imath} \qquad\Rightarrow\qquad \sum_{\vec{j}}\sum_{\vec{m}} \,,
\end{equation}
since we always use the largest volume eigenvalues in such sums. Note that we still need to make sure that each combination of $j_I$ included in the sum allows for a nonvanishing intertwiner.

\begin{figure}[ht]
	\begin{center}
		\includegraphics[width=0.27\textwidth]{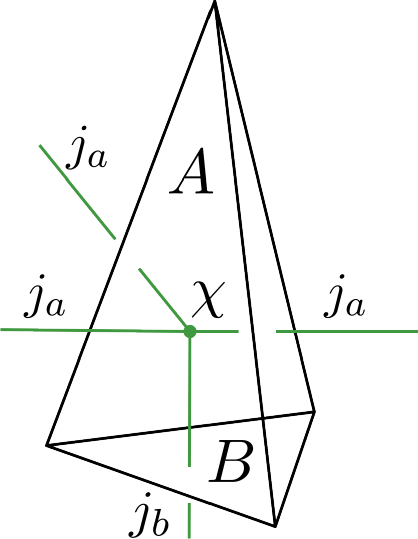}
	\end{center}
	\caption{\small The trisohedral tetrahedron has two types of areas, edges, or dihedral angles (between sides and between a side and the base), \textit{et cetera}. It generalises the regular tetrahedron, still remaining rather specific. We propose to associate a notion of anisotropy to such a non-equilateral shape.}
		\label{isosceles}
\end{figure}

A very minimal requirement for our definition of $\beta_\pm^{\vec{j}}$ (now implicitly associated to the intertwiner fixed by the $\vec{j}$) is that it should vanish for $\vec{j} = (j,j,j,j)$, but be nonzero if at least one of the four spins differs from the others. The simplest ``non-isotropic'' (but still orthocentric) building block one can think of is a tetrahedron we will refer to as {\em trisohedral}. As one can see in figure \ref{isosceles}, this too is a quite particular shape, with three isosceles triangles (called ``sides'' with area $A$) and an equilateral one (called ``base'' of the tetrahedron, with area $B$). 

Following the analogy from before, we then make the assumption that such a tetrahedron is represented by modes of the form
\begin{equation}\label{ourmode}
	\vec{j} = (j_a,j_a,j_a,j_b)\,,
\end{equation}
where the spin $j_a$ is associated with the area of the sides and $j_b$ with the base. Again, this assumption is partially justified by choosing an intertwiner that maximises the volume eigenvalue. It should be clear, however, that for given face areas this volume eigenvalue will not exactly match the classical volume of a trisohedral tetrahedron. We again refer to \cite{Bianchi_Haggard_vol} for more discussion of this in the context of LQG, and give more details in appendix \ref{AppA}.

In order to specify the numbers $\beta_\pm^{(j_a,j_a,j_a,j_b)}$, we now think of the trisohedral tetrahedron of figure \ref{isosceles} as embedded in a locally rotationally symmetric (LRS) Bianchi I spatial slice. This is a Bianchi I geometry with only one preferred direction in which the expansion (or contraction) differs from the other two directions. The (spatial) line element follows from \eqref{bianchimetric} and \eqref{misnervar}
\begin{equation}\label{LRSmetric}
	\dd l^2 = a_1^2(\dd x^2+ \dd y^2)+a_2^2 \dd z^2 = V^{2/3} e^{2\beta_+}( \dd x^2+\dd y^2)+V^{2/3}e^{-4\beta_+}\dd z^2\, .
\end{equation}
Because of the symmetry of our building block our model only needs two different scale factors, so we have set $\beta_- = 0$. Now consider a tetrahedron embedded in such a space, with one of its triangles lying on the $x-y$ plane. The tetrahedron is chosen such that it would be regular with respect to the background ``fiducial" coordinates $x$, $y$ and $z$; but its physical geometry depends on the dynamical variables\footnote{We fix the edge length $l = \sqrt{2}\sqrt[3]{3}$ such that the physical volume of the tetrahedron is equal to $V$ in \eqref{LRSmetric}.}. In particular, the areas of the equilateral base and the isosceles sides are 
\begin{equation}\label{AB}
	B = \frac{3\sqrt[6]{3}}{2}e^{2\beta_+}V^{2/3}\, , \qquad \qquad	A= \frac{\sqrt[6]{3}}{2}e^{2\beta_+}\sqrt{1+8e^{-6\beta_+}}V^{2/3}\, .
\end{equation}
We identify $B$ and $A$ with the area eigenvalues associated to the spins $j_b$ and $j_a$ (see figure \ref{isosceles}). We then see that a configuration in which $B$ and $A$ are different would be interpreted as anisotropy ($\beta_+ \neq 0$) whereas $B=A$ corresponds to $\beta_+=0$, as anticipated.

Inverting either one of equations \eqref{AB}, one can express the anisotropy parameter as a function of face areas and volume, and define this to be the ``anisotropy'' associated to the tetrahedron. In the quantum theory, $A$, $B$ and $V$ are represented as possible eigenvalues determined by the spins. This then leads us to possible proposals for how to define $\beta_+^{(j_a,j_a,j_a,j_b)}$ in (\ref{Vandbetaofchi}).

In appendix \ref{AppA} we compare different definitions for $\beta_+^{(j_a,j_a,j_a,j_b)}$ obtained from \eqref{AB}. They do not agree, because we fixed the volume of the tetrahedron to agree with the quantity $V$ in the metric assuming that we have an orthocentric tetrahedron, but there is no quantum eigenvalue satisfying exactly the same relations between volume and areas. Only one definition is simple and satisfies our desired property $\beta_+^{(j,j,j,j)}=0$; as one might have anticipated, this is a definition that does not use the volume eigenvalue at all. Indeed, from the ratio $A/B$ in \eqref{AB}, one finds $\beta_+ = -1/6 \log\left[(9A^2-B^2)/(8B^2)\right]$. Converting to LQG eigenvalues, we then define an effective local anisotropy associated to the mode \eqref{ourmode} of the quantum tetrahedron by
\begin{equation}\label{ourbeta}
	\beta_+^{(j_a,j_a,j_a,j_b)} =  -\frac{1}{6}\log \left(\frac{9j_a(j_a+1)-j_b(j_b+1)}{8j_b(j_b+1)}\right)\, .
\end{equation}
This gives zero if $j_a=j_b$ and if we assume $j \ge \frac{1}{2}$, \eqref{ourbeta} is always finite and well-defined, regardless of the volume eigenvalue for the given mode (this is not true for other definitions, see appendix \ref{AppA}) which may vanish for some spin configurations. For instance, $V_{\left(1,1,1,3\right)}=0$. Note that the anisotropy is conventionally negative when $j_a > j_b$.

With this definition, we are finally in a position to evaluate \eqref{Vandbetaofchi}. We now focus on the initial conditions needed for tackling the sums.

\subsection{Initial conditions}

Independently of the approach one follows, for a mode specified by \eqref{ourmode} the kinetic term \eqref{KKK} is characterised by
\begin{equation}\label{anisoK}
m_{\vec{j}}^2 = m^2 -M^2 \Big(3j_a(j_a+1)+j_b(j_b+1)\Big) \,.
\end{equation}
Since we would like to include a number of Peter--Weyl modes in \eqref{Vandbetaofchi} for which $m_{\vec{j}}^2>0$, the coupling $M$ needs to be small relative to the ``mass'' $m$. For any given ratio $M/m$ there will be maximum spins after which \eqref{anisoK} becomes negative; for instance, choosing $M/m = 0.1$ means that $3j_a(j_a+1)+j_b(j_b+1)<100$ for $m_{\vec{j}}^2>0$.  In the numerical analysis below, we will assume $M/m = 0.1$.

Given that we are not interested in the bounce or the connection between the contracting and expanding phases, we will be looking at $\chi$-symmetric solutions, for which the minimum of the volume is at $\chi=0$. Thus, in the general expressions in section \ref{reviewSec}, we set  $\alpha^+_J = \alpha^-_J\equiv\alpha_J$ in the algebraic approach (cf.~\eqref{solutionMF}-\eqref{volumeMF}) and $K(0)=0$ in the deparametrised formulation (cf.~\eqref{VDP}). The expectation values for single-mode number operators then become
\begin{equation}\label{NjMF}
	N^{\text{MF}}_{\vec{j},\vec{m}}(\chi) = |{\sigma_{\vec{j},\vec{m}}(\chi)}|^2 = 2 |\alpha_{\vec{j},\vec{m}}|^2 \left(1+\cosh(2m_{\vec{j}}\chi)\right) = 4 |\alpha_{\vec{j},\vec{m}}|^2 \cosh^2\left(m_{\vec{j}}\chi\right)
\end{equation}
for the algebraic method with a mean-field (MF) approximation, and
\begin{equation}\label{NjDP}
	N^{\text{DP}}_{\vec{j},\vec{m}}(\chi) = -\frac{1}{2} +\left(N^{\text{DP}}_{\vec{j},\vec{m}}(0)+\frac{1}{2}\right)\cosh(2 m_{\vec{j}}\chi) 
\end{equation}
for the deparametrised (DP) approach. In \eqref{NjMF}, $\alpha_{\vec{j},\vec{m}}$ is related to the number of quanta at $\chi = 0$ as $N^{\text{MF}}_{\vec{j},\vec{m}}(0)=4 |\alpha_{\vec{j},\vec{m}}|^2$. In the deparametrised approach, we then fix $N^{\text{DP}}_{\vec{j},\vec{m}}(0) = 4|\alpha_{\vec{j},\vec{m}}|^2$ so that the initial conditions are the same for both methods.  

We then need to fix the range of the sums in \eqref{sum}. We are interested in modes of the form \eqref{ourmode}, specified by two spins $j_a$ and $j_b$. In principle, the sums over $j_a$ and $j_b$ run from $\frac{1}{2}$ to $\infty$; but in practice, they have to be truncated because the eigenvalues are only computed numerically. Furthermore, not all combinations are allowed by $SU(2)$ recoupling theory, as not all of them allow for a nonvanishing intertwiner. Moreover, modes corresponding to zero volume are not really physically interesting and would not allow for a useful notion of anisotropy. In principle, all these considerations are to be taken into account. We will simplify these issues greatly by only keeping a few modes.

We also need to sum over magnetic indices $\vec{m}$. None of the geometric observables we are considering depend on $\vec{m}$, so the $\vec{m}$ index corresponds to a degeneracy factor for physically indistinguishable modes. Given this, we will assume that the initial condition parameters $N_{\vec{j},\vec{m}}(0)$ do not depend on $\vec{m}$, and so the coefficients $\alpha_{\vec{j},\vec{m}}\equiv\alpha_{\vec{j}}$ in \eqref{NjMF} and \eqref{NjDP} are independent of $\vec{m}$. The sums over $\vec{m}$ then just return additional multiplicative factors,
\begin{equation}\label{sumab}
 \sum_{\vec{j}}\sum_{m_a}\sum_{m_a}\sum_{m_a}\sum_{m_b} = \sum_{(j_a,j_b)} (2j_a+1)^3(2j_b+1) \, .
\end{equation}

We then still need to truncate these sums to a few chosen values for $j_a$ and $j_b$. One motivation for keeping only a small number of modes (but more than one) is the increasing arbitrariness coming from the need to specify initial conditions for each additional mode. This question of dependence on initial conditions generally affects cosmological models derived from fundamental quantum gravity, which tend to depend on some specific choice of (class of) initial states \cite{Alesci1,*Alesci2,*Alesci3,Dapor_effective_2017}.

There is then a choice of which modes to include, in particular whether $j_a$ and $j_b$ should be large or small. Here two main factors may come into play. At the \textit{kinematical} level, the quantum properties of tetrahedra show semiclassical geometrical features for large spins \cite{Brunnemann_Thiemann_Vol,PolyhedraIta,Bianchi_Haggard_vol}; large spins allow for more possible discrete (quantum) states, so that classical concepts such as orthocentric tetrahedra can be approximated closely. Indeed, the largest volume eigenvalues we choose are always smaller than the volume of the corresponding orthocentric tetrahedron, but the relative difference decreases as the spins grow. We discuss this feature in appendix \ref{AppA} for two specific examples. It would suggest keeping modes associated to large values of $(j_a,j_b)$, since the geometric picture of orthocentric tetrahedra would be closer to the volume eigenvalues used.

On the other hand, we know that for the type of GFT dynamics considered in this paper, the lowest spins will dominate the sum eventually \cite{Gielen_lowspin}, and larger spins give less important contributions to the dynamics at late times. This is why in previous work the sum was often trivialised to only the smallest spins, or to a few values. This truncation is analogous to loop quantum cosmology where the spins are usually all fixed to be $j=\frac{1}{2}$ (see \cite{BenAchour:2016ajk} for an attempt at generalisation).  Therefore, even though using large spins might make sense kinematically, when we include the dynamics we see that larger spins will quickly become insignificant for the evolution of the cosmological model. Finally, on general grounds one would expect that, for any fundamentally discrete (simplicial) approach to quantum gravity, a useful continuum limit is obtained for large simplicial complexes made of little simplices, rather than by magnified semiclassical building blocks. Thus, we will only consider a few relatively small spins. 

We now need to specify initial conditions given by the initial number of quanta $N_{j_a,j_b}(0)$. These initial conditions are of course in general arbitrary, but for concreteness we assume they follow a Gaussian distribution
\begin{equation}\label{gaussians}
	N_{j_a,j_b}(0) = \mathcal{N}  \exp\left({\frac{-(j_a-\overline{j}_a)^2-(j_b-\overline{j}_b)^2}{\sigma^2}}\right)\,,
\end{equation}
where we always set $N_{j_a,j_b}=0$ for modes with vanishing volume. For continuous parameters, such distributions allow for analytical integrations, as we will show in a toy model later. Here the spins are discrete, so we are considering a ``discrete Gaussian distribution'', assuming that the initial configuration is dominated by values around $(\overline{j}_a,\overline{j}_b)$. By setting these away from $j_a=j_b=\frac{1}{2}$, we will see some initial contribution coming from larger spins, before the lowest ones take over dynamically. Moreover, given that we effectively set to zero all terms after an arbitrary spin, it is reasonable to choose initial quanta having occupancy numbers that \textit{gradually} go to zero, as modes approach the last one. This motivates using a Gaussian also in the discrete case. All $N_{j_a,j_b}(0)$ are now determined by the peaks $\overline{j}_a$ and $\overline{j}_b$, standard deviation $\sigma$ and normalisation factor $\mathcal{N}$.

\subsection{Three modes with fixed base}\label{3modessection}

\begin{figure}[ht]
	\begin{center}
		\includegraphics[width=0.1\textwidth]{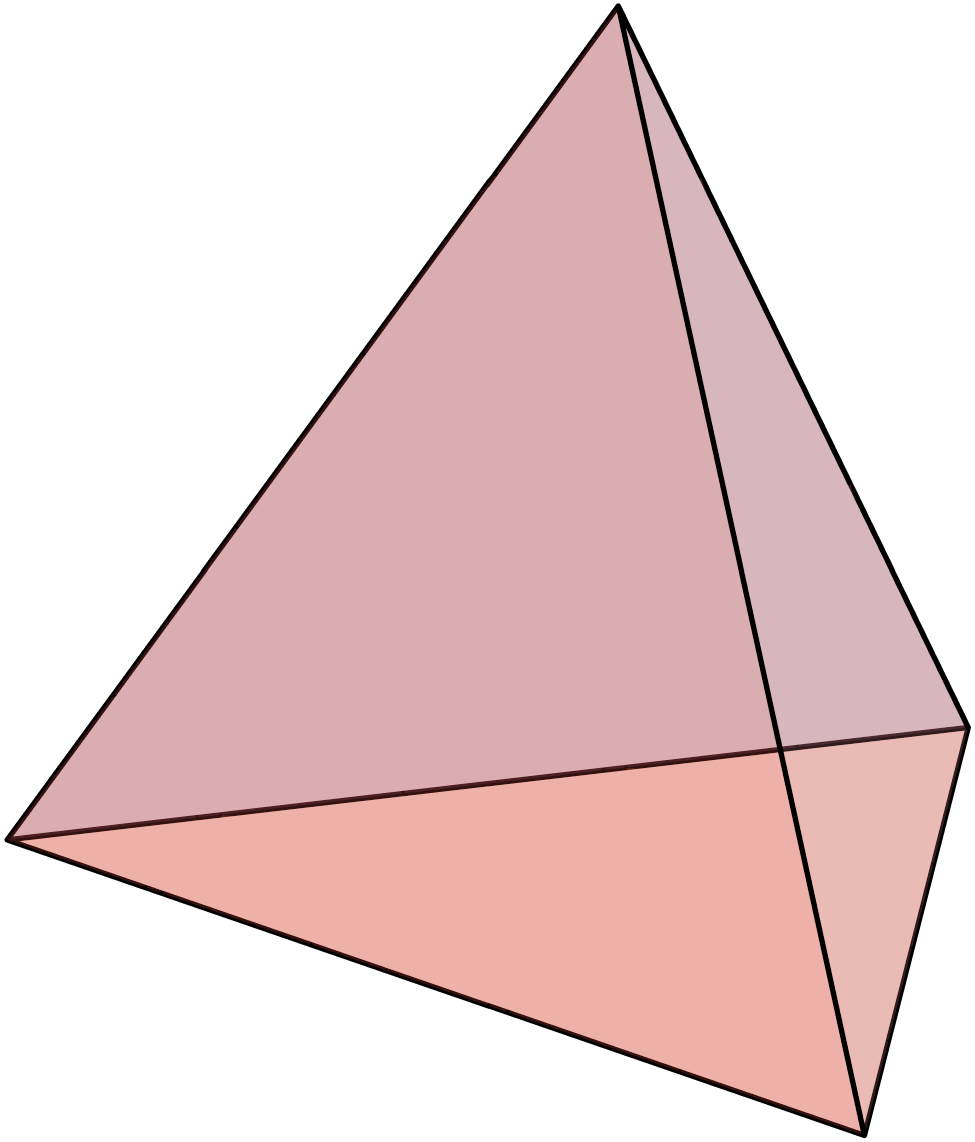}
		\hspace{2cm}
		\includegraphics[width=0.09\textwidth]{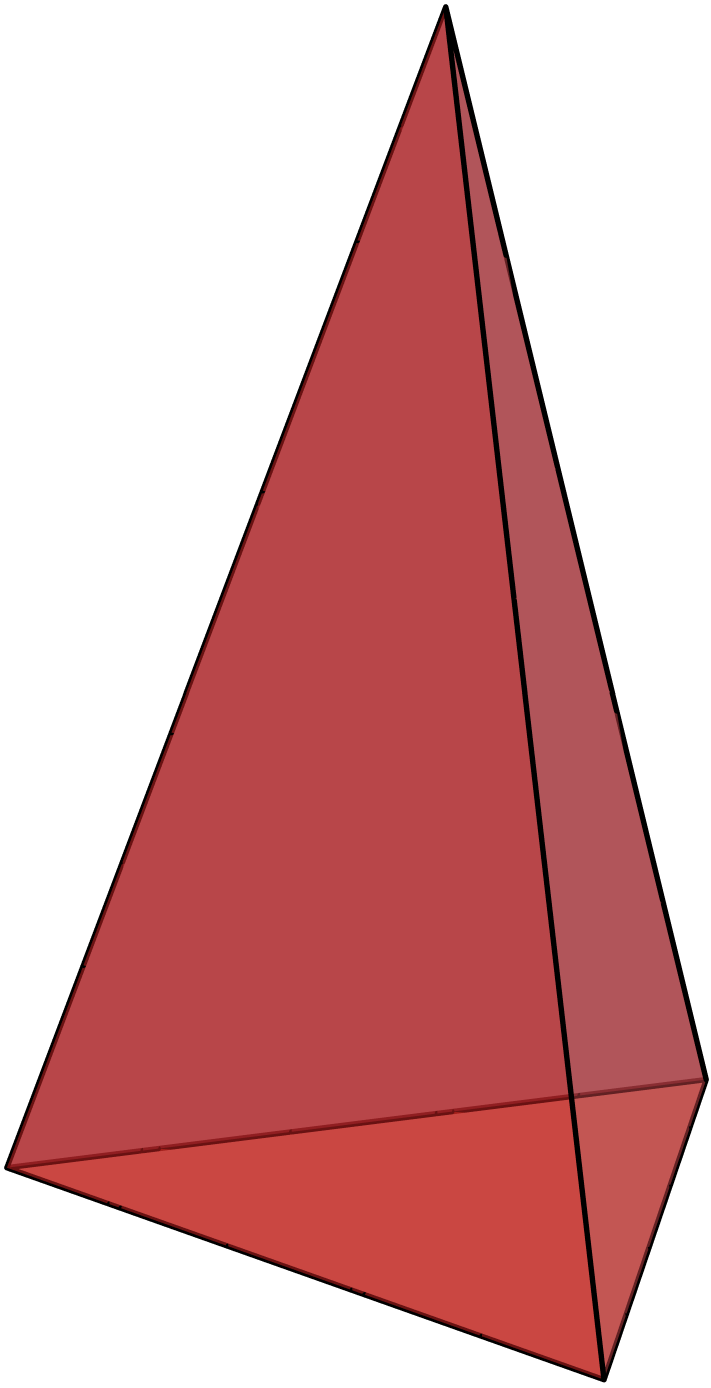}
		\hspace{2cm}
		\includegraphics[width=0.08\textwidth]{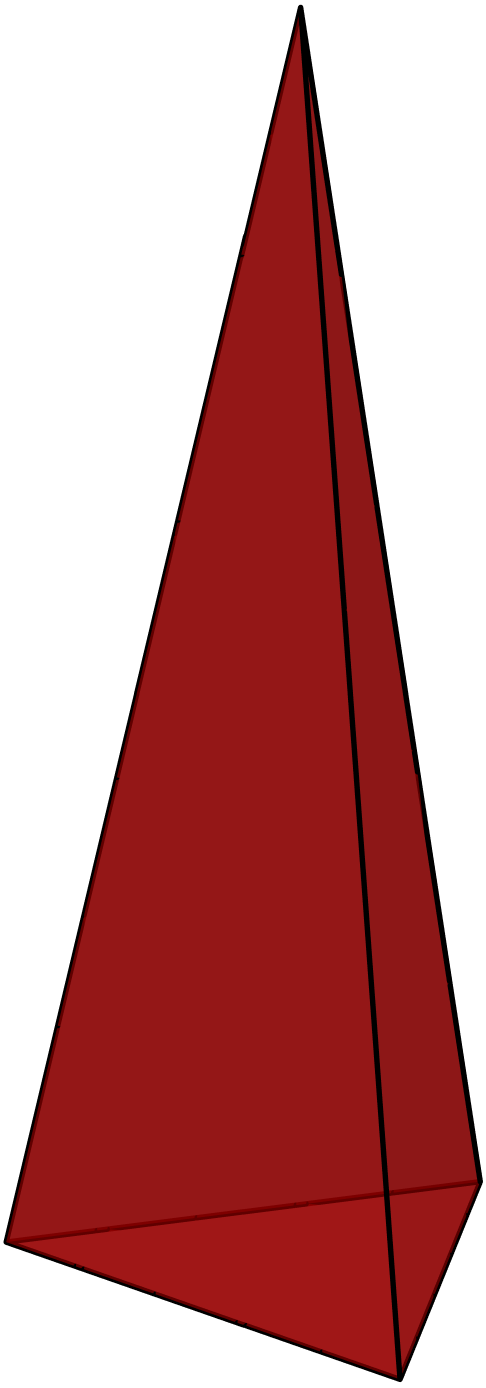}
	\end{center}
	{\small \hspace{0.3cm} $j_a=\frac{1}{2}$ \hspace{2.5cm}$j_a=\frac{3}{2}$ \hspace{2.5cm}$j_a=\frac{5}{2}$ }
	\caption{\small Three modes represent tetrahedra with the same base and three different area values for the sides. Note that the first mode corresponds to the regular tetrahedron.}
	\label{3tetra}
\end{figure}

We now focus on a simple model in which we assume the spin associated to the base of the tetrahedron to be fixed to its minimum value $j_b = \frac{1}{2}$. We keep three modes associated to the lowest (allowed) three spin values for the sides of the tetrahedron, $j_a= \{\frac{1}{2},\,\frac{3}{2},\,\frac{5}{2} \}$. This is a simple generalisation of the single-mode case, since we consider two modes which encode anisotropy (as defined in section \ref{definingbeta}) and one associated with equilateral tetrahedra (see figure \ref{3tetra}).

Denoting $V_{j_a,j_b} \equiv  V_{(j_a,j_a,j_a,j_b)}$ and $\beta_+^{j_a,j_b}\equiv  \beta_+^{(j_a,j_a,j_a,j_b)}$, and noticing that $(2j_b+1)=2$ in this case, we truncate \eqref{Vandbetaofchi} after the first three contributions obtaining
\begin{equation}\label{3V}
	V(\chi) =  2\left(2^3\, V_{\frac{1}{2},\frac{1}{2}} N_{\frac{1}{2},\frac{1}{2}}(\chi)+ 4^3\, V_{\frac{3}{2},\frac{1}{2}} N_{\frac{3}{2},\frac{1}{2}}(\chi) + 6^3\, V_{\frac{5}{2},\frac{1}{2}} N_{\frac{5}{2},\frac{1}{2}}(\chi)\right)
\end{equation}
and
\begin{equation}\label{3b}
	\beta_+(\chi)  =\frac{2}{N(\chi)}\left(\cancel{2^3\, \beta_+^{\frac{1}{2},\frac{1}{2}} N_{\frac{1}{2},\frac{1}{2}}(\chi)}+4^3\, \beta_+^{\frac{3}{2},\frac{1}{2}} N_{\frac{3}{2},\frac{1}{2}}(\chi) +6^3\,\beta_+^{\frac{5}{2},\frac{1}{2}} N_{\frac{5}{2},\frac{1}{2}}(\chi)\right)\,,
\end{equation}
where $N(\chi) = 2 \sum_{j_a=1/2}^{5/2} (2j_a+1)^3 N_{j_a,\frac{1}{2}}(\chi)$. Since $\beta_+^{\frac{1}{2},\frac{1}{2}}=0$, the first term in \eqref{3b} is zero. The other values of $V_{j_a,j_b}$ and $\beta_+^{j_a,j_b}$ can be found in table \ref{tab} at the end of the paper.

\begin{figure}[h!]
	\begin{center}
		\includegraphics[width=0.3\textwidth]{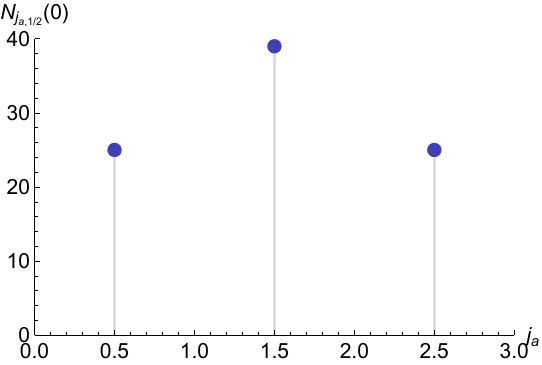}\hspace{.5cm}
		\includegraphics[width=0.3\textwidth]{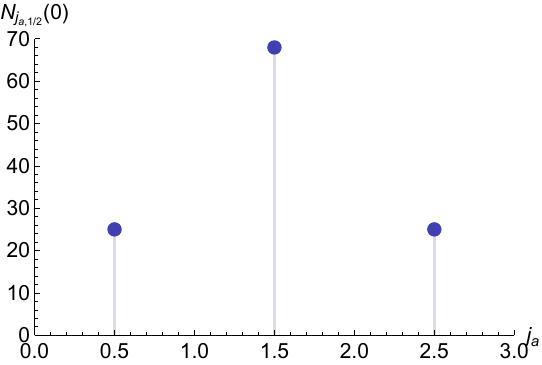}\hspace{.5cm}
		\includegraphics[width=0.3\textwidth]{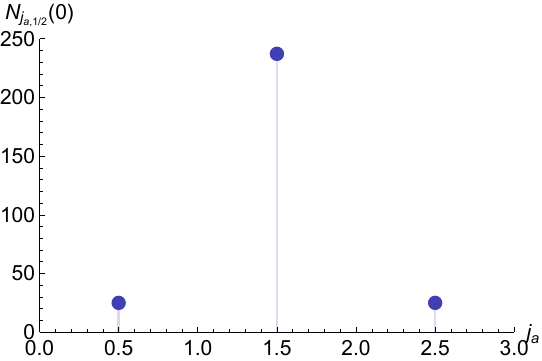}	
	\end{center}
	\caption{\small Initial conditions $N_{j_a,\frac{1}{2}}(0)$ for different standard deviations, $\sigma = \frac{3}{2}\,, 1 \,, \frac{2}{3}$. $\mathcal{N}$ is chosen such that the number of quanta at $\chi=0$ for the first and third mode is 25.}
	\label{initialconditions3}
\end{figure}

Our initial condition function \eqref{gaussians} now only depends on $j_a$. We fix the peak to be at $\overline{j}_a =\frac{3}{2}$; $\sigma$ is chosen to be around $1$ so that the initial distribution is peaked at $j_a=\frac{3}{2}$ but also includes some quanta in the other modes. $\mathcal{N}$ is fixed by the requirement that the initial state should be reasonably semiclassical; given that we only focus on expectation values of operators, quantum fluctuations should not be too large (see \cite{Gielen_2020} for a detailed analysis of relative fluctuations). As these fluctuations decrease with $N\gg 1$ (this is a generally expected behaviour also found, e.g., in \cite{LucaFluct}), we demand that initially all modes have an expected particle number of at least $25$. Figure \ref{initialconditions3} shows different $\sigma$ choices according to these criteria.

\begin{figure}[ht]
	\begin{center}
		\includegraphics[width=0.45\textwidth]{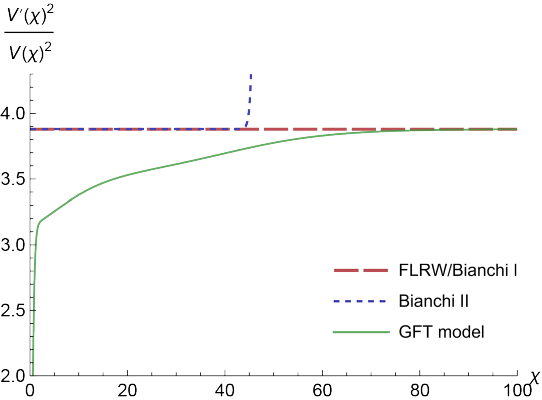}\hspace{.7cm}
		\includegraphics[width=0.46\textwidth]{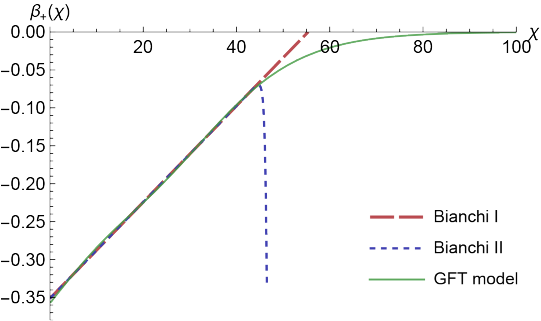}
	\end{center}
	\caption{\small Effective Friedmann equation $[V'(\chi)/V(\chi)]^2$ and evolution of anisotropy $\beta_+(\chi)$. GFT couplings are set to $m=1$, $M=0.1$, while the Gaussian parameters are fixed as in figure \ref{initialconditions3} with $\sigma=1$. Different parameter choices do not change the qualitative behaviour of these plots. 
	}
	\label{Classicalplot}
\end{figure}

In figure \ref{Classicalplot}, we plot the effective Friedmann equation $[V'(\chi)/V(\chi)]^2$ and the anisotropy evolution $\beta_+(\chi)$ stemming from \eqref{3V} and \eqref{3b}. We compare the effective Friedmann equation with the classical (relational) Bianchi dynamics reviewed in section \ref{relcosmology} (namely, solutions of \eqref{BianchiIEFE} and \eqref{BianchiIIEFE}), where we set $\beta_-=0$ because of the local rotational symmetry of our GFT model. While the Bianchi I model is a natural comparison, we also compare with Bianchi II as this is a less simple classical anisotropic cosmology, which deviates from Bianchi I only at late times (which is what the GFT description of the anisotropy does too, albeit in a different way). The other free parameters of the classical plots are fixed as follows: Newton's constant $G$ is determined by demanding that at late times we follow the isotropic solution $[V'(\chi)/V(\chi)]^2=12\pi G$; the slope $\dd\beta_+/\dd\chi$ is obtained from a linear fit of the initial part of the GFT anisotropy dynamics. 

The value of $p_\chi$ could be fixed from the fundamental GFT by using the quantity playing the role of a conjugate momentum of the scalar $\chi$. In the algebraic approach, $p_\chi = \sum_J Q_J$ with $Q_J=2m_{\vec{j}}\Im(\alpha_J^+\overline{\alpha_J^-})$ as discussed below \eqref{EFEORITI}; in the deparametrised formulation the conjugate momentum of $\chi$ is given by the (relational) Hamiltonian $p_\chi = \mathcal{H} \equiv \sum_J \langle  \hat{\mathcal{H}}_J \rangle$ as explained below \eqref{EFESTEFF}. For our states with $\alpha_J^+=\alpha_J^-$, the quantities $Q_J$ are actually zero, but one could introduce an arbitrary phase into $\alpha^+_J$ or $\alpha^-_J$ to obtain a nonvanishing $p_\chi$. In the deparametrised approach, $\mathcal{H} = \sum_{\vec{j}} m_{\vec{j}}N_{\vec{j}}(0)$ if we assume coherent states \eqref{coherentHAM} and real $\alpha_J$. In either case this gives a relatively small $p_\chi$ and in the plots Bianchi II would deviate from the linear Bianchi I behaviour very quickly, which is not what we see in the GFT model. In an attempt at obtaining a better fit, in figure \ref{Classicalplot} we fix $p_{\chi}$ by assuming that the nonlinear behaviour of Bianchi II appears roughly when the function $\beta_+(\chi)$ in GFT ceases to be linear.
 
  Notice that the constant dashed red curve in the left panel represents Bianchi I but is indistinguishable from the asymptotic FLRW limit because the classical anisotropy backreaction (cf.~(\ref{BianchiIEFE})) would be very small: it scales as $(\dd\beta_+/\dd\chi)^2 \sim 4 \times 10^{-5}$ if we take $\dd\beta_+/\dd\chi$ as the slope of the initially linear part of the GFT expression for $\beta_+(\chi)$. 

\begin{figure}[h!]
	\begin{center}
		\includegraphics[width=0.3\textwidth]{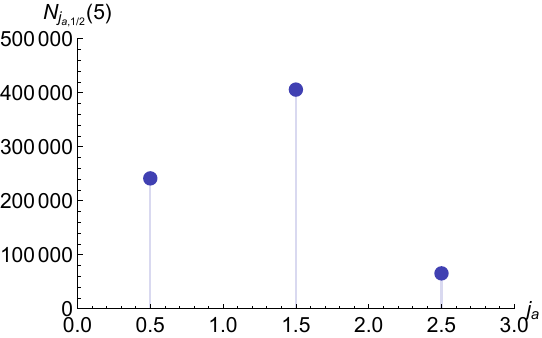}\hspace{.5cm}
		\includegraphics[width=0.3\textwidth]{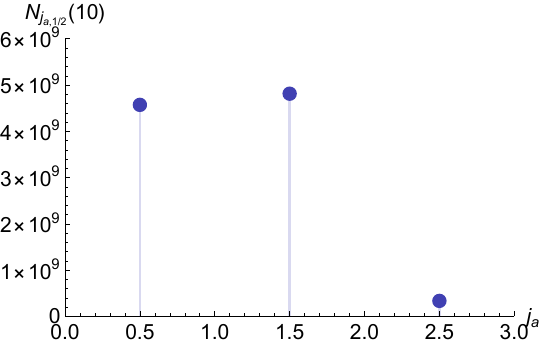}\hspace{.5cm}	
		\includegraphics[width=0.3\textwidth]{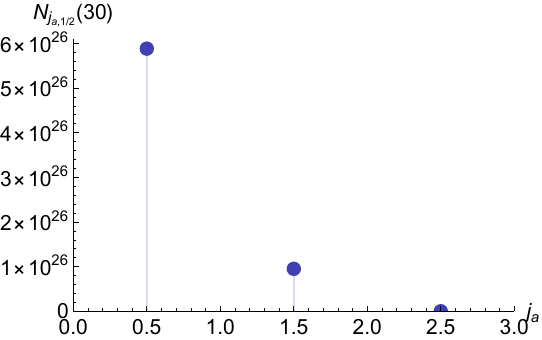}	
	\end{center}
	\caption{\small As time elapses, the peak moves towards the smallest spin until the mode with $j_a=j_b=\frac{1}{2}$ dominates forever. Soon after $\chi \sim 40$ the anisotropic contributions will become insignificant.}
	\label{domination}
\end{figure}

Figure \ref{Classicalplot} shows the dynamical ``isotropisation'' already mentioned in the literature \cite{Mairi_2017}. This late-time limit is inevitable given that, for a model involving multiple Peter--Weyl modes, the mode with the largest value that $m_{\vec{j}}$ can take will end up dominating the dynamics \cite{Gielen_lowspin}. In our case this is achieved by the smallest spins $j_a=j_b=\frac{1}{2}$; this mode dominates at some point and will then dominate indefinitely (see figure \ref{domination}). When this happens, the effective Friedmann equation reaches a constant plateau (described by the first term in \eqref{3V}) while the anisotropy gradually converges to zero (the crossed-out term in \eqref{3b} now dominates the average). The constant value taken by $(V'/V)^2$ in this asymptotic limit is then the one we would have obtained for a single-mode model corresponding to the FLRW universe. In the left-hand plot of figure \ref{Classicalplot} we label this constant value as ``FLRW/Bianchi I'' since, as we explained, the difference between FLRW and Bianchi I is too small to be noticeable in the plot.

The period before this asymptotic isotropisation (but after the bounce) can be compared with the classical Bianchi I model. In the left panel of figure \ref{Classicalplot} we see that the GFT curve does not show a constant $[V'(\chi)/V(\chi)]^2$ (with value higher than the asymptotic FLRW value), as the classical dynamics  \eqref{BianchiIEFE} would suggest. Instead, the green line shows the transition between different modes, which will eventually stop when the last mode takes over and dominates. This behaviour can be made more or less evident by changing the standard deviation $\sigma$ in \eqref{gaussians}. We will see an example with a manifest transition between modes later on.

In the right panel of figure \ref{Classicalplot} we see that $\beta(\chi)$, on the other hand, shows a much better agreement with classical (relational) cosmology. The anisotropy is approximately linear for a reasonably long time after the bounce, as would be expected from general relativity. Because of the tendency to isotropise, this agreement will obviously stop at some point as the anisotropy will approach zero; but before that point, $\beta(\chi)$ compares well with a classical Bianchi I model. It is worth reiterating here that in this construction we are ignoring GFT interactions (cf. section \ref{SecC}), which become important when $\chi$ becomes sufficiently large\footnote{When the total particle number is large, one expects correlations between the GFT quanta to be non-negligible. $N(\chi)$ grows quickly; with our choice of parameters, $N(5)\sim 10^7$, $N(25)\sim 10^{24}$ and $N(50)\sim 10^{45}$.}. Hence these plots are not to be trusted for too late times, when the weak-coupling assumption breaks down. In particular, we might never reach isotropisation but simply remain in a phase where $\beta(\chi)$ is approximatively linear, before interactions take over and change the picture completely.

Regardless of when $\beta_+(\chi)$ ceases to be linear, it is always monotonic; so in principle, as in classical cosmology, it could be used as a relational clock for GFT cosmology instead of the \textit{ad hoc} introduced scalar field $\chi$. Adding matter arbitrarily is often seen as an inescapable necessity plaguing any cosmological model coming from a background-independent quantum gravity theory, given the absence of (time) coordinates. In contrast, this model contains a gravitational degree of freedom which could be used as relational time, and perhaps help addressing issues such as the problem of time or clock dependence in quantum cosmology (for more on this see, e.g., \cite{Marchetti2021}).

\begin{figure}[ht]
	\begin{center}
		\includegraphics[width=0.4\textwidth]{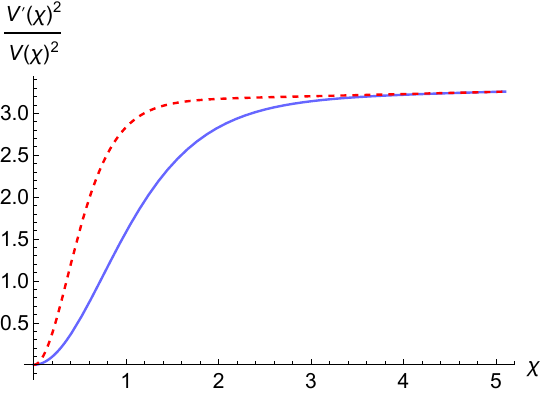}
		\includegraphics[scale=0.73]{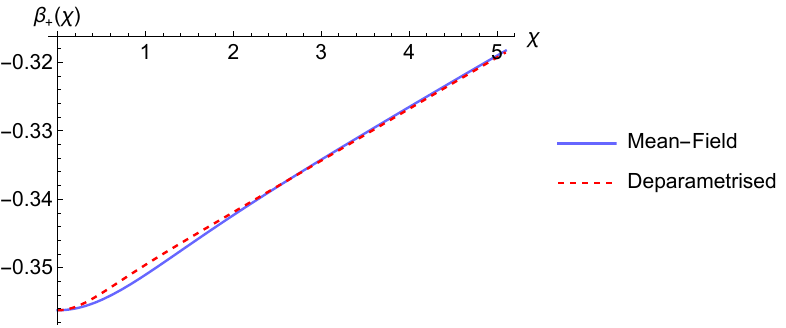}
	\end{center}
	\caption{\small Effective Friedmann equation and anisotropy dynamics at high energies directly after the bounce, for both approaches and with the same parameters (fixed as in figure \ref{Classicalplot}).}
	\label{2appro}
\end{figure}

So far, we have completely glossed over the question whether the time evolution of the particle number in each mode follows \eqref{NjMF} in the algebraic (mean-field) approach, or \eqref{NjDP} in the deparametrised approach. It turns out that, as one might have expected, the two approaches give identical results after a very short initial phase directly after the bounce, see figure \ref{2appro}. 

The discrepancy between the two approaches can be traced back to the different effective Friedmann equations for a single mode, as discussed in section \ref{SecC} (see \eqref{EFEORITI} and \eqref{EFESTEFF}). After the $1/V(\chi)^2$ contribution dominates (giving rise to the bounce), and before the constant term of the Friedmann equation takes over, there is a difference in the $1/V(\chi)$ term which in the mean-field approach depends on the arbitrary ``GFT energy'' $E_J$  (which is negative for \eqref{gaussians} and our choice of parameters), whereas in the deparametrised approach it is fixed and positive. The plot for $\beta_+(\chi)$ also shows a minor difference between the two approaches. Changing initial conditions changes the details of these differences, but the underlying features are always the same; and given that we are interested in comparing later times (larger volumes) with classical models, these differences do not play a role in our analysis. Already for these small values of $\chi$ we observe an almost-constant $[V'(\chi)/V(\chi)]^2$ and quasilinear $\beta_+(\chi)$.

\subsection{Mean-field continuous toy model}

As discussed, the general expressions \eqref{Vandbetaofchi} must be evaluated numerically, even in simple cases such as the model of section \ref{3modessection}. In order to obtain some analytical result, one can consider toy models which allow explicit solution due to some simplifying assumptions. The model presented in this section uses the solutions in the mean-field approximation of the algebraic approach (\ref{NjMF}), but drops the assumption of discrete spins $\vec{j}$. 

Recall that the classical relations \eqref{AB} relate the two types of area (for the base $B$ and the sides $A$ of the trisohedral tetrahedron) with the volume and the anisotropy parameter,
\begin{equation}
	B = B(V,\beta_+)\,, \qquad \qquad A = A(V,\beta_+)\, .
\end{equation}
Hence, classically the pair ($V$, $\beta_+$) is in one-to-one correspondence with the face areas ($A$, $B$), and it would be easier to take these variables as the basic characterisation of a tetrahedron, expressing $A$ and $B$ as functions of them. We will do this here, and forget about the fact that in LQG (and GFT) the fundamental variables are discrete areas $A$, $B$.

If we focus on $V$ and $\beta_+$, we can make another simplifying assumption: we assume that the volume per tetrahedron is simply fixed to be $V=V_0$, and that only $\beta_+$ varies \textit{continuously}. We follow the picture coming from single-mode truncations of GFT that the evolution of the total volume comes only from adding or removing building blocks, rather than changing their ``size''; but we still have a range of different ``shapes''. This avoids the previously inevitable mixing of effects coming from modes with both different volume eigenvalues and different values of $\beta_+$.

We want to work with the same GFT kinetic term, i.e., the same expression (\ref{anisoK}) for the effective couplings for each mode. But given that we no longer have discrete spins, we need to rewrite this expression using $V$ and $\beta_+$. We will do this by using the LQG relation $A_I^2 = l_0^4 j_I(j_I+1)$ for area eigenvalues (see footnote \ref{LQGareaFootnote}) and the classical relations \eqref{AB}. We can then write down a relation between the discrete and toy models for modes of the form \eqref{ourmode},
\begin{equation}\label{usefulrelation}
 3j_a(j_a+1)+j_b(j_b+1) \; \dot{=} \; \frac{ 3A^2 + B^2}{l_0^4} = \frac{(3V_0)^{4/3}}{l_0^4}\left(e^{4\beta_+}+2e^{-2\beta_+}\right)\,.
\end{equation}
The symbol $\dot{=}$ means that we are equating quantum eigenvalues with classical quantities using the LQG interpretation given to the spins. This relation is then substituted into the time-dependent expression of the average particle number \eqref{NjMF}, with initial conditions fixed by a suitable choice of $\alpha$ which we again take to be a Gaussian, $\alpha_{\beta_+} = \exp\{-(\beta_+-\overline{\beta_+})^2/(2\sigma^2)\}$. Given that we do not have discrete modes, we do not have to worry about a specific normalisation factor. This is now a continuous normal distribution, the analogue of \eqref{gaussians}. 

\begin{figure}[h!]
	\begin{center}
\includegraphics[width=0.4\textwidth]{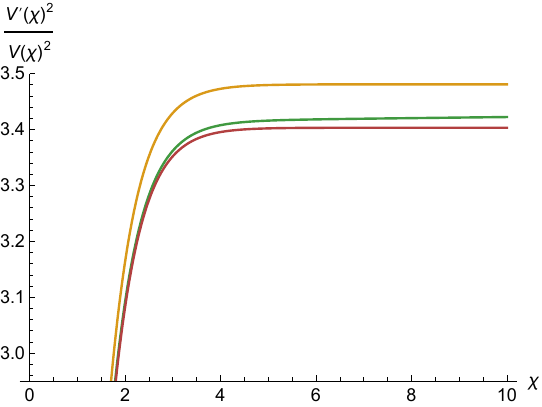}
\includegraphics[scale=0.78]{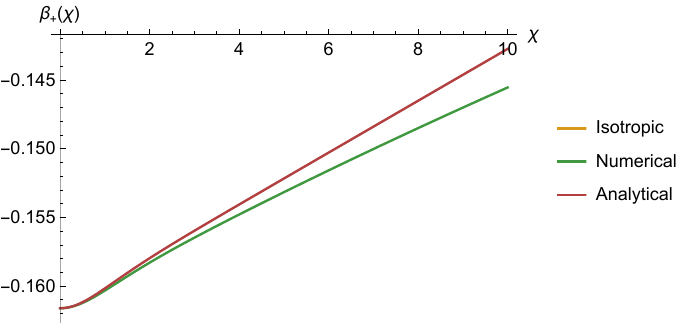}
	\end{center}
	\caption{\small The effective Friedmann equation $(V'/V)^2$ quickly becomes constant in the analytical approximation and grows slowly in the numerics, but always gives values lower than the isotropic FLRW case (orange line). The right panel shows that the anisotropy is well approximated by the linear behaviour seen in general relativity. The late-time limit shows isotropisation: $(V'/V)^2$ tends to the orange constant and $\beta_+(\chi)$ goes to zero. Here we fixed $m=1$, $M=0.1$, $V_0/l_0^3=1$, $\overline{\beta_+} = -0.1616$ and $\sigma = 0.15$.}
	\label{toyefe}
\end{figure}

The ``toy model counterparts" of \eqref{Vandbetaofchi} then read
\begin{equation}\label{intergalvolume}
	V (\chi) =  V_0 \int \dd\beta_+ \; \alpha_{\beta_+}^2 \cosh^2(\sqrt{m^2-M^2(3V_0/l_0^3)^{4/3}\left(e^{4\beta_+}+2e^{-2\beta_+}
		\right)}\;\chi)
\end{equation}
and
\begin{equation}\label{integralbeta}
	\beta_+(\chi) = \frac{V_0}{V(\chi)}\int \dd\beta_+ \;\beta_+  \alpha_{\beta_+}^2 \cosh^2(\sqrt{m^2-M^2(3V_0/l_0^3)^{4/3}\left(e^{4\beta_+}+2e^{-2\beta_+}
		\right)}\;\chi)\,.
\end{equation}
Numerical evaluation of $[V'(\chi)/V(\chi)]^2$ and $\beta_+(\chi)$ shows qualitatively similar behaviour to the plots in figure \ref{Classicalplot} derived in the previous three-mode model; see figure \ref{toyefe} for an example. Again we see an asymptotic ``isotropisation'' leading to the effective FLRW limit at late times, and an approximately linear evolution for the parameter $\beta_+$. We can now also strengthen these results by analytical approximations.

In order to find analytical solutions to the integrals \eqref{intergalvolume} and \eqref{integralbeta} we assume small anisotropy, i.e., a Gaussian distribution with $\overline{\beta_+}\ll 1$. This justifies approximating the $\cosh$ function under the integral up to second order in $\beta_+$, 
\be
\cosh^2(\sqrt{\ldots}\,\chi)\approx \cosh^2\left({\bf m}\chi\right) - 12 \beta_+^2 \frac{M^2(3V_0/l_0^3)^{4/3}\cosh({\bf m}\chi)\sinh({\bf m}\chi)}{{\bf m}}\,\chi
\ee
where ${\bf m}:=\sqrt{m^2-3M^2(3V_0/l_0^3)^{4/3}}$. This approximation leads to integrals over $\beta_+$ that can be done immediately, and relatively simple expressions such as
\be
V(\chi)\approx \sqrt{\pi}\,V_0\,\sigma\left(\cosh^2\left({\bf m}\,\chi\right)-\frac{3M^2(3V_0/l_0^3)^{4/3}(2\overline{\beta_+}+\sigma^2)\,\chi\,\sinh(2\,{\bf m}\,\chi)}{{\bf m}}\right)\,.
\ee
To simplify these further we then use the fact that $M/m$ is small (see discussion below \eqref{anisoK}) and expand $\beta_+(\chi)$ and $\left[V'(\chi)/V(\chi)\right]^2$ up to second order in $M/m$. This yields a simple analytical expression for the anisotropy,
\begin{equation}\label{betaM}
	\beta_+(\chi)\approx	\overline{\beta_+} -\frac{ 12 M^2  {(3 V_0/l_0^3)}^{4/3} \overline{\beta_+}\sigma ^2}{m} \chi  \tanh (m \chi )\, ,
\end{equation}
and the effective Friedmann equation
	\begin{equation}
	\left(\frac{V'(\chi)}{V(\chi)}\right)^2 \approx 4m^2\tanh^2(m\chi) -6 M^2 (3V_0/l_0^3)^{4/3} \left(4 \overline{\beta_+}^2+2 \sigma ^2+1\right) \mathcal{F}(m\chi) \,,
\end{equation}
where $\mathcal{F}(m\chi) := \text{sech}^2(m\chi)\left(2m\chi+\sinh(2m\chi)\right)\tanh(m\chi)$.
We can see that the anisotropy \eqref{betaM} is essentially a linear function of $\chi$ very soon after the bounce, with a slope dependent on parameters of our Gaussian such as $\overline{\beta_+}$ and $\sigma$. Moreover, noticing that $\mathcal{F}(m\chi)\overset{|\chi|\rightarrow\infty}{\rightarrow} 2$, we also obtain a constant late-time limit for the effective Friedmann equation,
\begin{equation}\label{efetoy}
	\left(\frac{V'(\chi)}{V(\chi)}\right)^2 \; \overset{|\chi|\rightarrow\infty}{\sim} \;	4 \left[m^2-3 M^2 {(3V_0/l_0^3)}^{4/3} \left(4 \overline{\beta_+}^2+2 \sigma ^2+1\right)\right]\, .
\end{equation}
By comparing with exact numerical results we can see that these approximations cannot be trusted for too large $\chi$, see figure \ref{toyefe}.

In the limit in which we ``switch off'' anisotropic contributions, $\overline{\beta_+}\rightarrow 0$ and $\sigma\rightarrow 0$, \eqref{betaM} vanishes and in \eqref{efetoy} we have $(V'/V)^2 \rightarrow 4m^2- 12M^2(3V_0/l_0^3)^{4/3}$ which is nothing but the orange line in figure \ref{toyefe}. In \eqref{usefulrelation} this corresponds to $4j(j+1) = 3(3V_0/l_0^3)^{4/3}$ when the spins are all equal.

As a final comment, recall that in the classical Bianchi I model the Friedmann equation gets a constant contribution $(\dd\beta_+/\dd\chi)^2$ compared to the FLRW Universe. We can ask whether the terms in \eqref{efetoy} that depend on $\overline{\beta_+}$ and $\sigma$ are related to the derivative of \eqref{betaM}. But we see that
\begin{equation}
	\left(\frac{\dd\beta_+}{\dd\chi}\right)^2 \; \overset{|\chi|\rightarrow\infty}{\sim} \;  \frac{144 M^4 (3V_0/l_0^3)^{8/3} \overline{\beta_+}^2 \sigma^4}{m^2} \neq -	\frac{4}{3} {M^2 \,{(3V_0/l_0^3)}^{4/3}} \left(4 \overline{\beta_+}^2+2 \sigma ^2\right)\, .
\end{equation}
As already noticed in the full GFT model, contrary to what happens in general relativity, in our toy model the presence of anisotropy decreases $(V'/V)^2$. The two quantities we are comparing also are of different orders of magnitude, in particular different powers of the small ratio $M/m$. 

Our toy model could reproduce the main qualitative features seen in the full GFT analysis, in particular a nearly linear growth in the anisotropy for a range of $\chi$ and a negative contribution to the effective Friedmann equation.

\subsection{Including more modes into GFT models}

\begin{figure}[h!]
	\begin{center}
		\includegraphics[width=0.48\textwidth]{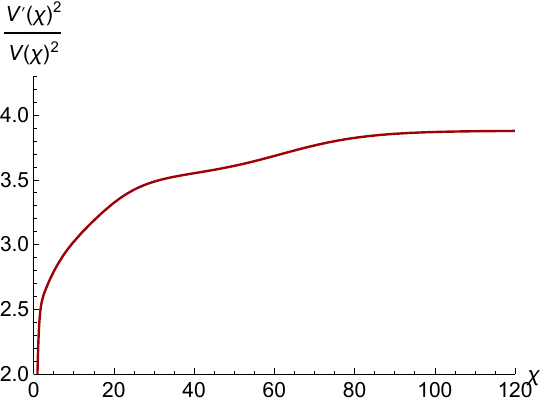}\hspace{2mm}
		\includegraphics[width=0.48\textwidth]{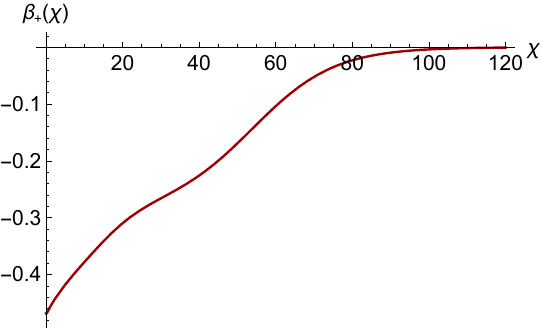}
	\end{center}
	\caption{\small The main features encountered in figure \ref{Classicalplot} are reproduced in this model: $\left[V'(\chi)/V(\chi)\right]^2$ shows shifts between modes before reaching the plateau for $j_a=j_b=\frac{1}{2}$, and $\beta_+(\chi)$ can be approximated by a linear function for some time before going to zero. Initial conditions and parameter choices are shown in figure \ref{icmm}.}
	\label{9modes}
\end{figure}

We now return to the setting of GFT, presenting results for models that go beyond the simple case described in section \ref{3modessection}. The easiest extension of what we showed before is to include more than three modes, but keep the assumption of a fixed base area (i.e., $j_b=\frac{1}{2}$). This means we add shapes to figure \ref{3tetra} for greater $j_a$ which are increasingly more stretched (``anisotropic''). We find that such an extension gives results that are not qualitatively different from the previous case, regardless of how many modes of this form we add. In figure \ref{9modes} we show the case of five modes, letting $j_a$ range between $\frac{1}{2}$ and $\frac{9}{2}$.

Since $j_b=\frac{1}{2}$ for all modes, we can see from (\ref{ourbeta}) that the  values of $\beta_+$ for the modes we consider all have the same sign, and are increasingly more negative as $j_a$ grows. This is why $\beta_+(\chi)$ does not lose its monotonicity property in figure \ref{9modes}. From the geometrical point of view, the shapes we are adding are progressively more stretched along one axis so that their local anisotropy never flips the direction in which it changes.

\begin{figure}[h!]
	\begin{center}
		\includegraphics[width=0.4\textwidth]{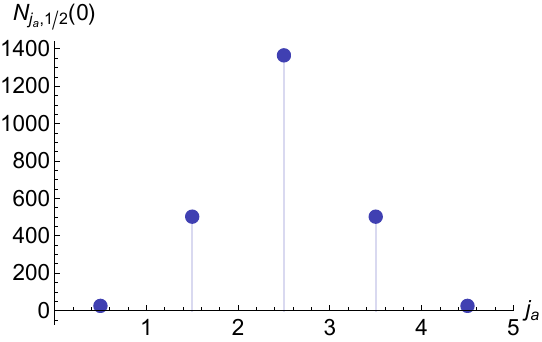}\hspace{.5cm}
		\includegraphics[width=0.5\textwidth]{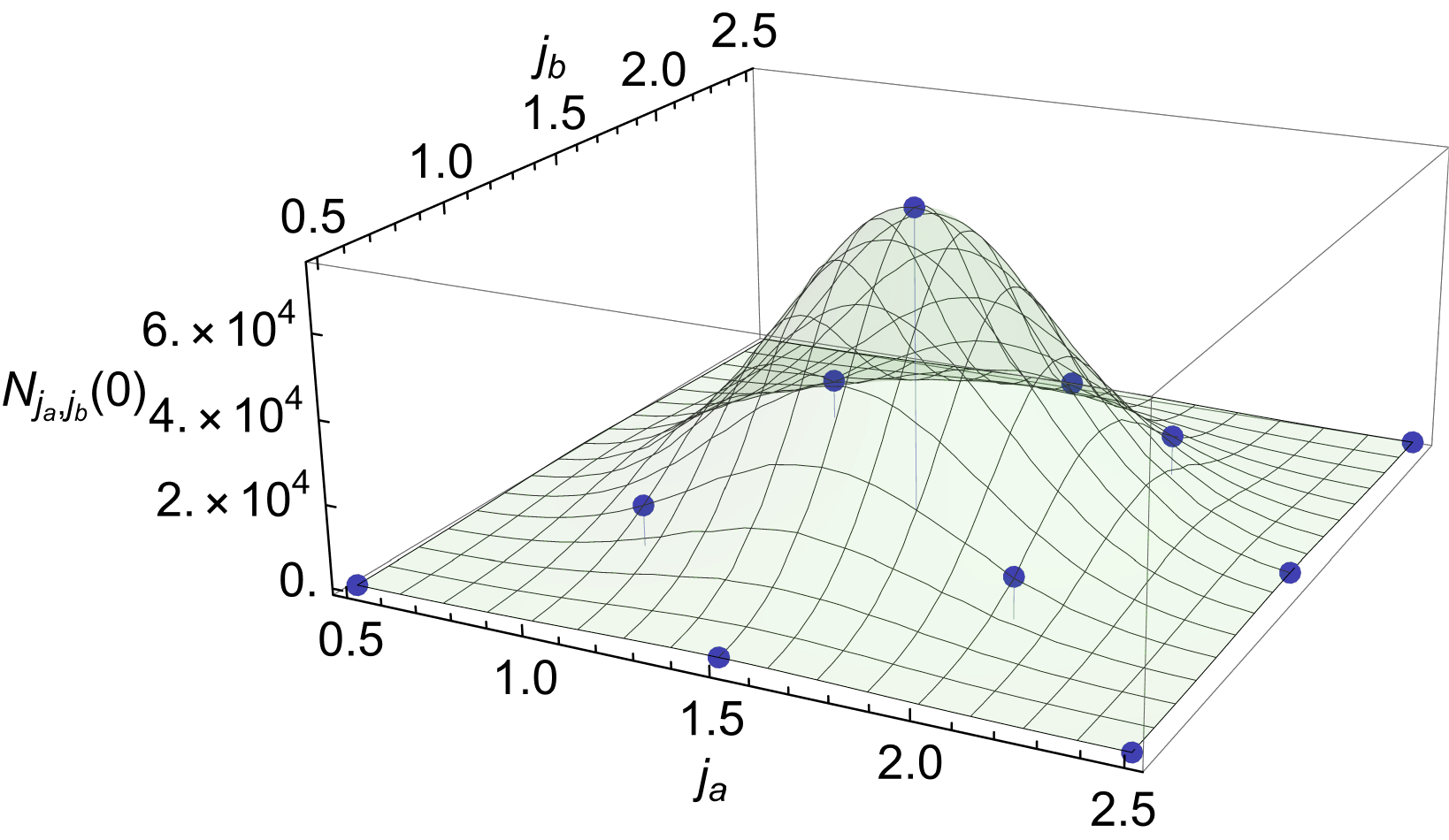}
	\end{center}
	\caption{\small Initial conditions (defined as in \eqref{gaussians}) for five and for eleven modes. As in figure \ref{initialconditions3}, $\mathcal{N}$ is such that the number of quanta is $25$ in the modes which are the farthest from the peak. Left: $j_b=\frac{1}{2}$, $j_a\in \{\frac{1}{2}, \dots,  \frac{9}{2} \}$, $\overline{j_a}=\frac{5}{2}$ and $\sigma=1$. Right: $(j_a,j_b)\in \{\frac{1}{2}, \dots, \frac{5}{2} \}$, $\overline{j_a}=\overline{j_b}=\frac{3}{2}$ and $\sigma=\frac{1}{2}$. Combinations of spins with zero volume are not included.}
	\label{icmm}
\end{figure}

\begin{figure}[h!]
	\begin{center}
		\includegraphics[width=0.45\textwidth]{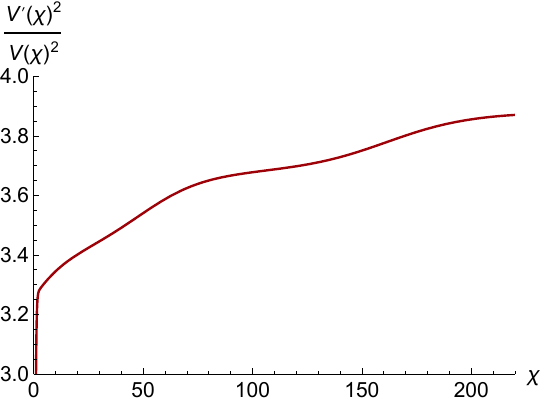}\hspace{2mm}
		\includegraphics[scale=.92]{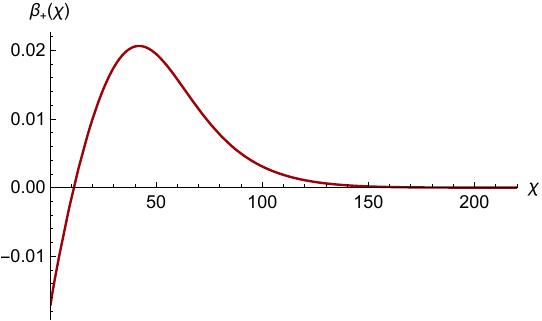}
	\end{center}
	\caption{\small Shifts between modes in $\left[V'(\chi)/V(\chi)\right]^2$ are accentuated thanks to a smaller standard deviation in \eqref{gaussians}, $\sigma=\frac{1}{2}$. Anisotropy can decrease because these modes have a non-monotonic sequence of dominant $\beta_+^{j_a,j_b}$ values as $\chi$ elapses. Initial conditions and parameter choices are shown in figure \ref{icmm}.}
	\label{17modes}
\end{figure}

A more important generalisation of our model can be obtained by relaxing the assumption that $j_b$ is fixed to $\frac{1}{2}$. If $j_b$ can vary, the first minor novelty comes from the fact that we now have some combinations of spins which need to be removed; these are spin configurations for which volume eigenvalues are zero. For instance, when $j_b=3j_a$ the volume eigenvalue vanishes (see table \ref{tab} and appendix \ref{AppA}) as one might expect from classical arguments (the tetrahedron of figure \ref{isosceles} flattens into a plane when $B=3A$). A second key novelty comes from the fact that not all the $\beta_+^{j_a,j_b}$ parameters have the same sign: in the dynamical evolution, we now no longer obtain a strictly monotonically increasing sequence of $\beta_+^{j_a,j_b}$ values associated to the modes dominating at different times. Hence, we generically find a non-monotonic $\beta_+(\chi)$. We show in figure \ref{17modes} an example with eleven generic modes, defined by letting both spins vary in the range $\{\frac{1}{2},\dots,\frac{5}{2}\}$ and excluding the ones not allowed by $SU(2)$ recoupling theory. See right panel of figure \ref{icmm} for a depiction of the initial number of quanta in the allowed modes. One could change initial conditions such that $\beta_+(\chi)$ is monotonic even with variable base spin $j_b$, by choosing \textit{ad hoc} modes such that the succession of dominant $\beta_+^{j_a,j_b}$ values goes to zero monotonically.

While any number of modes can be included without any computational obstacle, we do not report additional details on these many-mode scenarios because they are characterised by a larger arbitrariness encoded in further initial conditions, without introducing important novelties.

\section{Conclusions}
\label{conclusions}

This paper is a first attempt at characterising homogeneous but anisotropic (Bianchi) cosmologies in group field theory (GFT), focusing on the simplest case given by a Bianchi I model with an additional local rotational symmetry. Previous work on GFT cosmology had included anisotropies perturbatively, showing that for a certain class of kinetic terms they undergo a ``decay process'' leading to isotropisation. However, what was largely missing was a precise definition that would allow to discern whether an effective cosmology in GFT is isotropic or not, and to quantify the amount of anisotropy. Our work fills this gap in the literature by proposing one particular measure of anisotropy and by studying its dynamics. This is done within the context of simplified models which share the main setup and basic premises (such as neglecting interactions and making assumptions on the form of kinetic term) of previous work in GFT cosmology.

Inspired by Misner's parametrisation of the Bianchi I model, which introduces anisotropy parameters $\beta_\pm$ behaving as free massless scalar fields on a flat FLRW background, we defined an analogue of the $\beta_\pm$ variables in GFT. Given that GFT inherits from loop quantum gravity notions of geometry which are discrete, the construction of analogue anisotropies requires some care. Our proposal is to define a quantity which to some extent follows the structure of expectation values of GFT operators, assigning a microscopic value to anisotropy at the level of each Peter--Weyl mode which is multiplied by the number operator for this mode, but then dividing by the total number of quanta in order to obtain an ``intensive'' quantity. The question is then what function of LQG eigenvalues for areas and volumes should be used to effectively describe the anisotropy of a quantum building block. This function is analogous to eigenvalues of LQG operators used, e.g., to define a total volume in the usual treatment.

Having discussed various options for such an effective notion of anisotropy, and defined one which satisfies the demand that a tetrahedron with equal face areas is considered isotropic, we then studied the dynamics of the expectation value of the total volume and of the newly introduced anisotropy observable (due to the additional rotational symmetry, there is only one anisotropy degree of freedom). We were mostly interested in a relative late-time regime far away from the GFT bounce, where one would hope to see reasonable semiclassical physics emerge. At very late times, previous results already suggested a process of isotropisation and reduction to effective FLRW cosmology, so the regime we are interested here is before this final stage.

For isotropic cosmology, restricting the analysis to a single Peter--Weyl mode is enough to obtain a simple cosmology that reduces to general relativity at large volume while resolving the singularity by a bounce. However, such a restriction is too drastic to allow for dynamically evolving anisotropies. We hence considered various simple models which include multiple modes.

One model includes three Peter--Weyl modes, one describing equilateral (isotropic) and the other two describing trisohedral (anisotropic) tetrahedra. Here we found partial agreement with general relativity: while the evolution of the volume does not show the anisotropy backreaction that one would expect classically, the dynamics of anisotropy match the linear evolution dictated by classical relativity quite well. The behaviour one would expect for the volume classically is a faster rate of expansion when anisotropies are present compared to the isotropic case; but this can never be matched in the type of GFT model we consider, with a certain kind of kinetic term, since the fastest rate of expansion is reached for equilateral (isotropic) tetrahedra. This is precisely also why asymptotically only equilateral tetrahedra dominate and we see isotropisation at very late times. Hence, something more drastic than simply changing initial conditions or, e.g., considering more modes would be needed to find agreement with classical relativity. For instance, one could use different types of kinetic term, or include the effects of GFT interactions. Interactions will always dominate at some point as the Universe grows to a certain volume, and so the asymptotic isotropisation seen in a noninteracting approximation may not be physically relevant. Despite all this, the anisotropy observable of our model shows exactly the classically expected behaviour for some period of time before isotropisation, which is promising.

We showed that one can include more modes into the model, and depending on the additional assumptions one makes this can lead to slightly different results. In particular, if one of the four spins (associated to the base of the tetrahedron) is fixed, including many modes gives qualitatively similar results; but if one relaxes that assumption, one can in general obtain a non-monotonic evolution for the anisotropy.

Our main results could only be obtained numerically, and the origin of the exactly linear evolution of the anisotropy was not transparent. To gain further insights we also studied an analytical toy model with further simplifications; assuming that anisotropy is a continuous parameter while the volume per tetrahedron is fixed, we could find expressions for the effective Friedmann equation and anisotropy dynamics which reproduce the main features of the discrete GFT models. Here we found that the toy model expression for $\beta_+$ shows a linear behaviour soon after the cosmological bounce, as required by general relativity and as seen in the more refined numerical analysis.

Let us mention possible developments stemming from this work. As already pointed out, one could ask whether the effective dynamics of anisotropies change if GFT interactions are taken into account. This would complicate the resulting calculations further, beyond the need for multiple Peter--Weyl modes (see, e.g., \cite{Gielen_2020} for an already quite involved numerical study of a single mode with interactions). A different direction would be to analyse the detailed effects of anisotropy on the bounce phase. In this work we focused on a post-bounce regime because we aimed to compare with classical relativity, but in general one might ask whether the bounce itself could be spoiled (or in general modified) by anisotropies, as is often a main worry in bounce scenarios in which anisotropies dominate asymptotically on approach to the singularity. In our model, we have a massless scalar field as matter, which would classically prevent the domination of anisotropies. We saw that anisotropies are present at early times and disappear at late times, but singularity avoidance does not seem affected by the inclusion of anisotropies. The details of the bounce may still be altered by their presence. Another line of investigation, closely related to the previous one, would be to compare the exact details of the cosmological bounce with the similar singularity resolution of loop quantum cosmology. In the Bianchi I context, this would mean to investigate whether the anisotropic nature of the model has the same influence (if any) on the bounce in loop quantum cosmology \cite{AshtekarEdBianchiI,*BojowaldMisner} and in our work.

One restriction of this paper was that we were studying the GFT analogue of a Bianchi I Universe with an additional rotational symmetry, so that there is only one $\beta$ variable rather than two. To lift this restriction, one could study more general types of tetrahedra rather than trisohedral ones, discuss different proposals for $\beta_\pm$ variables in this more general context, and study their dynamics along the lines we have discussed. We would anticipate additional ambiguities in such a process on top of the ones mentioned in the appendix, and perhaps not many additional insights beyond the ones here, given that both $\beta_\pm$ variables appear on the same footing in the Bianchi I model.

Finally, given the monotonic evolution of the new anisotropy observable $\beta_+$ in GFT, one may hope to use such a gravitational degree of freedom as relational clock. This would be similar to what happens at the classical level, where in the simplest Bianchi I Universe without matter the classical Friedmann equation can be written as
\begin{equation}
	\left(\frac{1}{3V}\frac{\dd V}{\dd\beta_+}\right)^2 = 1\,,
\end{equation}	
so that $\beta_+$ can be a relational clock with no need for a separate matter field. One might hope to incorporate this idea into GFT, and describe relational evolution without coupling to the somewhat arbitrary massless scalar field. Such a relational formalism would appear to require using an expectation value as a clock parameter, perhaps along the lines suggested in \cite{Marchetti2021}. The availability of multiple candidate clock degrees of freedom would also help identifying a GFT analogue of the symmetry of continuum general relativity under time reparametrisations. We leave these open questions for future work.

\let\oldaddcontentsline\addcontentsline
\renewcommand{\addcontentsline}[3]{}
\begin{acknowledgments}
	We would like to thank Luca Marchetti and Edward Wilson-Ewing for helpful comments on the manuscript, and Yili Wang for insights on $SU(2)$ recoupling theory. The work of SG was funded by the Royal Society through a University Research Fellowship (UF160622).
\end{acknowledgments}
\let\addcontentsline\oldaddcontentsline

\appendix
\section{Classical and quantum geometry of the tetrahedron }\label{AppA}

In general, the shape of a tetrahedron requires the knowledge of six quantities to be determined unambiguously; if only the areas of the four faces are known the space of possible configurations forms a two-sphere \cite{Bianchi_Haggard_vol}. However, there is a unique \textit{orthocentric} tetrahedron for given face areas, which is the one of maximal volume \cite{Gerber_Ortho}. A tetrahedron is orthocentric if and only if all three pairs of opposite edges are perpendicular, or equivalently
\begin{equation}\label{ortho}
	e_1^2+e_2^2=e_3^2+e_4^2=e_5^2+e_6^2\, ,
\end{equation}
where $e_1, \ldots, e_6$ are the six edge lengths (such that $e_1$ is opposite to $e_2$, and so on). 
For such three-simplices, an analogue of Heron's formula
\begin{equation}\label{Heron1}
	16\Delta^2 = (a+b+c)(a-b+c)(a+b-c)(-a+b+c)\, ,
\end{equation}
which gives the area $\Delta$ of a triangle given its sides $(a,b,c$), can be derived. In the cases of interest for this paper one finds that face areas $A$ or $(A,B)$ and the volume $V$ are related by
\begin{equation}\label{equiclassicalV}
	V^2= \frac{8}{27\sqrt{3}}A^3 \qquad \text{and}\qquad 	V^2= 
	\frac{1}{27\sqrt{3}}B(9A^2-B^2)
\end{equation}
for the regular and the trisohedral tetrahedron of figure \ref{isosceles}, respectively. In the quantum theory, we take the analogue of an orthocentric tetrahedron to be the $SU(2)$ intertwiner with largest volume eigenvalue; we explain the concept of LQG volume eigenvalues in what follows.

Following the terminology introduced in section \ref{reviewSec}, we focus a single four-valent node, with spins labelling its links denoted by $\vec{j}=(j_1,j_2,j_3,j_4)$ (see figure \ref{tetra+5graph}). To each representation $j_I$, with $I=1,\dots,4$, we associate a vector space $\mathcal{H}_{j_I}$ that carries the action of the $SU(2)$ generators $\vec{J}_I$. The Hilbert space of the quantum tetrahedron then reads $\mathcal{H}_4=\textmd{Inv}\left[\mathcal{H}_{j_1}\otimes \ldots \otimes \mathcal{H}_{j_4}\right]$, and objects that live in this space are called intertwiners. A nonvanishing intertwiner can only exist if the $j_I$ sum to an integer. We introduce a basis labelled by $k$ in the recoupling channel $\mathcal{H}_{j_1}\otimes \mathcal{H}_{j_2}$, where the index $k$ ranges between $k_{min}=\max \left\{|j_1-j_2|,|j_3-j_4|\right\}$ and $k_{max}=\min \left\{j_1+j_2,j_3+j_4\right\}$ in integer steps. The Hilbert space $\mathcal{H}_4$ is $d$-dimensional with $d=k_{max}-k_{min}+1$. States on this space can be understood as describing quantum tetrahedra, as firstly pointed out in \cite{Barbieri_1997,Baez_1999}.

We refer to the literature (see, e.g., \cite{DePietriGEO}) for derivations of the geometrical eigenvalues in the context of LQG. Here we only recall that the area operator takes the form $\widehat{A} = l_0^2 \sqrt{{\vec{J}}\cdot \vec{{J}}}$, where $l_0$ is a fundamental quantum gravity length scale, usually taken to depend on the Barbero--Immirzi parameter of LQG. States in $\mathcal{H}_4$ are eigenstates of the operator $\widehat{A}_I$, which measures the area of the $I$-th face of the quantum tetrahedron, with eigenvalues $l_0^2 \sqrt{j_I(j_I+1)}$. Finally, the volume operator introduced for a tetrahedron in LQG reads \cite{RovelliSmolin_vol,AshtekarLewandowski_vol}
\begin{equation}\label{volumelqg}
	\widehat{V} =	\frac{\sqrt{2}}{3}l_0^3\sqrt{|\epsilon_{ijk} J^i_1J^j_2J^k_3|}\, .
\end{equation}
We can focus on the radicand in \eqref{volumelqg}. Without showing all the details of the calculation (which can be found in \cite{Brunnemann_Thiemann_Vol,Bianchi_Haggard_vol}), one defines the operator $\widehat{Q}: = \vec{J}_1 \cdot (\vec{J}_2\times \vec{J}_3) $ with matrix element $Q_{k,k'}\equiv \langle k | \widehat{Q} | k'\rangle$, where $k$ and $k'$ label intertwiner states as explained above. One can then show that nonvanishing matrix elements are obtained only if $k$ and $k'$ differ by $1$. Thus one can denote $a_k := {\rm i}Q_{k,k-1}$, so that
\begin{equation}\label{qmatrix}
	Q=\begin{pmatrix}
		0 & {\rm i} a_1&0&\cdots& \\-  {\rm i} a_1&0&  {\rm i} a_2&&\\
		0&- {\rm i} a_2&0&&\\
		\vdots&&&\ddots& 
	\end{pmatrix} 
\end{equation} 
is a $d\times d$ matrix. With the above conventions, the matrix elements are found to be \cite{Brunnemann1_2008,*Brunnemann2_2008}
\begin{equation}\label{matrixelement}
	\begin{aligned}[t]
		a_k = \frac{1}{4}&\frac{\sqrt{(j_1+j_2+k+1)(-j_1+j_2+k)(j_1-j_2+k)(j_1+j_2-k+1)}}{\sqrt{2k+1}}\\& \quad\times \frac{\sqrt{(j_3+j_4+k+1)(-j_3+j_4+k)(j_3-j_4+k)(j_3+j_4-k+1)}}{\sqrt{2k-1}}\, .
	\end{aligned}
\end{equation}
To obtain a compact expression, \eqref{matrixelement} can be cast in terms of Heron's formula \eqref{Heron1} as
\begin{equation}\label{matrixelementheron}
	a_k = \frac{4}{\sqrt{4k^2-1}}\Delta\left(j_1+\frac{1}{2},j_2+\frac{1}{2},k\right)\Delta\left(j_3+\frac{1}{2},j_4+\frac{1}{2},k\right)\, .
\end{equation}

Hence, for given spins $\vec{j}$ computing the spectrum of the volume operator amounts to finding the $d$ eigenvalues of \eqref{qmatrix} (let us denote them $q_{\vec{j},k}$) and then computing, according to \eqref{volumelqg},
\begin{equation}\label{Veigenvalue}
	V^k_{\vec{j}}	= \frac{\sqrt{2}}{3}l_0^3\sqrt{q_{\vec{j},k}} \,.
\end{equation}
Note that if $\vec{j}=(j,j,j,j)$ the matrix elements \eqref{matrixelementheron} simplify to
\begin{equation}\label{aisok}
	a^{(j)}_k 
	= \frac{(2 j k+k)^2-k^4}{4 \sqrt{4 k^2-1}} \,,
\end{equation}
while for $\vec{j}=(j_a,j_a,j_a,j_b)$ (as in \eqref{ourmode}) they reduce to
\begin{equation}\label{oura}
	a_k^{(j_a,j_b)}= \frac{\sqrt{(2 j_a k+k)^2-k^4} }{4 \sqrt{4 k^2-1}}\sqrt{(j_b-j_a+k) (j_a+j_b-k+1) (j_a-j_b+k) (j_a+j_b+k+1)}\, .
\end{equation}
These matrix elements can be used to find the maximal volume eigenvalues $V^{\text{max}}_j$ and $V^{\text{max}}_{j_a,j_b}$ for the quantum shapes corresponding to regular and trisohedral tetrahedra, respectively; see table \ref{tab} for examples.

For large spins, these volume eigenvalues show semiclassical behaviour (see, e.g., \cite{Brunnemann_Thiemann_Vol,PolyhedraIta,Bianchi_Haggard_vol}), and so the ``eigenvalue-counterparts'' of equations \eqref{equiclassicalV},
\begin{equation}\label{vj}
	(V_j^{\text{max}})^2/ l_0^6 \, \approx \,\frac{8}{27\sqrt{3}}[j(j+1)]^{3/2}\, , \qquad (V_{j_a,j_b}^{\text{max}})^2/l_0^6\, \approx\, \frac{1}{27\sqrt{3}}\sqrt{j_b(j_b+1)}\left[9j_a(j_a+1)-j_b(j_b+1)\right]\,,
\end{equation}
become more accurate as the spins grow. Here $V^{\text{max}}$ means that we fix the intertwiner $k$ so as to obtain the largest volume eigenvalue.

\begin{figure}[ht]
	\begin{center}
		\includegraphics[width=0.45\textwidth]{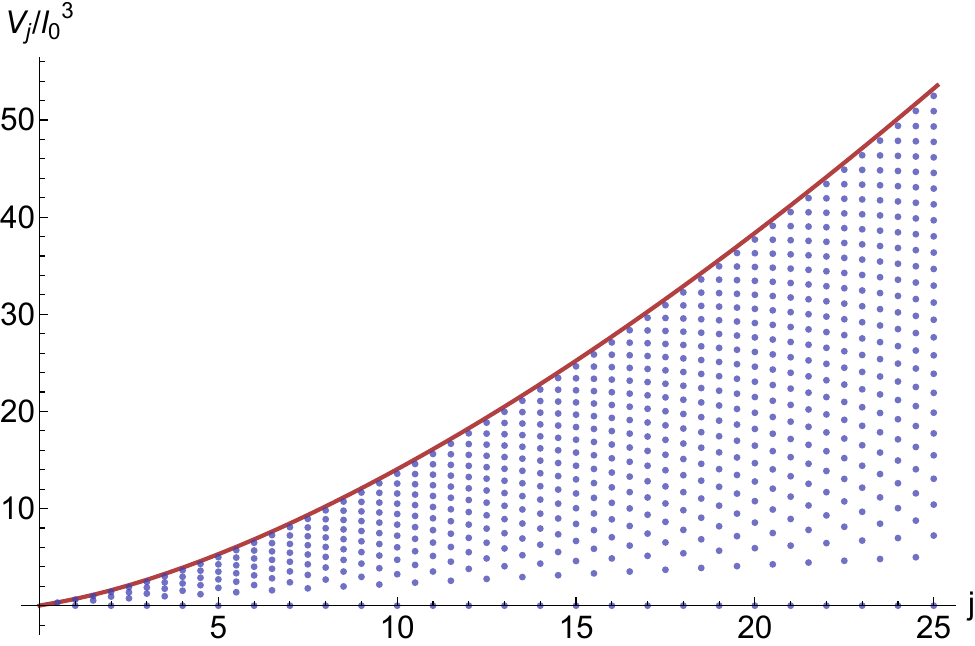}
		\includegraphics[width=0.45\textwidth]{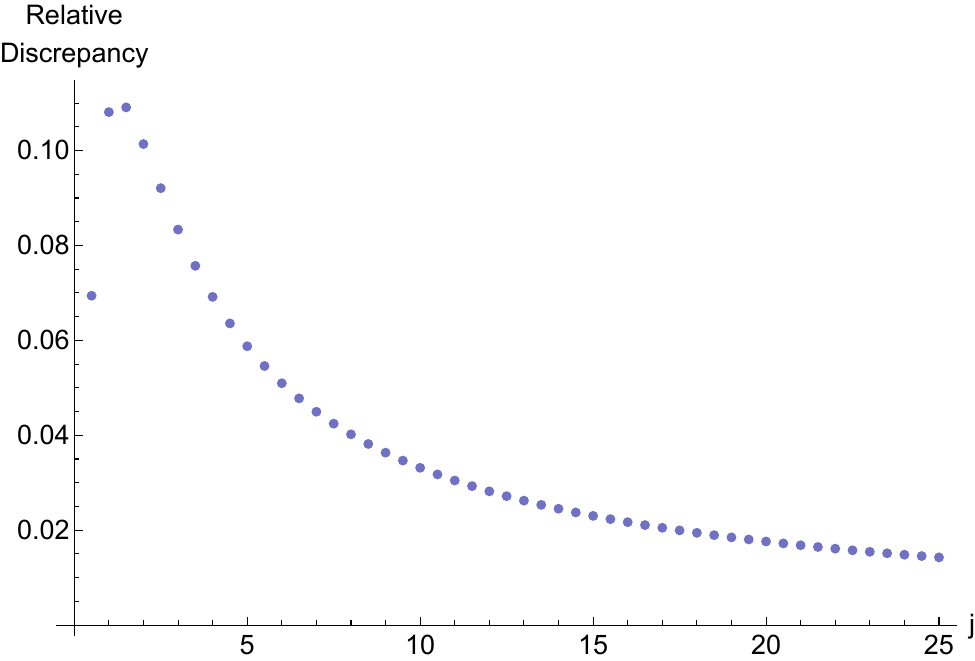}
	\end{center}
	\caption{\small Left: comparison between LQG volume eigenvalues $V^k_j/l_0^3$ (dots) and the classical volume of an equilateral tetrahedron as function of the area (curve). Right: relative difference between the classical volume and largest LQG eigenvalue. After two initial anomalies the mismatch decreases indefinitely: it becomes reasonably small ($\sim1\%$) when $j \sim 40$, and goes down to $\sim 0.1\%$ when $j\sim 350$. }
	\label{eigen}
\end{figure}

\begin{figure}[ht]
	\begin{center}
		\begin{subfigure}[c]{0.5\textwidth}
			\includegraphics[width=\textwidth]{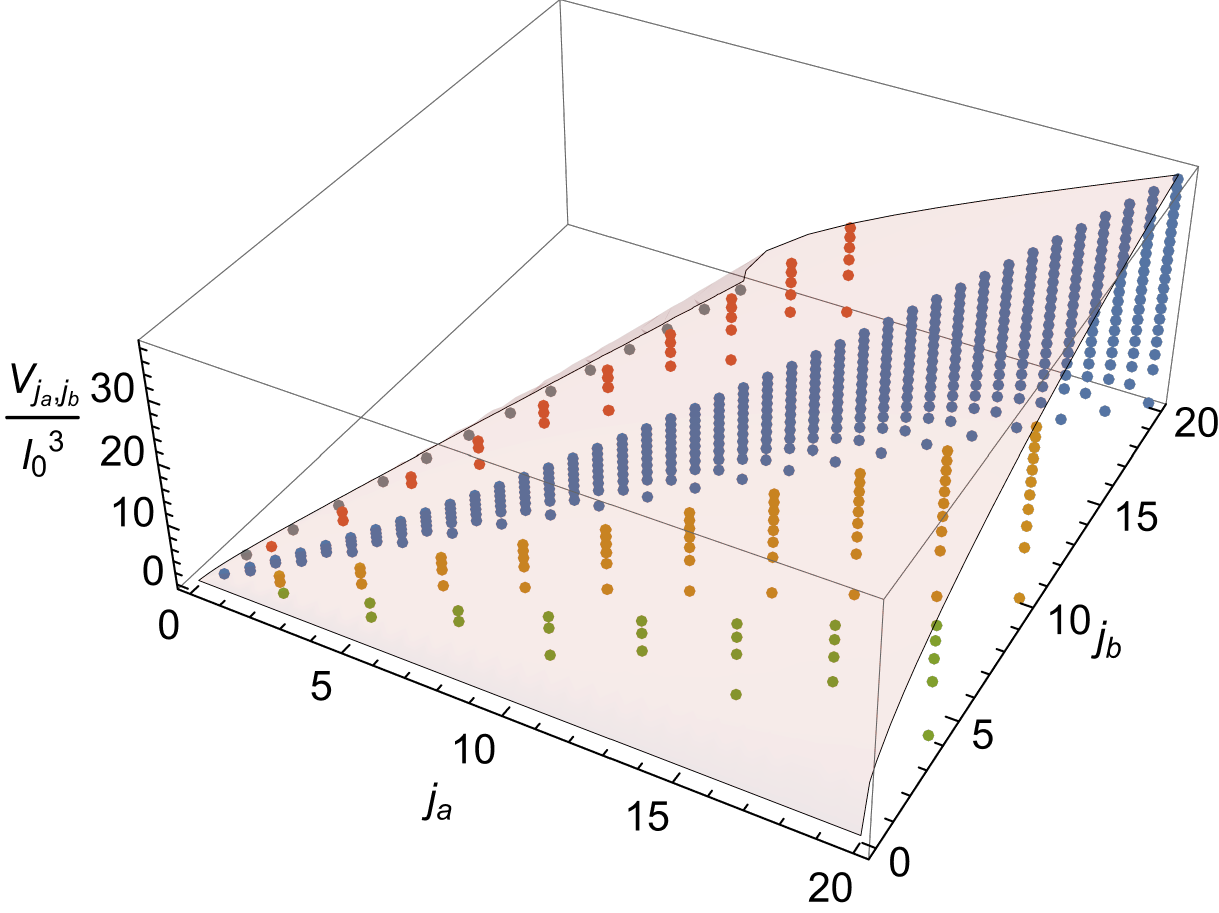}
		\end{subfigure} \hspace{5mm}
		\begin{subfigure}[c]{0.45\textwidth}
			\includegraphics[width=\textwidth]{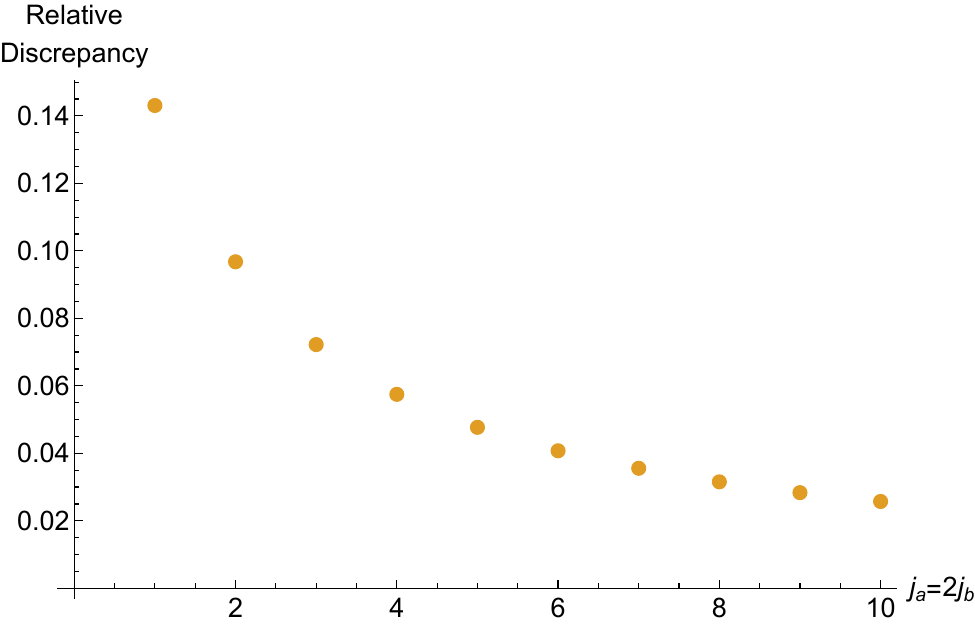}
		\end{subfigure}
	\end{center}
	\caption{\small Left: comparison between some of the LQG volume eigenvalues $V^k_{j_a,j_b}/l_0^3$ (dots) and the classical volume of a trisohedral tetrahedron (transparent surface) as function of the two areas. We plot eigenvalues along specific planes for which $j_a=j_b$ (blue), $j_a=2j_b$ (orange), $j_a=5j_b$ (green), $2j_a=j_b$ (red) and $3j_a=j_b$ (grey). The latter gives vanishing eigenvalues only and represents the limiting case as there are no tetrahedra (no nonzero intertwiners) for $3j_a<j_b$. Right: relative discrepancy along the plane $j_a=2j_b$.}
	\label{keel}
\end{figure}

In figures \ref{eigen} and \ref{keel} we show the possible volume eigenvalues $V^k_j$ and $V^k_{j_a,j_b}$ in units of $l_0^3$ (dots), compared with the maximal classically allowed volume (curve and surface) for the same face areas (given by $l_0^2\sqrt{j(j+1)}$). We also show how the relative difference between this classical volume (for orthocentric tetrahedra) and the highest LQG volume eigenvalue decreases as the spins increase. Figure \ref{eigen} focuses on the case of equal areas $j_a=j_b$, whereas figure \ref{keel} shows various other combinations of $j_a$ and $j_b$. Notice that both the classical and quantum volume of a tetrahedron are no longer well-defined for $j_b>3j_a$. In the quantum theory, this constraint comes about because the dimension $d$ of the Hilbert space $\mathcal{H}_4$ needs to be positive. The right panel of figure \ref{keel} shows, as an illustrative example, the relative distances between the maximum eigenvalues and the surface along the plane $j_a=2j_b$, but the qualitative behaviour is similar for other ratios. 

We now finally turn  to the question of how to define a notion of anisotropy at the level of quantum geometry, in terms of a function $\beta_+^{j_a,j_b}$. In particular, we show that though the definition \eqref{ourbeta} is not unique, it is the best candidate to play such a role. 

Starting from \eqref{AB}, there are three ways to classically define the anisotropy of a tetrahedron in terms of two other geometric quantities: by inverting any one of \eqref{AB} or combining them, one can obtain classical expressions such as $\beta_+(V,B)$, $\beta_+(V,A)$ or $\beta_+(A,B)$. Once we pick a favoured definition, the strategy is to replace classical expressions by LQG eigenvalues to obtain an effective (discrete) notion of anisotropy as a function of the spins associated to the faces of a tetrahedron.

For instance, inverting the first equation in \eqref{AB}, one can readily obtain 
\begin{equation}\label{BV}
	\beta_+ (V,B) = \frac{1}{2}\log\left(\frac{2}{3\sqrt[6]{3}}\frac{B}{V^{2/3}}\right) \qquad\Rightarrow \qquad 	\beta_+^{j_a,j_b} = \frac{1}{2}\log \left(\frac{2}{3\sqrt[6]{3}}\frac{l_0^2\sqrt{j_b(j_b+1)}}{({V^{\text{max}}_{j_a,j_b}})^{2/3}}\right)\,,
\end{equation}
where we simply replaced the classical area $B$ with $l_0^2\sqrt{j_b(j_b+1)}$ and the classical volume with the maximal eigenvalue $V^{\text{max}}_{j_a,j_b}$. Notice that by definition $\beta_+^{j_a,j_b}>0$ when $j_a<j_b$ (naively, for volumes smaller than the equilateral ones). \eqref{BV} could be used as a definition, but we can identify a number of issues. First, it cannot be written more explicitly, since the volume eigenvalue $V^{\text{max}}_{j_a,j_b}$ is only obtained numerically, as outlined in the previous section. Moreover, \eqref{BV} is not well-defined when the volume is exactly zero, and it is nonvanishing when $j_a=j_b$.

The same arguments would apply if we defined the anisotropy starting from the second relation in \eqref{AB}. In this case we would obtain an even more cumbersome expression $\beta_+(V,A)$, which could then be turned into a $\beta_+^{j_a,j_b}$ as a function of $j_a$ and volume eigenvalues $V^k_{j_a,j_b}$. 

A third definition is obtained by taking the ratio of equations \eqref{AB}. The advantage of this choice is that one can get rid of the volume:
\begin{equation}
	\frac{A}{B}= \frac{1}{3}\sqrt{1+8e^{-6\beta_+}}\, .
\end{equation}
We can now rearrange to obtain a simpler definition of the anisotropy parameter as a function of $A$ and $B$, which straightforwardly turn into the spins $j_a$ and $j_b$. Indeed, our candidate reads
\begin{equation}\label{ourbeta2}
	\beta_+ (A, B) = -\frac{1}{6}\ln \left(\frac{9A^2-B^2}{8B^2}\right) \qquad\Rightarrow  \qquad	\beta_+^{j_a,j_b} =  -\frac{1}{6}\ln \left(\frac{9j_a(j_a+1)-j_b(j_b+1)}{8j_b(j_b+1)}\right)\,,
\end{equation}
which is nothing but \eqref{ourbeta}. This definition again gives a negative value for $j_a>j_b$, but it also gives $\beta_+^{j_a,j_b} =  0$ for ``isotropic'' configurations, as we would like to demand. Moreover, it is always finite and well-defined regardless of the volume eigenvalue associated to the same pair $(j_a,j_b)$.

\begin{figure}[h!]
	\begin{center}
		\begin{subfigure}[]{0.45\textwidth}\includegraphics[width=\textwidth]{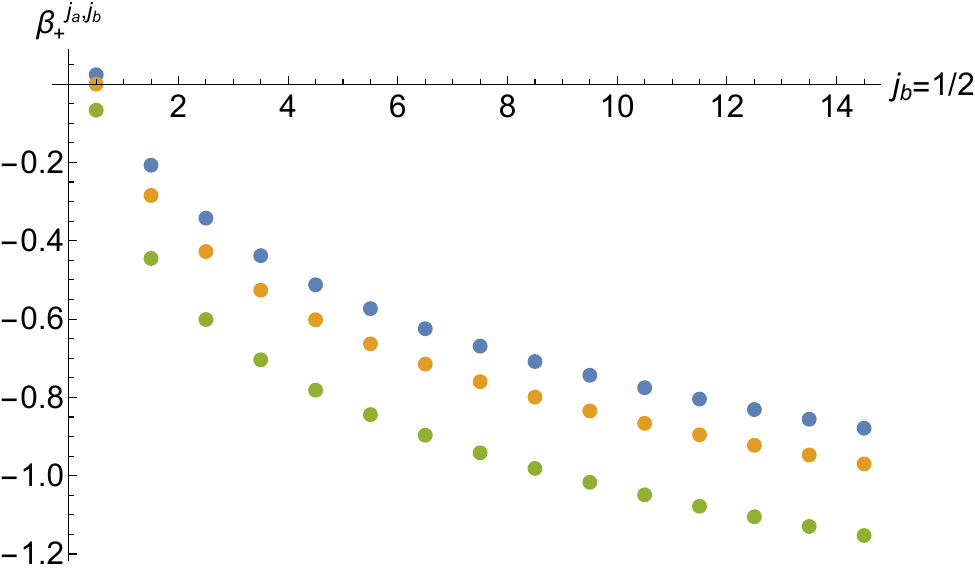}\end{subfigure}
		\hspace{5mm}
		\begin{subfigure}[]{0.5\textwidth}		\includegraphics[width=\textwidth]{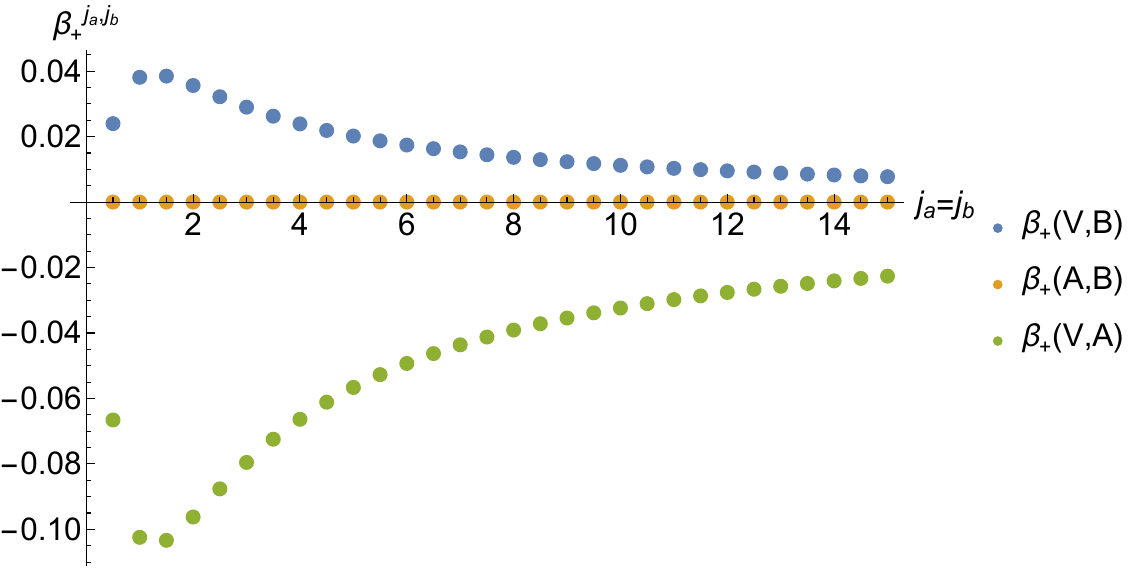}\end{subfigure}
	\end{center}
	\caption{\small Left: For constant $j_b$, we see that the discrepancy between the three $\beta_+^{j_a,j_b}$ definitions is larger away from $j_a=j_b$. Right: Choosing $j_a=j_b$ shows how the three definitions agree more and more for increasing spin. It is also clear that only one definition satisfies $\beta_+^{j_a,j_b}=0$ for equal spins. }
	\label{last}
\end{figure}

Even though we consider the definition (\ref{ourbeta2}) to be well-motivated for GFT cosmology, the ambiguity we have highlighted here introduces, at least in principle, an additional uncertainty into the cosmological interpretation of GFT models. To quantify this uncertainty we can compare the various definitions we have discussed.  To do this we plot in figure \ref{last} two-dimensional slices of the anisotropy dependence on the spins $j_a$ and $j_b$, first for constant $j_b=\frac{1}{2}$ and along the plane $j_a=j_b$. 

The table below shows the maximum volume eigenvalues \eqref{Veigenvalue}, expressed in units of $l_0^3$, and the corresponding (dimensionless) anisotropies \eqref{ourbeta} for given spins $\vec{j} = (j_a,j_a,j_a,j_b)$. Even though we may use larger spins in some of our calculations, we only give values up to $j_a=j_b=7/2$.

\begin{table}[h!]
       \caption{Maximal volume eigenvalues and anisotropy parameters for a range of spins.}
	\label{tab}
	\centering
	\begin{tabular}{c c | c c}
		$j_a$ & $j_b$ & $V^{\text{max}}_{j_a,j_b}$ & $\beta_+^{j_a,j_b}$ \\ [0.5ex] 
		\hline\hline
		1/2 & 1/2 & 0.310 & 0 \\
		\hline
		{ 1/2}&{3/2}& 0 & 0.384 \\
		\hline
		1 & 1 & 0.620 & 0 \\
		\hline
		1 & 2 & 0.620 & 0.231 \\
		\hline
		{	1} & { 3} & 0 & 0.462 \\
		\hline
		3/2 & 1/2 & 0.620 & -0.284 \\
		\hline
		3/2 & 3/2 & 0.993 & 0 \\
		\hline
		3/2 & 5/2  & 1.075 & 0.172 \\
		\hline
		3/2 & 7/2  & 0.931 & 0.324 \\
		\hline
		2 & 1 & 1.075 & -0.196 \\
		\hline
		2 & 2 & 1.425 & 0 \\
	\end{tabular}
	\qquad\qquad \begin{tabular}{c c | c c} 				
		\hline\hline		
			2 & 3  & 1.560 & 0.138 \\
		\hline
		5/2 & 1/2& 0.931 & -0.427 \\
		\hline
		5/2 & 3/2& 1.560 & -0.153 \\
		\hline
		5/2 & 5/2& 1.910 & 0 \\
		\hline
		5/2 & 7/2& 2.086 & 0.116 \\
		\hline
		3 & 1 & 1.520 & -0.315 \\
		\hline
		3 & 2 & 2.086 & -0.126 \\
		\hline
		3 & 3 & 2.444 & 0 \\
		\hline
		7/2 & 1/2 & 1.241 & -0.526 \\
		\hline
		7/2 & 3/2 & 2.111 & -0.254 \\
		\hline
		7/2 & 5/2 & 2.653 & -0.107 \\
		\hline
		7/2 & 7/2 & 3.022 & 0 \\
	\end{tabular}
\end{table}
 
\let\oldaddcontentsline\addcontentsline
\renewcommand{\addcontentsline}[3]{}
 \bibliography{bianGFT}

\begin{thebibliography}{76}%
\makeatletter
\providecommand \@ifxundefined [1]{%
 \@ifx{#1\undefined}
}%
\providecommand \@ifnum [1]{%
 \ifnum #1\expandafter \@firstoftwo
 \else \expandafter \@secondoftwo
 \fi
}%
\providecommand \@ifx [1]{%
 \ifx #1\expandafter \@firstoftwo
 \else \expandafter \@secondoftwo
 \fi
}%
\providecommand \natexlab [1]{#1}%
\providecommand \enquote  [1]{``#1''}%
\providecommand \bibnamefont  [1]{#1}%
\providecommand \bibfnamefont [1]{#1}%
\providecommand \citenamefont [1]{#1}%
\providecommand \href@noop [0]{\@secondoftwo}%
\providecommand \href [0]{\begingroup \@sanitize@url \@href}%
\providecommand \@href[1]{\@@startlink{#1}\@@href}%
\providecommand \@@href[1]{\endgroup#1\@@endlink}%
\providecommand \@sanitize@url [0]{\catcode `\\12\catcode `\$12\catcode
  `\&12\catcode `\#12\catcode `\^12\catcode `\_12\catcode `\%12\relax}%
\providecommand \@@startlink[1]{}%
\providecommand \@@endlink[0]{}%
\providecommand \url  [0]{\begingroup\@sanitize@url \@url }%
\providecommand \@url [1]{\endgroup\@href {#1}{\urlprefix }}%
\providecommand \urlprefix  [0]{URL }%
\providecommand \Eprint [0]{\href }%
\providecommand \doibase [0]{https://doi.org/}%
\providecommand \selectlanguage [0]{\@gobble}%
\providecommand \bibinfo  [0]{\@secondoftwo}%
\providecommand \bibfield  [0]{\@secondoftwo}%
\providecommand \translation [1]{[#1]}%
\providecommand \BibitemOpen [0]{}%
\providecommand \bibitemStop [0]{}%
\providecommand \bibitemNoStop [0]{.\EOS\space}%
\providecommand \EOS [0]{\spacefactor3000\relax}%
\providecommand \BibitemShut  [1]{\csname bibitem#1\endcsname}%
\let\auto@bib@innerbib\@empty
\bibitem [{\citenamefont {Bombelli}\ \emph {et~al.}(2009)\citenamefont
  {Bombelli}, \citenamefont {Corichi},\ and\ \citenamefont
  {Winkler}}]{manifoldGraph}%
  \BibitemOpen
  \bibfield  {author} {\bibinfo {author} {\bibfnamefont {L.}~\bibnamefont
  {Bombelli}}, \bibinfo {author} {\bibfnamefont {A.}~\bibnamefont {Corichi}},\
  and\ \bibinfo {author} {\bibfnamefont {O.}~\bibnamefont {Winkler}},\
  }\bibfield  {title} {\bibinfo {title} {{Semiclassical quantum gravity:
  obtaining manifolds from graphs}},\ }\href
  {https://doi.org/10.1088/0264-9381/26/24/245012} {\bibfield  {journal}
  {\bibinfo  {journal} {Class. Quant. Grav.}\ }\textbf {\bibinfo {volume}
  {26}},\ \bibinfo {pages} {245012} (\bibinfo {year} {2009})},\ \Eprint
  {https://arxiv.org/abs/0905.3492} {arXiv:0905.3492 [gr-qc]} \BibitemShut
  {NoStop}%
\bibitem [{\citenamefont {Dittrich}(2011)}]{BiancaDiffeo}%
  \BibitemOpen
  \bibfield  {author} {\bibinfo {author} {\bibfnamefont {B.}~\bibnamefont
  {Dittrich}},\ }\bibfield  {title} {\bibinfo {title} {{How to construct
  diffeomorphism symmetry on the lattice}},\ }\href
  {https://doi.org/10.22323/1.140.0012} {\bibfield  {journal} {\bibinfo
  {journal} {PoS}\ }\textbf {\bibinfo {volume} {QGQGS2011}},\ \bibinfo {pages}
  {012} (\bibinfo {year} {2011})},\ \Eprint {https://arxiv.org/abs/1201.3840}
  {arXiv:1201.3840 [gr-qc]} \BibitemShut {NoStop}%
\bibitem [{\citenamefont {Adelberger}\ \emph {et~al.}(2009)\citenamefont
  {Adelberger}, \citenamefont {Gundlach}, \citenamefont {Heckel}, \citenamefont
  {Hoedl},\ and\ \citenamefont {Schlamminger}}]{ExpConstraints}%
  \BibitemOpen
  \bibfield  {author} {\bibinfo {author} {\bibfnamefont {E.~G.}\ \bibnamefont
  {Adelberger}}, \bibinfo {author} {\bibfnamefont {J.~H.}\ \bibnamefont
  {Gundlach}}, \bibinfo {author} {\bibfnamefont {B.~R.}\ \bibnamefont
  {Heckel}}, \bibinfo {author} {\bibfnamefont {S.}~\bibnamefont {Hoedl}},\ and\
  \bibinfo {author} {\bibfnamefont {S.}~\bibnamefont {Schlamminger}},\
  }\bibfield  {title} {\bibinfo {title} {{Torsion balance experiments: A
  low-energy frontier of particle physics}},\ }\href
  {https://doi.org/https://doi.org/10.1016/j.ppnp.2008.08.002} {\bibfield
  {journal} {\bibinfo  {journal} {Prog. Part. Nucl. Phys.}\ }\textbf {\bibinfo
  {volume} {62}},\ \bibinfo {pages} {102} (\bibinfo {year} {2009})}\BibitemShut
  {NoStop}%
\bibitem [{\citenamefont {Bojowald}(2005)}]{LQCBojo}%
  \BibitemOpen
  \bibfield  {author} {\bibinfo {author} {\bibfnamefont {M.}~\bibnamefont
  {Bojowald}},\ }\bibfield  {title} {\bibinfo {title} {{Loop Quantum
  Cosmology}},\ }\href {https://doi.org/10.12942/lrr-2005-11} {\bibfield
  {journal} {\bibinfo  {journal} {Living Rev. Rel.}\ }\textbf {\bibinfo
  {volume} {8}},\ \bibinfo {pages} {11} (\bibinfo {year} {2005})},\ \Eprint
  {https://arxiv.org/abs/gr-qc/0601085} {arXiv:gr-qc/0601085} \BibitemShut
  {NoStop}%
\bibitem [{\citenamefont {Ashtekar}\ and\ \citenamefont
  {Singh}(2011)}]{Ashtekar+Singh_LQC}%
  \BibitemOpen
  \bibfield  {author} {\bibinfo {author} {\bibfnamefont {A.}~\bibnamefont
  {Ashtekar}}\ and\ \bibinfo {author} {\bibfnamefont {P.}~\bibnamefont
  {Singh}},\ }\bibfield  {title} {\bibinfo {title} {{Loop quantum cosmology: a
  status report}},\ }\href {https://doi.org/10.1088/0264-9381/28/21/213001}
  {\bibfield  {journal} {\bibinfo  {journal} {Class. Quant. Grav.}\ }\textbf
  {\bibinfo {volume} {28}},\ \bibinfo {pages} {213001} (\bibinfo {year}
  {2011})},\ \Eprint {https://arxiv.org/abs/1108.0893} {arXiv:1108.0893
  [gr-qc]} \BibitemShut {NoStop}%
\bibitem [{\citenamefont {Banerjee}\ \emph {et~al.}(2012)\citenamefont
  {Banerjee}, \citenamefont {Calcagni},\ and\ \citenamefont
  {Mart{\'i}n-Benito}}]{Banerjee_2012}%
  \BibitemOpen
  \bibfield  {author} {\bibinfo {author} {\bibfnamefont {K.}~\bibnamefont
  {Banerjee}}, \bibinfo {author} {\bibfnamefont {G.}~\bibnamefont {Calcagni}},\
  and\ \bibinfo {author} {\bibfnamefont {M.}~\bibnamefont
  {Mart{\'i}n-Benito}},\ }\bibfield  {title} {\bibinfo {title} {{Introduction
  to Loop Quantum Cosmology}},\ }\href {https://doi.org/10.3842/SIGMA.2012.016}
  {\bibfield  {journal} {\bibinfo  {journal} {SIGMA}\ }\textbf {\bibinfo
  {volume} {8}},\ \bibinfo {pages} {016} (\bibinfo {year} {2012})},\ \Eprint
  {https://arxiv.org/abs/1109.6801} {arXiv:1109.6801 [gr-qc]} \BibitemShut
  {NoStop}%
\bibitem [{\citenamefont {Gambini}\ and\ \citenamefont {Pullin}(2013)}]{LQBH1}%
  \BibitemOpen
  \bibfield  {author} {\bibinfo {author} {\bibfnamefont {R.}~\bibnamefont
  {Gambini}}\ and\ \bibinfo {author} {\bibfnamefont {J.}~\bibnamefont
  {Pullin}},\ }\bibfield  {title} {\bibinfo {title} {{Loop Quantization of the
  Schwarzschild Black Hole}},\ }\href
  {https://doi.org/10.1103/PhysRevLett.110.211301} {\bibfield  {journal}
  {\bibinfo  {journal} {Phys. Rev. Lett.}\ }\textbf {\bibinfo {volume} {110}},\
  \bibinfo {pages} {211301} (\bibinfo {year} {2013})},\ \Eprint
  {https://arxiv.org/abs/1302.5265} {arXiv:1302.5265 [gr-qc]} \BibitemShut
  {NoStop}%
\bibitem [{\citenamefont {Bojowald}\ \emph {et~al.}(2018)\citenamefont
  {Bojowald}, \citenamefont {Brahma},\ and\ \citenamefont {Yeom}}]{LQBH2}%
  \BibitemOpen
  \bibfield  {author} {\bibinfo {author} {\bibfnamefont {M.}~\bibnamefont
  {Bojowald}}, \bibinfo {author} {\bibfnamefont {S.}~\bibnamefont {Brahma}},\
  and\ \bibinfo {author} {\bibfnamefont {D.-h.}\ \bibnamefont {Yeom}},\
  }\bibfield  {title} {\bibinfo {title} {{Effective line elements and
  black-hole models in canonical loop quantum gravity}},\ }\href
  {https://doi.org/10.1103/PhysRevD.98.046015} {\bibfield  {journal} {\bibinfo
  {journal} {Phys. Rev. D}\ }\textbf {\bibinfo {volume} {98}},\ \bibinfo
  {pages} {046015} (\bibinfo {year} {2018})},\ \Eprint
  {https://arxiv.org/abs/1803.01119} {arXiv:1803.01119 [gr-qc]} \BibitemShut
  {NoStop}%
\bibitem [{\citenamefont {Kelly}\ \emph {et~al.}(2020)\citenamefont {Kelly},
  \citenamefont {Santacruz},\ and\ \citenamefont {Wilson-Ewing}}]{LQBH3}%
  \BibitemOpen
  \bibfield  {author} {\bibinfo {author} {\bibfnamefont {J.~G.}\ \bibnamefont
  {Kelly}}, \bibinfo {author} {\bibfnamefont {R.}~\bibnamefont {Santacruz}},\
  and\ \bibinfo {author} {\bibfnamefont {E.}~\bibnamefont {Wilson-Ewing}},\
  }\bibfield  {title} {\bibinfo {title} {{Effective loop quantum gravity
  framework for vacuum spherically symmetric spacetimes}},\ }\href
  {https://doi.org/10.1103/PhysRevD.102.106024} {\bibfield  {journal} {\bibinfo
   {journal} {Phys. Rev. D}\ }\textbf {\bibinfo {volume} {102}},\ \bibinfo
  {pages} {106024} (\bibinfo {year} {2020})},\ \Eprint
  {https://arxiv.org/abs/2006.09302} {arXiv:2006.09302 [gr-qc]} \BibitemShut
  {NoStop}%
\bibitem [{\citenamefont {Freidel}(2005)}]{FreidelGFT}%
  \BibitemOpen
  \bibfield  {author} {\bibinfo {author} {\bibfnamefont {L.}~\bibnamefont
  {Freidel}},\ }\bibfield  {title} {\bibinfo {title} {{Group field theory: An
  Overview}},\ }\href {https://doi.org/10.1007/s10773-005-8894-1} {\bibfield
  {journal} {\bibinfo  {journal} {Int. J. Theor. Phys.}\ }\textbf {\bibinfo
  {volume} {44}},\ \bibinfo {pages} {1769} (\bibinfo {year} {2005})},\ \Eprint
  {https://arxiv.org/abs/hep-th/0505016} {arXiv:hep-th/0505016} \BibitemShut
  {NoStop}%
\bibitem [{\citenamefont {Oriti}(2017{\natexlab{a}})}]{Oriti_GFTandLQG}%
  \BibitemOpen
  \bibfield  {author} {\bibinfo {author} {\bibfnamefont {D.}~\bibnamefont
  {Oriti}},\ }\bibinfo {title} {{Group Field Theory and Loop Quantum
  Gravity}},\ in\ \href@noop {} {\emph {\bibinfo {booktitle} {Loop Quantum
  Gravity: The First 30 Years}}},\ \bibinfo {editor} {edited by\ \bibinfo
  {editor} {\bibfnamefont {A.}~\bibnamefont {Ashtekar}}\ and\ \bibinfo {editor}
  {\bibfnamefont {J.}~\bibnamefont {Pullin}}}\ (\bibinfo  {publisher} {World
  Scientific},\ \bibinfo {year} {2017})\ pp.\ \bibinfo {pages} {125--151},\
  \Eprint {https://arxiv.org/abs/1408.7112} {arXiv:1408.7112 [gr-qc]}
  \BibitemShut {NoStop}%
\bibitem [{\citenamefont {Perez}(2013)}]{PerezSFQG}%
  \BibitemOpen
  \bibfield  {author} {\bibinfo {author} {\bibfnamefont {A.}~\bibnamefont
  {Perez}},\ }\bibfield  {title} {\bibinfo {title} {{The Spin-Foam Approach to
  Quantum Gravity}},\ }\href {https://doi.org/10.12942/lrr-2013-3} {\bibfield
  {journal} {\bibinfo  {journal} {Living Rev. Rel.}\ }\textbf {\bibinfo
  {volume} {16}},\ \bibinfo {pages} {3} (\bibinfo {year} {2013})},\ \Eprint
  {https://arxiv.org/abs/1205.2019} {arXiv:1205.2019 [gr-qc]} \BibitemShut
  {NoStop}%
\bibitem [{\citenamefont {Di~Francesco}(2006)}]{matrix}%
  \BibitemOpen
  \bibfield  {author} {\bibinfo {author} {\bibfnamefont {P.}~\bibnamefont
  {Di~Francesco}},\ }\bibinfo {title} {{2D quantum gravity, matrix models and
  graph combinatorics}},\ in\ \href
  {https://doi.org/10.48550/ARXIV.MATH-PH/0406013} {\emph {\bibinfo {booktitle}
  {{Proceedings of the NATO Advanced Study Institute on Applications of Random
  Matrices in Physics, Les Houches, France, 6--25 June 2004}}}}\ (\bibinfo
  {year} {2006})\ pp.\ \bibinfo {pages} {33--88},\ \Eprint
  {https://arxiv.org/abs/math-ph/0406013} {arXiv:math-ph/0406013} \BibitemShut
  {NoStop}%
\bibitem [{\citenamefont {Gurau}\ and\ \citenamefont
  {Ryan}(2012)}]{ColourTensor}%
  \BibitemOpen
  \bibfield  {author} {\bibinfo {author} {\bibfnamefont {R.}~\bibnamefont
  {Gurau}}\ and\ \bibinfo {author} {\bibfnamefont {J.~P.}\ \bibnamefont
  {Ryan}},\ }\bibfield  {title} {\bibinfo {title} {{Colored Tensor Models - a
  review}},\ }\href {https://doi.org/10.3842/SIGMA.2012.020} {\bibfield
  {journal} {\bibinfo  {journal} {SIGMA}\ }\textbf {\bibinfo {volume} {8}},\
  \bibinfo {pages} {020} (\bibinfo {year} {2012})},\ \Eprint
  {https://arxiv.org/abs/1109.4812} {arXiv:1109.4812 [hep-th]} \BibitemShut
  {NoStop}%
\bibitem [{\citenamefont {Gielen}\ \emph {et~al.}(2014)\citenamefont {Gielen},
  \citenamefont {Oriti},\ and\ \citenamefont {Sindoni}}]{GFTcosmoLONGpaper}%
  \BibitemOpen
  \bibfield  {author} {\bibinfo {author} {\bibfnamefont {S.}~\bibnamefont
  {Gielen}}, \bibinfo {author} {\bibfnamefont {D.}~\bibnamefont {Oriti}},\ and\
  \bibinfo {author} {\bibfnamefont {L.}~\bibnamefont {Sindoni}},\ }\bibfield
  {title} {\bibinfo {title} {{Homogeneous cosmologies as group field theory
  condensates}},\ }\href {https://doi.org/10.1007/JHEP06(2014)013} {\bibfield
  {journal} {\bibinfo  {journal} {{JHEP}}\ }\textbf {\bibinfo {volume} {06}},\
  \bibinfo {pages} {013} (\bibinfo {year} {2014})},\ \Eprint
  {https://arxiv.org/abs/1311.1238} {arXiv:1311.1238 [gr-qc]} \BibitemShut
  {NoStop}%
\bibitem [{\citenamefont {Gielen}\ and\ \citenamefont
  {Sindoni}(2016)}]{Gielen_2016}%
  \BibitemOpen
  \bibfield  {author} {\bibinfo {author} {\bibfnamefont {S.}~\bibnamefont
  {Gielen}}\ and\ \bibinfo {author} {\bibfnamefont {L.}~\bibnamefont
  {Sindoni}},\ }\bibfield  {title} {\bibinfo {title} {{Quantum Cosmology from
  Group Field Theory Condensates: a Review}},\ }\href
  {https://doi.org/10.3842/SIGMA.2016.082} {\bibfield  {journal} {\bibinfo
  {journal} {SIGMA}\ }\textbf {\bibinfo {volume} {12}},\ \bibinfo {pages} {082}
  (\bibinfo {year} {2016})},\ \Eprint {https://arxiv.org/abs/1602.08104}
  {arXiv:1602.08104 [gr-qc]} \BibitemShut {NoStop}%
\bibitem [{\citenamefont {Oriti}\ \emph {et~al.}(2016)\citenamefont {Oriti},
  \citenamefont {Sindoni},\ and\ \citenamefont {Wilson-Ewing}}]{Oriti_2016}%
  \BibitemOpen
  \bibfield  {author} {\bibinfo {author} {\bibfnamefont {D.}~\bibnamefont
  {Oriti}}, \bibinfo {author} {\bibfnamefont {L.}~\bibnamefont {Sindoni}},\
  and\ \bibinfo {author} {\bibfnamefont {E.}~\bibnamefont {Wilson-Ewing}},\
  }\bibfield  {title} {\bibinfo {title} {{Emergent Friedmann dynamics with a
  quantum bounce from quantum gravity condensates}},\ }\href
  {https://doi.org/10.1088/0264-9381/33/22/224001} {\bibfield  {journal}
  {\bibinfo  {journal} {Class. Quant. Grav.}\ }\textbf {\bibinfo {volume}
  {33}},\ \bibinfo {pages} {224001} (\bibinfo {year} {2016})},\ \Eprint
  {https://arxiv.org/abs/1602.05881} {arXiv:1602.05881 [gr-qc]} \BibitemShut
  {NoStop}%
\bibitem [{\citenamefont {Oriti}\ \emph {et~al.}(2017)\citenamefont {Oriti},
  \citenamefont {Sindoni},\ and\ \citenamefont {Wilson-Ewing}}]{BOriti_2017}%
  \BibitemOpen
  \bibfield  {author} {\bibinfo {author} {\bibfnamefont {D.}~\bibnamefont
  {Oriti}}, \bibinfo {author} {\bibfnamefont {L.}~\bibnamefont {Sindoni}},\
  and\ \bibinfo {author} {\bibfnamefont {E.}~\bibnamefont {Wilson-Ewing}},\
  }\bibfield  {title} {\bibinfo {title} {{Bouncing cosmologies from quantum
  gravity condensates}},\ }\href {https://doi.org/10.1088/1361-6382/aa549a}
  {\bibfield  {journal} {\bibinfo  {journal} {Class. Quant. Grav.}\ }\textbf
  {\bibinfo {volume} {34}},\ \bibinfo {pages} {04LT01} (\bibinfo {year}
  {2017})},\ \Eprint {https://arxiv.org/abs/1602.08271} {arXiv:1602.08271
  [gr-qc]} \BibitemShut {NoStop}%
\bibitem [{\citenamefont {Adjei}\ \emph {et~al.}(2018)\citenamefont {Adjei},
  \citenamefont {Gielen},\ and\ \citenamefont {Wieland}}]{toy}%
  \BibitemOpen
  \bibfield  {author} {\bibinfo {author} {\bibfnamefont {E.}~\bibnamefont
  {Adjei}}, \bibinfo {author} {\bibfnamefont {S.}~\bibnamefont {Gielen}},\ and\
  \bibinfo {author} {\bibfnamefont {W.}~\bibnamefont {Wieland}},\ }\bibfield
  {title} {\bibinfo {title} {{Cosmological evolution as squeezing: a toy model
  for group field cosmology}},\ }\href
  {https://doi.org/10.1088/1361-6382/aaba11} {\bibfield  {journal} {\bibinfo
  {journal} {Class. Quant. Grav.}\ }\textbf {\bibinfo {volume} {35}},\ \bibinfo
  {pages} {105016} (\bibinfo {year} {2018})},\ \Eprint
  {https://arxiv.org/abs/1712.07266} {arXiv:1712.07266 [gr-qc]} \BibitemShut
  {NoStop}%
\bibitem [{\citenamefont {Wilson-Ewing}(2019)}]{relham_Wilson_Ewing_2019}%
  \BibitemOpen
  \bibfield  {author} {\bibinfo {author} {\bibfnamefont {E.}~\bibnamefont
  {Wilson-Ewing}},\ }\bibfield  {title} {\bibinfo {title} {{Relational
  Hamiltonian for group field theory}},\ }\href
  {https://doi.org/10.1103/PhysRevD.99.086017} {\bibfield  {journal} {\bibinfo
  {journal} {Phys. Rev. D}\ }\textbf {\bibinfo {volume} {99}},\ \bibinfo
  {pages} {086017} (\bibinfo {year} {2019})},\ \Eprint
  {https://arxiv.org/abs/1810.01259} {arXiv:1810.01259 [gr-qc]} \BibitemShut
  {NoStop}%
\bibitem [{\citenamefont {Gielen}\ \emph {et~al.}(2019)\citenamefont {Gielen},
  \citenamefont {Polaczek},\ and\ \citenamefont {Wilson-Ewing}}]{relhamadd}%
  \BibitemOpen
  \bibfield  {author} {\bibinfo {author} {\bibfnamefont {S.}~\bibnamefont
  {Gielen}}, \bibinfo {author} {\bibfnamefont {A.}~\bibnamefont {Polaczek}},\
  and\ \bibinfo {author} {\bibfnamefont {E.}~\bibnamefont {Wilson-Ewing}},\
  }\bibfield  {title} {\bibinfo {title} {{Addendum to ``Relational Hamiltonian
  for group field theory''}},\ }\href
  {https://doi.org/10.1103/PhysRevD.100.106002} {\bibfield  {journal} {\bibinfo
   {journal} {Phys. Rev. D}\ }\textbf {\bibinfo {volume} {100}},\ \bibinfo
  {pages} {106002} (\bibinfo {year} {2019})},\ \Eprint
  {https://arxiv.org/abs/1908.09850} {arXiv:1908.09850 [gr-qc]} \BibitemShut
  {NoStop}%
\bibitem [{\citenamefont {Gielen}\ and\ \citenamefont
  {Polaczek}(2020)}]{Gielen_2020}%
  \BibitemOpen
  \bibfield  {author} {\bibinfo {author} {\bibfnamefont {S.}~\bibnamefont
  {Gielen}}\ and\ \bibinfo {author} {\bibfnamefont {A.}~\bibnamefont
  {Polaczek}},\ }\bibfield  {title} {\bibinfo {title} {{Generalised effective
  cosmology from group field theory}},\ }\href
  {https://doi.org/10.1088/1361-6382/ab8f67} {\bibfield  {journal} {\bibinfo
  {journal} {Class. Quant. Grav.}\ }\textbf {\bibinfo {volume} {37}},\ \bibinfo
  {pages} {165004} (\bibinfo {year} {2020})},\ \Eprint
  {https://arxiv.org/abs/1912.06143} {arXiv:1912.06143 [gr-qc]} \BibitemShut
  {NoStop}%
\bibitem [{\citenamefont {Marchetti}\ and\ \citenamefont
  {Oriti}(2021{\natexlab{a}})}]{Marchetti2021}%
  \BibitemOpen
  \bibfield  {author} {\bibinfo {author} {\bibfnamefont {L.}~\bibnamefont
  {Marchetti}}\ and\ \bibinfo {author} {\bibfnamefont {D.}~\bibnamefont
  {Oriti}},\ }\bibfield  {title} {\bibinfo {title} {{Effective relational
  cosmological dynamics from quantum gravity}},\ }\href
  {https://doi.org/10.1007/JHEP05(2021)025} {\bibfield  {journal} {\bibinfo
  {journal} {JHEP}\ }\textbf {\bibinfo {volume} {05}},\ \bibinfo {pages}
  {025}},\ \Eprint {https://arxiv.org/abs/2008.02774} {arXiv:2008.02774
  [gr-qc]} \BibitemShut {NoStop}%
\bibitem [{\citenamefont {Gerhardt}\ \emph {et~al.}(2018)\citenamefont
  {Gerhardt}, \citenamefont {Oriti},\ and\ \citenamefont
  {Wilson-Ewing}}]{GFTsepUniv}%
  \BibitemOpen
  \bibfield  {author} {\bibinfo {author} {\bibfnamefont {F.}~\bibnamefont
  {Gerhardt}}, \bibinfo {author} {\bibfnamefont {D.}~\bibnamefont {Oriti}},\
  and\ \bibinfo {author} {\bibfnamefont {E.}~\bibnamefont {Wilson-Ewing}},\
  }\bibfield  {title} {\bibinfo {title} {{Separate universe framework in group
  field theory condensate cosmology}},\ }\href
  {https://doi.org/10.1103/PhysRevD.98.066011} {\bibfield  {journal} {\bibinfo
  {journal} {Phys. Rev. D}\ }\textbf {\bibinfo {volume} {98}},\ \bibinfo
  {pages} {066011} (\bibinfo {year} {2018})},\ \Eprint
  {https://arxiv.org/abs/1805.03099} {arXiv:1805.03099 [gr-qc]} \BibitemShut
  {NoStop}%
\bibitem [{\citenamefont {Marchetti}\ and\ \citenamefont
  {Oriti}(2022)}]{LucaGFTpert}%
  \BibitemOpen
  \bibfield  {author} {\bibinfo {author} {\bibfnamefont {L.}~\bibnamefont
  {Marchetti}}\ and\ \bibinfo {author} {\bibfnamefont {D.}~\bibnamefont
  {Oriti}},\ }\bibfield  {title} {\bibinfo {title} {{Effective dynamics of
  scalar cosmological perturbations from quantum gravity}},\ }\href
  {https://doi.org/10.1088/1475-7516/2022/07/004} {\bibfield  {journal}
  {\bibinfo  {journal} {JCAP}\ }\textbf {\bibinfo {volume} {07}}\bibfield
  {number} {\bibinfo  {number} { (07)},\ \bibinfo {pages} {004}},\ }\Eprint
  {https://arxiv.org/abs/2112.12677} {arXiv:2112.12677 [gr-qc]} \BibitemShut
  {NoStop}%
\bibitem [{\citenamefont {de~Cesare}\ \emph {et~al.}(2018)\citenamefont
  {de~Cesare}, \citenamefont {Oriti}, \citenamefont {Pithis},\ and\
  \citenamefont {Sakellariadou}}]{de_Cesare_2017}%
  \BibitemOpen
  \bibfield  {author} {\bibinfo {author} {\bibfnamefont {M.}~\bibnamefont
  {de~Cesare}}, \bibinfo {author} {\bibfnamefont {D.}~\bibnamefont {Oriti}},
  \bibinfo {author} {\bibfnamefont {A.~G.~A.}\ \bibnamefont {Pithis}},\ and\
  \bibinfo {author} {\bibfnamefont {M.}~\bibnamefont {Sakellariadou}},\
  }\bibfield  {title} {\bibinfo {title} {{Dynamics of anisotropies close to a
  cosmological bounce in quantum gravity}},\ }\href
  {https://doi.org/10.1088/1361-6382/aa986a} {\bibfield  {journal} {\bibinfo
  {journal} {Class. Quant. Grav.}\ }\textbf {\bibinfo {volume} {35}},\ \bibinfo
  {pages} {015014} (\bibinfo {year} {2018})},\ \Eprint
  {https://arxiv.org/abs/1709.00994} {arXiv:1709.00994 [gr-qc]} \BibitemShut
  {NoStop}%
\bibitem [{\citenamefont {Pithis}\ and\ \citenamefont
  {Sakellariadou}(2017)}]{Mairi_2017}%
  \BibitemOpen
  \bibfield  {author} {\bibinfo {author} {\bibfnamefont {A.~G.~A.}\
  \bibnamefont {Pithis}}\ and\ \bibinfo {author} {\bibfnamefont
  {M.}~\bibnamefont {Sakellariadou}},\ }\bibfield  {title} {\bibinfo {title}
  {{Relational evolution of effectively interacting group field theory quantum
  gravity condensates}},\ }\href {https://doi.org/10.1103/PhysRevD.95.064004}
  {\bibfield  {journal} {\bibinfo  {journal} {Phys. Rev. D}\ }\textbf {\bibinfo
  {volume} {95}},\ \bibinfo {pages} {064004} (\bibinfo {year} {2017})},\
  \Eprint {https://arxiv.org/abs/1612.02456} {arXiv:1612.02456 [gr-qc]}
  \BibitemShut {NoStop}%
\bibitem [{\citenamefont {Bojowald}(2010)}]{BojoBook}%
  \BibitemOpen
  \bibfield  {author} {\bibinfo {author} {\bibfnamefont {M.}~\bibnamefont
  {Bojowald}},\ }\href {https://doi.org/10.1017/CBO9780511921759} {\emph
  {\bibinfo {title} {Canonical Gravity and Applications: Cosmology, Black
  Holes, and Quantum Gravity}}}\ (\bibinfo  {publisher} {Cambridge University
  Press},\ \bibinfo {year} {2010})\BibitemShut {NoStop}%
\bibitem [{\citenamefont {Ashtekar}\ and\ \citenamefont
  {Wilson-Ewing}(2009)}]{AshtekarEdBianchiI}%
  \BibitemOpen
  \bibfield  {author} {\bibinfo {author} {\bibfnamefont {A.}~\bibnamefont
  {Ashtekar}}\ and\ \bibinfo {author} {\bibfnamefont {E.}~\bibnamefont
  {Wilson-Ewing}},\ }\bibfield  {title} {\bibinfo {title} {{Loop quantum
  cosmology of Bianchi type I models}},\ }\href
  {https://doi.org/10.1103/PhysRevD.79.083535} {\bibfield  {journal} {\bibinfo
  {journal} {Phys. Rev. D}\ }\textbf {\bibinfo {volume} {79}},\ \bibinfo
  {pages} {083535} (\bibinfo {year} {2009})},\ \Eprint
  {https://arxiv.org/abs/0903.3397} {arXiv:0903.3397 [gr-qc]} \BibitemShut
  {NoStop}%
\bibitem [{\citenamefont {Bojowald}\ \emph {et~al.}(2007)\citenamefont
  {Bojowald}, \citenamefont {Cartin},\ and\ \citenamefont
  {Khanna}}]{BojowaldMisner}%
  \BibitemOpen
  \bibfield  {author} {\bibinfo {author} {\bibfnamefont {M.}~\bibnamefont
  {Bojowald}}, \bibinfo {author} {\bibfnamefont {D.}~\bibnamefont {Cartin}},\
  and\ \bibinfo {author} {\bibfnamefont {G.}~\bibnamefont {Khanna}},\
  }\bibfield  {title} {\bibinfo {title} {{Lattice refining loop quantum
  cosmology, anisotropic models, and stability}},\ }\href
  {https://doi.org/10.1103/PhysRevD.76.064018} {\bibfield  {journal} {\bibinfo
  {journal} {Phys. Rev. D}\ }\textbf {\bibinfo {volume} {76}},\ \bibinfo
  {pages} {064018} (\bibinfo {year} {2007})},\ \Eprint
  {https://arxiv.org/abs/0704.1137} {arXiv:0704.1137 [gr-qc]} \BibitemShut
  {NoStop}%
\bibitem [{\citenamefont {Gielen}\ and\ \citenamefont
  {Polaczek}(2021)}]{Axel_Steffen_scalars}%
  \BibitemOpen
  \bibfield  {author} {\bibinfo {author} {\bibfnamefont {S.}~\bibnamefont
  {Gielen}}\ and\ \bibinfo {author} {\bibfnamefont {A.}~\bibnamefont
  {Polaczek}},\ }\bibfield  {title} {\bibinfo {title} {{Hamiltonian group field
  theory with multiple scalar matter fields}},\ }\href
  {https://doi.org/10.1103/PhysRevD.103.086011} {\bibfield  {journal} {\bibinfo
   {journal} {Phys. Rev. D}\ }\textbf {\bibinfo {volume} {103}},\ \bibinfo
  {pages} {086011} (\bibinfo {year} {2021})},\ \Eprint
  {https://arxiv.org/abs/2009.00615} {arXiv:2009.00615 [gr-qc]} \BibitemShut
  {NoStop}%
\bibitem [{\citenamefont {Barbero~G.}(1995)}]{barbero}%
  \BibitemOpen
  \bibfield  {author} {\bibinfo {author} {\bibfnamefont {J.~F.}\ \bibnamefont
  {Barbero~G.}},\ }\bibfield  {title} {\bibinfo {title} {{Real Ashtekar
  variables for Lorentzian signature space-times}},\ }\href
  {https://doi.org/10.1103/PhysRevD.51.5507} {\bibfield  {journal} {\bibinfo
  {journal} {Phys. Rev. D}\ }\textbf {\bibinfo {volume} {51}},\ \bibinfo
  {pages} {5507} (\bibinfo {year} {1995})},\ \Eprint
  {https://arxiv.org/abs/gr-qc/9410014} {arXiv:gr-qc/9410014} \BibitemShut
  {NoStop}%
\bibitem [{\citenamefont {Li}\ \emph {et~al.}(2017)\citenamefont {Li},
  \citenamefont {Oriti},\ and\ \citenamefont {Zhang}}]{Li_2017}%
  \BibitemOpen
  \bibfield  {author} {\bibinfo {author} {\bibfnamefont {Y.}~\bibnamefont
  {Li}}, \bibinfo {author} {\bibfnamefont {D.}~\bibnamefont {Oriti}},\ and\
  \bibinfo {author} {\bibfnamefont {M.}~\bibnamefont {Zhang}},\ }\bibfield
  {title} {\bibinfo {title} {{Group field theory for quantum gravity minimally
  coupled to a scalar field}},\ }\href
  {https://doi.org/10.1088/1361-6382/aa85d2} {\bibfield  {journal} {\bibinfo
  {journal} {Class. Quant. Grav.}\ }\textbf {\bibinfo {volume} {34}},\ \bibinfo
  {pages} {195001} (\bibinfo {year} {2017})},\ \Eprint
  {https://arxiv.org/abs/1701.08719} {arXiv:1701.08719 [gr-qc]} \BibitemShut
  {NoStop}%
\bibitem [{\citenamefont {Reisenberger}\ and\ \citenamefont
  {Rovelli}(2001)}]{Reisenberger_2000}%
  \BibitemOpen
  \bibfield  {author} {\bibinfo {author} {\bibfnamefont {M.~P.}\ \bibnamefont
  {Reisenberger}}\ and\ \bibinfo {author} {\bibfnamefont {C.}~\bibnamefont
  {Rovelli}},\ }\bibfield  {title} {\bibinfo {title} {{Spacetime as a Feynman
  diagram: the connection formulation}},\ }\href
  {https://doi.org/10.1088/0264-9381/18/1/308} {\bibfield  {journal} {\bibinfo
  {journal} {Class. Quant. Grav.}\ }\textbf {\bibinfo {volume} {18}},\ \bibinfo
  {pages} {121} (\bibinfo {year} {2001})},\ \Eprint
  {https://arxiv.org/abs/gr-qc/0002095} {arXiv:gr-qc/0002095} \BibitemShut
  {NoStop}%
\bibitem [{\citenamefont {Ben~Geloun}(2013)}]{Geloun_2013}%
  \BibitemOpen
  \bibfield  {author} {\bibinfo {author} {\bibfnamefont {J.}~\bibnamefont
  {Ben~Geloun}},\ }\bibfield  {title} {\bibinfo {title} {{On the finite
  amplitudes for open graphs in Abelian dynamical colored
  Boulatov\textendash{}Ooguri models}},\ }\href
  {https://doi.org/10.1088/1751-8113/46/40/402002} {\bibfield  {journal}
  {\bibinfo  {journal} {J. Phys. A}\ }\textbf {\bibinfo {volume} {46}},\
  \bibinfo {pages} {402002} (\bibinfo {year} {2013})},\ \Eprint
  {https://arxiv.org/abs/1307.8299} {arXiv:1307.8299 [hep-th]} \BibitemShut
  {NoStop}%
\bibitem [{\citenamefont {Gielen}(2016)}]{Gielen_lowspin}%
  \BibitemOpen
  \bibfield  {author} {\bibinfo {author} {\bibfnamefont {S.}~\bibnamefont
  {Gielen}},\ }\bibfield  {title} {\bibinfo {title} {{Emergence of a low spin
  phase in group field theory condensates}},\ }\href
  {https://doi.org/10.1088/0264-9381/33/22/224002} {\bibfield  {journal}
  {\bibinfo  {journal} {Class. Quant. Grav.}\ }\textbf {\bibinfo {volume}
  {33}},\ \bibinfo {pages} {224002} (\bibinfo {year} {2016})},\ \Eprint
  {https://arxiv.org/abs/1604.06023} {arXiv:1604.06023 [gr-qc]} \BibitemShut
  {NoStop}%
\bibitem [{\citenamefont {Regge}\ and\ \citenamefont
  {Williams}(2000)}]{ReggeGFT}%
  \BibitemOpen
  \bibfield  {author} {\bibinfo {author} {\bibfnamefont {T.}~\bibnamefont
  {Regge}}\ and\ \bibinfo {author} {\bibfnamefont {R.~M.}\ \bibnamefont
  {Williams}},\ }\bibfield  {title} {\bibinfo {title} {{Discrete structures in
  gravity}},\ }\href {https://doi.org/10.1063/1.533333} {\bibfield  {journal}
  {\bibinfo  {journal} {J. Math. Phys.}\ }\textbf {\bibinfo {volume} {41}},\
  \bibinfo {pages} {3964} (\bibinfo {year} {2000})},\ \Eprint
  {https://arxiv.org/abs/gr-qc/0012035} {arXiv:gr-qc/0012035} \BibitemShut
  {NoStop}%
\bibitem [{\citenamefont {Ooguri}(1992)}]{Ooguri}%
  \BibitemOpen
  \bibfield  {author} {\bibinfo {author} {\bibfnamefont {H.}~\bibnamefont
  {Ooguri}},\ }\bibfield  {title} {\bibinfo {title} {{Topological lattice
  models in four dimensions}},\ }\href
  {https://doi.org/10.1142/S0217732392004171} {\bibfield  {journal} {\bibinfo
  {journal} {Mod. Phys. Lett. A}\ }\textbf {\bibinfo {volume} {07}},\ \bibinfo
  {pages} {2799} (\bibinfo {year} {1992})},\ \Eprint
  {https://arxiv.org/abs/hep-th/9205090} {arXiv:hep-th/9205090} \BibitemShut
  {NoStop}%
\bibitem [{\citenamefont {De~Pietri}\ \emph {et~al.}(2000)\citenamefont
  {De~Pietri}, \citenamefont {Freidel}, \citenamefont {Krasnov},\ and\
  \citenamefont {Rovelli}}]{BCmodelfromGFT}%
  \BibitemOpen
  \bibfield  {author} {\bibinfo {author} {\bibfnamefont {R.}~\bibnamefont
  {De~Pietri}}, \bibinfo {author} {\bibfnamefont {L.}~\bibnamefont {Freidel}},
  \bibinfo {author} {\bibfnamefont {K.}~\bibnamefont {Krasnov}},\ and\ \bibinfo
  {author} {\bibfnamefont {C.}~\bibnamefont {Rovelli}},\ }\bibfield  {title}
  {\bibinfo {title} {{Barrett-Crane model from a Boulatov-Ooguri field theory
  over a homogeneous space}},\ }\href
  {https://doi.org/10.1016/S0550-3213(00)00005-5} {\bibfield  {journal}
  {\bibinfo  {journal} {Nucl. Phys. B}\ }\textbf {\bibinfo {volume} {574}},\
  \bibinfo {pages} {785} (\bibinfo {year} {2000})},\ \Eprint
  {https://arxiv.org/abs/hep-th/9907154} {arXiv:hep-th/9907154} \BibitemShut
  {NoStop}%
\bibitem [{\citenamefont {Lahoche}\ and\ \citenamefont
  {Samary}(2019)}]{Lahoche:2018hou}%
  \BibitemOpen
  \bibfield  {author} {\bibinfo {author} {\bibfnamefont {V.}~\bibnamefont
  {Lahoche}}\ and\ \bibinfo {author} {\bibfnamefont {D.~O.}\ \bibnamefont
  {Samary}},\ }\bibfield  {title} {\bibinfo {title} {{Progress in Solving the
  Nonperturbative Renormalization Group for Tensorial Group Field Theory}},\
  }\href {https://doi.org/10.3390/universe5030086} {\bibfield  {journal}
  {\bibinfo  {journal} {Universe}\ }\textbf {\bibinfo {volume} {5}},\ \bibinfo
  {pages} {86} (\bibinfo {year} {2019})},\ \Eprint
  {https://arxiv.org/abs/1812.00905} {arXiv:1812.00905 [hep-th]} \BibitemShut
  {NoStop}%
\bibitem [{\citenamefont {Oriti}(2016)}]{Oriti_GFT2ndLQG}%
  \BibitemOpen
  \bibfield  {author} {\bibinfo {author} {\bibfnamefont {D.}~\bibnamefont
  {Oriti}},\ }\bibfield  {title} {\bibinfo {title} {{Group field theory as the
  second quantization of loop quantum gravity}},\ }\href
  {https://doi.org/10.1088/0264-9381/33/8/085005} {\bibfield  {journal}
  {\bibinfo  {journal} {Class. Quant. Grav.}\ }\textbf {\bibinfo {volume}
  {33}},\ \bibinfo {pages} {085005} (\bibinfo {year} {2016})},\ \Eprint
  {https://arxiv.org/abs/1310.7786} {arXiv:1310.7786 [gr-qc]} \BibitemShut
  {NoStop}%
\bibitem [{\citenamefont {Thiemann}(2007)}]{ThiemannBook}%
  \BibitemOpen
  \bibfield  {author} {\bibinfo {author} {\bibfnamefont {T.}~\bibnamefont
  {Thiemann}},\ }\href {https://doi.org/10.1017/CBO9780511755682} {\emph
  {\bibinfo {title} {Modern Canonical Quantum General Relativity}}},\ Cambridge
  Monographs on Mathematical Physics\ (\bibinfo  {publisher} {Cambridge
  University Press},\ \bibinfo {year} {2007})\BibitemShut {NoStop}%
\bibitem [{\citenamefont {Giesel}\ and\ \citenamefont
  {Thiemann}(2007)}]{GieselThiemann}%
  \BibitemOpen
  \bibfield  {author} {\bibinfo {author} {\bibfnamefont {K.}~\bibnamefont
  {Giesel}}\ and\ \bibinfo {author} {\bibfnamefont {T.}~\bibnamefont
  {Thiemann}},\ }\bibfield  {title} {\bibinfo {title} {{Algebraic quantum
  gravity (AQG): II. Semiclassical analysis}},\ }\href
  {https://doi.org/10.1088/0264-9381/24/10/004} {\bibfield  {journal} {\bibinfo
   {journal} {Class. Quant. Grav.}\ }\textbf {\bibinfo {volume} {24}},\
  \bibinfo {pages} {2499} (\bibinfo {year} {2007})},\ \Eprint
  {https://arxiv.org/abs/gr-qc/0607100} {arXiv:gr-qc/0607100} \BibitemShut
  {NoStop}%
\bibitem [{\citenamefont {Alesci}\ and\ \citenamefont
  {Cianfrani}(2015{\natexlab{a}})}]{Alesci1}%
  \BibitemOpen
  \bibfield  {author} {\bibinfo {author} {\bibfnamefont {E.}~\bibnamefont
  {Alesci}}\ and\ \bibinfo {author} {\bibfnamefont {F.}~\bibnamefont
  {Cianfrani}},\ }\bibfield  {title} {\bibinfo {title} {{Loop quantum cosmology
  from quantum reduced loop gravity}},\ }\href
  {https://doi.org/10.1209/0295-5075/111/40002} {\bibfield  {journal} {\bibinfo
   {journal} {EPL}\ }\textbf {\bibinfo {volume} {111}},\ \bibinfo {pages}
  {40002} (\bibinfo {year} {2015}{\natexlab{a}})},\ \Eprint
  {https://arxiv.org/abs/1410.4788} {arXiv:1410.4788 [gr-qc]} \BibitemShut
  {NoStop}%
\bibitem [{\citenamefont {Alesci}\ and\ \citenamefont
  {Cianfrani}(2014)}]{Alesci2}%
  \BibitemOpen
  \bibfield  {author} {\bibinfo {author} {\bibfnamefont {E.}~\bibnamefont
  {Alesci}}\ and\ \bibinfo {author} {\bibfnamefont {F.}~\bibnamefont
  {Cianfrani}},\ }\bibfield  {title} {\bibinfo {title} {{Quantum reduced loop
  gravity: Semiclassical limit}},\ }\href
  {https://doi.org/10.1103/PhysRevD.90.024006} {\bibfield  {journal} {\bibinfo
  {journal} {Phys. Rev. D}\ }\textbf {\bibinfo {volume} {90}},\ \bibinfo
  {pages} {024006} (\bibinfo {year} {2014})},\ \Eprint
  {https://arxiv.org/abs/1402.3155} {arXiv:1402.3155 [gr-qc]} \BibitemShut
  {NoStop}%
\bibitem [{\citenamefont {Alesci}\ and\ \citenamefont
  {Cianfrani}(2015{\natexlab{b}})}]{Alesci3}%
  \BibitemOpen
  \bibfield  {author} {\bibinfo {author} {\bibfnamefont {E.}~\bibnamefont
  {Alesci}}\ and\ \bibinfo {author} {\bibfnamefont {F.}~\bibnamefont
  {Cianfrani}},\ }\bibfield  {title} {\bibinfo {title} {{Quantum reduced loop
  gravity: Universe on a lattice}},\ }\href
  {https://doi.org/10.1103/PhysRevD.92.084065} {\bibfield  {journal} {\bibinfo
  {journal} {Phys. Rev. D}\ }\textbf {\bibinfo {volume} {92}},\ \bibinfo
  {pages} {084065} (\bibinfo {year} {2015}{\natexlab{b}})},\ \Eprint
  {https://arxiv.org/abs/1506.07835} {arXiv:1506.07835 [gr-qc]} \BibitemShut
  {NoStop}%
\bibitem [{\citenamefont {Dapor}\ and\ \citenamefont
  {Liegener}(2018)}]{Dapor_effective_2017}%
  \BibitemOpen
  \bibfield  {author} {\bibinfo {author} {\bibfnamefont {A.}~\bibnamefont
  {Dapor}}\ and\ \bibinfo {author} {\bibfnamefont {K.}~\bibnamefont
  {Liegener}},\ }\bibfield  {title} {\bibinfo {title} {{Cosmological effective
  Hamiltonian from full loop quantum gravity dynamics}},\ }\href
  {https://doi.org/10.1016/j.physletb.2018.09.005} {\bibfield  {journal}
  {\bibinfo  {journal} {Phys. Lett. B}\ }\textbf {\bibinfo {volume} {785}},\
  \bibinfo {pages} {506} (\bibinfo {year} {2018})},\ \Eprint
  {https://arxiv.org/abs/1706.09833} {arXiv:1706.09833 [gr-qc]} \BibitemShut
  {NoStop}%
\bibitem [{\citenamefont {Brown}\ and\ \citenamefont
  {Kuchar}(1995)}]{LQG_dust1}%
  \BibitemOpen
  \bibfield  {author} {\bibinfo {author} {\bibfnamefont {J.~D.}\ \bibnamefont
  {Brown}}\ and\ \bibinfo {author} {\bibfnamefont {K.~V.}\ \bibnamefont
  {Kuchar}},\ }\bibfield  {title} {\bibinfo {title} {{Dust as a standard of
  space and time in canonical quantum gravity}},\ }\href
  {https://doi.org/10.1103/PhysRevD.51.5600} {\bibfield  {journal} {\bibinfo
  {journal} {Phys. Rev. D}\ }\textbf {\bibinfo {volume} {51}},\ \bibinfo
  {pages} {5600} (\bibinfo {year} {1995})},\ \Eprint
  {https://arxiv.org/abs/gr-qc/9409001} {arXiv:gr-qc/9409001} \BibitemShut
  {NoStop}%
\bibitem [{\citenamefont {Giesel}\ and\ \citenamefont
  {Thiemann}(2010)}]{AQG_thiemanngiesel}%
  \BibitemOpen
  \bibfield  {author} {\bibinfo {author} {\bibfnamefont {K.}~\bibnamefont
  {Giesel}}\ and\ \bibinfo {author} {\bibfnamefont {T.}~\bibnamefont
  {Thiemann}},\ }\bibfield  {title} {\bibinfo {title} {{Algebraic quantum
  gravity (AQG). IV. Reduced phase space quantisation of loop quantum
  gravity}},\ }\href {https://doi.org/10.1088/0264-9381/27/17/175009}
  {\bibfield  {journal} {\bibinfo  {journal} {Class. Quant. Grav.}\ }\textbf
  {\bibinfo {volume} {27}},\ \bibinfo {pages} {175009} (\bibinfo {year}
  {2010})},\ \Eprint {https://arxiv.org/abs/0711.0119} {arXiv:0711.0119
  [gr-qc]} \BibitemShut {NoStop}%
\bibitem [{\citenamefont {Domagala}\ \emph {et~al.}(2010)\citenamefont
  {Domagala}, \citenamefont {Giesel}, \citenamefont {Kaminski},\ and\
  \citenamefont {Lewandowski}}]{gravityquantised_Lewandowski}%
  \BibitemOpen
  \bibfield  {author} {\bibinfo {author} {\bibfnamefont {M.}~\bibnamefont
  {Domagala}}, \bibinfo {author} {\bibfnamefont {K.}~\bibnamefont {Giesel}},
  \bibinfo {author} {\bibfnamefont {W.}~\bibnamefont {Kaminski}},\ and\
  \bibinfo {author} {\bibfnamefont {J.}~\bibnamefont {Lewandowski}},\
  }\bibfield  {title} {\bibinfo {title} {{Gravity quantized: Loop quantum
  gravity with a scalar field}},\ }\href
  {https://doi.org/10.1103/PhysRevD.82.104038} {\bibfield  {journal} {\bibinfo
  {journal} {Phys. Rev. D}\ }\textbf {\bibinfo {volume} {82}},\ \bibinfo
  {pages} {104038} (\bibinfo {year} {2010})},\ \Eprint
  {https://arxiv.org/abs/1009.2445} {arXiv:1009.2445 [gr-qc]} \BibitemShut
  {NoStop}%
\bibitem [{\citenamefont {Husain}\ and\ \citenamefont
  {Pawlowski}(2012)}]{LQG_de-parametrised}%
  \BibitemOpen
  \bibfield  {author} {\bibinfo {author} {\bibfnamefont {V.}~\bibnamefont
  {Husain}}\ and\ \bibinfo {author} {\bibfnamefont {T.}~\bibnamefont
  {Pawlowski}},\ }\bibfield  {title} {\bibinfo {title} {{Time and a Physical
  Hamiltonian for Quantum Gravity}},\ }\href
  {https://doi.org/10.1103/PhysRevLett.108.141301} {\bibfield  {journal}
  {\bibinfo  {journal} {Phys. Rev. Lett.}\ }\textbf {\bibinfo {volume} {108}},\
  \bibinfo {pages} {141301} (\bibinfo {year} {2012})},\ \Eprint
  {https://arxiv.org/abs/1108.1145} {arXiv:1108.1145 [gr-qc]} \BibitemShut
  {NoStop}%
\bibitem [{\citenamefont {Giesel}\ and\ \citenamefont
  {Thiemann}(2015)}]{Giesel_scalarmaterialreference}%
  \BibitemOpen
  \bibfield  {author} {\bibinfo {author} {\bibfnamefont {K.}~\bibnamefont
  {Giesel}}\ and\ \bibinfo {author} {\bibfnamefont {T.}~\bibnamefont
  {Thiemann}},\ }\bibfield  {title} {\bibinfo {title} {{Scalar material
  reference systems and loop quantum gravity}},\ }\href
  {https://doi.org/10.1088/0264-9381/32/13/135015} {\bibfield  {journal}
  {\bibinfo  {journal} {Class. Quant. Grav.}\ }\textbf {\bibinfo {volume}
  {32}},\ \bibinfo {pages} {135015} (\bibinfo {year} {2015})},\ \Eprint
  {https://arxiv.org/abs/1206.3807} {arXiv:1206.3807 [gr-qc]} \BibitemShut
  {NoStop}%
\bibitem [{\citenamefont {Assanioussi}\ and\ \citenamefont
  {Kotecha}(2020)}]{Isha_thermalGFT}%
  \BibitemOpen
  \bibfield  {author} {\bibinfo {author} {\bibfnamefont {M.}~\bibnamefont
  {Assanioussi}}\ and\ \bibinfo {author} {\bibfnamefont {I.}~\bibnamefont
  {Kotecha}},\ }\bibfield  {title} {\bibinfo {title} {{Thermal representations
  in group field theory: squeezed vacua and quantum gravity condensates}},\
  }\href {https://doi.org/10.1007/JHEP02(2020)173} {\bibfield  {journal}
  {\bibinfo  {journal} {{JHEP}}\ }\textbf {\bibinfo {volume} {02}},\ \bibinfo
  {pages} {173} (\bibinfo {year} {2020})},\ \Eprint
  {https://arxiv.org/abs/1910.06889} {arXiv:1910.06889 [gr-qc]} \BibitemShut
  {NoStop}%
\bibitem [{\citenamefont {Oriti}(2017{\natexlab{b}})}]{ORITI2017235}%
  \BibitemOpen
  \bibfield  {author} {\bibinfo {author} {\bibfnamefont {D.}~\bibnamefont
  {Oriti}},\ }\bibfield  {title} {\bibinfo {title} {{The universe as a quantum
  gravity condensate}},\ }\href {https://doi.org/10.1016/j.crhy.2017.02.003}
  {\bibfield  {journal} {\bibinfo  {journal} {Comptes Rendus Physique}\
  }\textbf {\bibinfo {volume} {18}},\ \bibinfo {pages} {235} (\bibinfo {year}
  {2017}{\natexlab{b}})},\ \Eprint {https://arxiv.org/abs/1612.09521}
  {arXiv:1612.09521 [gr-qc]} \BibitemShut {NoStop}%
\bibitem [{\citenamefont {Barbieri}(1998)}]{Barbieri_1997}%
  \BibitemOpen
  \bibfield  {author} {\bibinfo {author} {\bibfnamefont {A.}~\bibnamefont
  {Barbieri}},\ }\bibfield  {title} {\bibinfo {title} {{Quantum tetrahedra and
  simplicial spin networks}},\ }\href
  {https://doi.org/10.1016/S0550-3213(98)00093-5} {\bibfield  {journal}
  {\bibinfo  {journal} {Nucl. Phys. B}\ }\textbf {\bibinfo {volume} {518}},\
  \bibinfo {pages} {714} (\bibinfo {year} {1998})},\ \Eprint
  {https://arxiv.org/abs/gr-qc/9707010} {arXiv:gr-qc/9707010} \BibitemShut
  {NoStop}%
\bibitem [{\citenamefont {Bianchi}\ \emph {et~al.}(2011)\citenamefont
  {Bianchi}, \citenamefont {Don\`a},\ and\ \citenamefont
  {Speziale}}]{PolyhedraIta}%
  \BibitemOpen
  \bibfield  {author} {\bibinfo {author} {\bibfnamefont {E.}~\bibnamefont
  {Bianchi}}, \bibinfo {author} {\bibfnamefont {P.}~\bibnamefont {Don\`a}},\
  and\ \bibinfo {author} {\bibfnamefont {S.}~\bibnamefont {Speziale}},\
  }\bibfield  {title} {\bibinfo {title} {{Polyhedra in loop quantum gravity}},\
  }\href {https://doi.org/10.1103/PhysRevD.83.044035} {\bibfield  {journal}
  {\bibinfo  {journal} {Phys. Rev. D}\ }\textbf {\bibinfo {volume} {83}},\
  \bibinfo {pages} {044035} (\bibinfo {year} {2011})},\ \Eprint
  {https://arxiv.org/abs/1009.3402} {arXiv:1009.3402 [gr-qc]} \BibitemShut
  {NoStop}%
\bibitem [{\citenamefont {Baez}\ and\ \citenamefont
  {Barrett}(1999)}]{Baez_1999}%
  \BibitemOpen
  \bibfield  {author} {\bibinfo {author} {\bibfnamefont {J.~C.}\ \bibnamefont
  {Baez}}\ and\ \bibinfo {author} {\bibfnamefont {J.~W.}\ \bibnamefont
  {Barrett}},\ }\bibfield  {title} {\bibinfo {title} {{The quantum tetrahedron
  in 3 and 4 dimensions}},\ }\href {https://doi.org/10.4310/ATMP.1999.v3.n4.a3}
  {\bibfield  {journal} {\bibinfo  {journal} {Adv. Theor. Math. Phys.}\
  }\textbf {\bibinfo {volume} {3}},\ \bibinfo {pages} {815} (\bibinfo {year}
  {1999})},\ \Eprint {https://arxiv.org/abs/gr-qc/9903060}
  {arXiv:gr-qc/9903060} \BibitemShut {NoStop}%
\bibitem [{\citenamefont {Gielen}(2021)}]{frozenf}%
  \BibitemOpen
  \bibfield  {author} {\bibinfo {author} {\bibfnamefont {S.}~\bibnamefont
  {Gielen}},\ }\bibfield  {title} {\bibinfo {title} {{Frozen formalism and
  canonical quantization in group field theory}},\ }\href
  {https://doi.org/10.1103/PhysRevD.104.106011} {\bibfield  {journal} {\bibinfo
   {journal} {Phys. Rev. D}\ }\textbf {\bibinfo {volume} {104}},\ \bibinfo
  {pages} {106011} (\bibinfo {year} {2021})},\ \Eprint
  {https://arxiv.org/abs/2105.01100} {arXiv:2105.01100 [hep-th]} \BibitemShut
  {NoStop}%
\bibitem [{\citenamefont {Konopka}\ \emph {et~al.}(2006)\citenamefont
  {Konopka}, \citenamefont {Markopoulou},\ and\ \citenamefont
  {Smolin}}]{Graphity}%
  \BibitemOpen
  \bibfield  {author} {\bibinfo {author} {\bibfnamefont {T.}~\bibnamefont
  {Konopka}}, \bibinfo {author} {\bibfnamefont {F.}~\bibnamefont
  {Markopoulou}},\ and\ \bibinfo {author} {\bibfnamefont {L.}~\bibnamefont
  {Smolin}},\ }\bibfield  {title} {\bibinfo {title} {{Quantum Graphity}},\
  }\Eprint {https://arxiv.org/abs/hep-th/0611197} {arXiv:hep-th/0611197}
  (\bibinfo {year} {2006})\BibitemShut {NoStop}%
\bibitem [{\citenamefont {Oriti}(2007)}]{GFTquantumST_Oriti}%
  \BibitemOpen
  \bibfield  {author} {\bibinfo {author} {\bibfnamefont {D.}~\bibnamefont
  {Oriti}},\ }\bibfield  {title} {\bibinfo {title} {{Group field theory as the
  microscopic description of the quantum spacetime fluid: a new perspective on
  the continuum in quantum gravity}},\ }\href@noop {} {\bibfield  {journal}
  {\bibinfo  {journal} {PoS}\ }\textbf {\bibinfo {volume} {QG-PH}},\ \bibinfo
  {pages} {030} (\bibinfo {year} {2007})},\ \Eprint
  {https://arxiv.org/abs/0710.3276} {arXiv:0710.3276 [gr-qc]} \BibitemShut
  {NoStop}%
\bibitem [{\citenamefont {Oriti}(2014)}]{DisappEmergenceSpaceTime}%
  \BibitemOpen
  \bibfield  {author} {\bibinfo {author} {\bibfnamefont {D.}~\bibnamefont
  {Oriti}},\ }\bibfield  {title} {\bibinfo {title} {{Disappearance and
  emergence of space and time in quantum gravity}},\ }\href
  {https://doi.org/10.1016/j.shpsb.2013.10.006} {\bibfield  {journal} {\bibinfo
   {journal} {Stud. Hist. Phil. Sci. B}\ }\textbf {\bibinfo {volume} {46}},\
  \bibinfo {pages} {186} (\bibinfo {year} {2014})},\ \Eprint
  {https://arxiv.org/abs/1302.2849} {arXiv:1302.2849 [physics.hist-ph]}
  \BibitemShut {NoStop}%
\bibitem [{\citenamefont {Kotecha}\ and\ \citenamefont
  {Oriti}(2018)}]{IshaDaniele}%
  \BibitemOpen
  \bibfield  {author} {\bibinfo {author} {\bibfnamefont {I.}~\bibnamefont
  {Kotecha}}\ and\ \bibinfo {author} {\bibfnamefont {D.}~\bibnamefont
  {Oriti}},\ }\bibfield  {title} {\bibinfo {title} {{Statistical equilibrium in
  quantum gravity: Gibbs states in group field theory}},\ }\href
  {https://doi.org/10.1088/1367-2630/aacbbd} {\bibfield  {journal} {\bibinfo
  {journal} {New J. Phys.}\ }\textbf {\bibinfo {volume} {20}},\ \bibinfo
  {pages} {073009} (\bibinfo {year} {2018})},\ \Eprint
  {https://arxiv.org/abs/1801.09964} {arXiv:1801.09964 [gr-qc]} \BibitemShut
  {NoStop}%
\bibitem [{\citenamefont {Gielen}(2018)}]{Gielen_GFT_matterframes}%
  \BibitemOpen
  \bibfield  {author} {\bibinfo {author} {\bibfnamefont {S.}~\bibnamefont
  {Gielen}},\ }\bibfield  {title} {\bibinfo {title} {{Group field theory and
  its cosmology in a matter reference frame}},\ }\href
  {https://doi.org/10.3390/universe4100103} {\bibfield  {journal} {\bibinfo
  {journal} {Universe}\ }\textbf {\bibinfo {volume} {4}},\ \bibinfo {pages}
  {103} (\bibinfo {year} {2018})},\ \Eprint {https://arxiv.org/abs/1808.10469}
  {arXiv:1808.10469 [gr-qc]} \BibitemShut {NoStop}%
\bibitem [{\citenamefont {Baytas}\ \emph {et~al.}(2019)\citenamefont {Baytas},
  \citenamefont {Bojowald},\ and\ \citenamefont {Crowe}}]{Bojo_2019}%
  \BibitemOpen
  \bibfield  {author} {\bibinfo {author} {\bibfnamefont {B.}~\bibnamefont
  {Baytas}}, \bibinfo {author} {\bibfnamefont {M.}~\bibnamefont {Bojowald}},\
  and\ \bibinfo {author} {\bibfnamefont {S.}~\bibnamefont {Crowe}},\ }\bibfield
   {title} {\bibinfo {title} {{Equivalence of models in loop quantum cosmology
  and group field theory}},\ }\href {https://doi.org/10.3390/universe5020041}
  {\bibfield  {journal} {\bibinfo  {journal} {Universe}\ }\textbf {\bibinfo
  {volume} {5}},\ \bibinfo {pages} {41} (\bibinfo {year} {2019})},\ \Eprint
  {https://arxiv.org/abs/1811.11156} {arXiv:1811.11156 [gr-qc]} \BibitemShut
  {NoStop}%
\bibitem [{\citenamefont {Gielen}\ \emph {et~al.}(2022)\citenamefont {Gielen},
  \citenamefont {Marchetti}, \citenamefont {Oriti},\ and\ \citenamefont
  {Polaczek}}]{LastAxel}%
  \BibitemOpen
  \bibfield  {author} {\bibinfo {author} {\bibfnamefont {S.}~\bibnamefont
  {Gielen}}, \bibinfo {author} {\bibfnamefont {L.}~\bibnamefont {Marchetti}},
  \bibinfo {author} {\bibfnamefont {D.}~\bibnamefont {Oriti}},\ and\ \bibinfo
  {author} {\bibfnamefont {A.}~\bibnamefont {Polaczek}},\ }\bibfield  {title}
  {\bibinfo {title} {{Effective cosmology from one-body operators in group
  field theory}},\ }\href {https://doi.org/10.1088/1361-6382/ac5052} {\bibfield
   {journal} {\bibinfo  {journal} {Class. Quant. Grav.}\ }\textbf {\bibinfo
  {volume} {39}},\ \bibinfo {pages} {075002} (\bibinfo {year} {2022})},\
  \Eprint {https://arxiv.org/abs/2110.11176} {arXiv:2110.11176 [gr-qc]}
  \BibitemShut {NoStop}%
\bibitem [{\citenamefont {Bianchi}\ and\ \citenamefont
  {Haggard}(2012)}]{Bianchi_Haggard_vol}%
  \BibitemOpen
  \bibfield  {author} {\bibinfo {author} {\bibfnamefont {E.}~\bibnamefont
  {Bianchi}}\ and\ \bibinfo {author} {\bibfnamefont {H.~M.}\ \bibnamefont
  {Haggard}},\ }\bibfield  {title} {\bibinfo {title} {{Bohr-Sommerfeld
  quantization of space}},\ }\href {https://doi.org/10.1103/PhysRevD.86.124010}
  {\bibfield  {journal} {\bibinfo  {journal} {Phys. Rev. D}\ }\textbf {\bibinfo
  {volume} {86}},\ \bibinfo {pages} {124010} (\bibinfo {year} {2012})},\
  \Eprint {https://arxiv.org/abs/1208.2228} {arXiv:1208.2228 [gr-qc]}
  \BibitemShut {NoStop}%
\bibitem [{\citenamefont {Bianchi}\ and\ \citenamefont
  {Haggard}(2011)}]{Bianchi_Haggard_letter}%
  \BibitemOpen
  \bibfield  {author} {\bibinfo {author} {\bibfnamefont {E.}~\bibnamefont
  {Bianchi}}\ and\ \bibinfo {author} {\bibfnamefont {H.~M.}\ \bibnamefont
  {Haggard}},\ }\bibfield  {title} {\bibinfo {title} {{Discreteness of the
  volume of space from Bohr-Sommerfeld quantization}},\ }\href
  {https://doi.org/10.1103/PhysRevLett.107.011301} {\bibfield  {journal}
  {\bibinfo  {journal} {Phys. Rev. Lett.}\ }\textbf {\bibinfo {volume} {107}},\
  \bibinfo {pages} {011301} (\bibinfo {year} {2011})},\ \Eprint
  {https://arxiv.org/abs/1102.5439} {arXiv:1102.5439 [gr-qc]} \BibitemShut
  {NoStop}%
\bibitem [{\citenamefont {Gerber}(1975)}]{Gerber_Ortho}%
  \BibitemOpen
  \bibfield  {author} {\bibinfo {author} {\bibfnamefont {L.}~\bibnamefont
  {Gerber}},\ }\bibfield  {title} {\bibinfo {title} {{The orthocentric simplex
  as an extreme simplex}},\ }\href {https://doi.org/10.2140/pjm.1975.56.97}
  {\bibfield  {journal} {\bibinfo  {journal} {Pacific J. Math.}\ }\textbf
  {\bibinfo {volume} {56}},\ \bibinfo {pages} {97} (\bibinfo {year}
  {1975})}\BibitemShut {NoStop}%
\bibitem [{\citenamefont {Brunnemann}\ and\ \citenamefont
  {Thiemann}(2006)}]{Brunnemann_Thiemann_Vol}%
  \BibitemOpen
  \bibfield  {author} {\bibinfo {author} {\bibfnamefont {J.}~\bibnamefont
  {Brunnemann}}\ and\ \bibinfo {author} {\bibfnamefont {T.}~\bibnamefont
  {Thiemann}},\ }\bibfield  {title} {\bibinfo {title} {{Simplification of the
  spectral analysis of the volume operator in loop quantum gravity}},\ }\href
  {https://doi.org/10.1088/0264-9381/23/4/014} {\bibfield  {journal} {\bibinfo
  {journal} {Class. Quant. Grav.}\ }\textbf {\bibinfo {volume} {23}},\ \bibinfo
  {pages} {1289} (\bibinfo {year} {2006})},\ \Eprint
  {https://arxiv.org/abs/gr-qc/0405060} {arXiv:gr-qc/0405060} \BibitemShut
  {NoStop}%
\bibitem [{\citenamefont {Ben~Achour}\ \emph {et~al.}(2017)\citenamefont
  {Ben~Achour}, \citenamefont {Brahma},\ and\ \citenamefont
  {Geiller}}]{BenAchour:2016ajk}%
  \BibitemOpen
  \bibfield  {author} {\bibinfo {author} {\bibfnamefont {J.}~\bibnamefont
  {Ben~Achour}}, \bibinfo {author} {\bibfnamefont {S.}~\bibnamefont {Brahma}},\
  and\ \bibinfo {author} {\bibfnamefont {M.}~\bibnamefont {Geiller}},\
  }\bibfield  {title} {\bibinfo {title} {{New Hamiltonians for loop quantum
  cosmology with arbitrary spin representations}},\ }\href
  {https://doi.org/10.1103/PhysRevD.95.086015} {\bibfield  {journal} {\bibinfo
  {journal} {Phys. Rev. D}\ }\textbf {\bibinfo {volume} {95}},\ \bibinfo
  {pages} {086015} (\bibinfo {year} {2017})},\ \Eprint
  {https://arxiv.org/abs/1612.07615} {arXiv:1612.07615 [gr-qc]} \BibitemShut
  {NoStop}%
\bibitem [{\citenamefont {Marchetti}\ and\ \citenamefont
  {Oriti}(2021{\natexlab{b}})}]{LucaFluct}%
  \BibitemOpen
  \bibfield  {author} {\bibinfo {author} {\bibfnamefont {L.}~\bibnamefont
  {Marchetti}}\ and\ \bibinfo {author} {\bibfnamefont {D.}~\bibnamefont
  {Oriti}},\ }\bibfield  {title} {\bibinfo {title} {{Quantum Fluctuations in
  the Effective Relational GFT Cosmology}},\ }\href
  {https://doi.org/10.3389/fspas.2021.683649} {\bibfield  {journal} {\bibinfo
  {journal} {Front. Astron. Space Sci.}\ }\textbf {\bibinfo {volume} {8}},\
  \bibinfo {pages} {683649} (\bibinfo {year} {2021}{\natexlab{b}})},\ \Eprint
  {https://arxiv.org/abs/2010.09700} {arXiv:2010.09700 [gr-qc]} \BibitemShut
  {NoStop}%
\bibitem [{\citenamefont {De~Pietri}\ and\ \citenamefont
  {Rovelli}(1996)}]{DePietriGEO}%
  \BibitemOpen
  \bibfield  {author} {\bibinfo {author} {\bibfnamefont {R.}~\bibnamefont
  {De~Pietri}}\ and\ \bibinfo {author} {\bibfnamefont {C.}~\bibnamefont
  {Rovelli}},\ }\bibfield  {title} {\bibinfo {title} {{Geometry eigenvalues and
  the scalar product from recoupling theory in loop quantum gravity}},\ }\href
  {https://doi.org/10.1103/PhysRevD.54.2664} {\bibfield  {journal} {\bibinfo
  {journal} {Phys. Rev. D}\ }\textbf {\bibinfo {volume} {54}},\ \bibinfo
  {pages} {2664} (\bibinfo {year} {1996})},\ \Eprint
  {https://arxiv.org/abs/gr-qc/9602023} {arXiv:gr-qc/9602023} \BibitemShut
  {NoStop}%
\bibitem [{\citenamefont {Rovelli}\ and\ \citenamefont
  {Smolin}(1995)}]{RovelliSmolin_vol}%
  \BibitemOpen
  \bibfield  {author} {\bibinfo {author} {\bibfnamefont {C.}~\bibnamefont
  {Rovelli}}\ and\ \bibinfo {author} {\bibfnamefont {L.}~\bibnamefont
  {Smolin}},\ }\bibfield  {title} {\bibinfo {title} {{Discreteness of area and
  volume in quantum gravity}},\ }\href
  {https://doi.org/10.1016/0550-3213(95)00150-Q} {\bibfield  {journal}
  {\bibinfo  {journal} {Nucl. Phys. B}\ }\textbf {\bibinfo {volume} {442}},\
  \bibinfo {pages} {593} (\bibinfo {year} {1995})},\ \bibinfo {note} {[Erratum:
  Nucl.Phys.B 456, 753--754 (1995)]},\ \Eprint
  {https://arxiv.org/abs/gr-qc/9411005} {arXiv:gr-qc/9411005} \BibitemShut
  {NoStop}%
\bibitem [{\citenamefont {Ashtekar}\ and\ \citenamefont
  {Lewandowski}(1998)}]{AshtekarLewandowski_vol}%
  \BibitemOpen
  \bibfield  {author} {\bibinfo {author} {\bibfnamefont {A.}~\bibnamefont
  {Ashtekar}}\ and\ \bibinfo {author} {\bibfnamefont {J.}~\bibnamefont
  {Lewandowski}},\ }\bibfield  {title} {\bibinfo {title} {{Quantum theory of
  geometry II: Volume operators}},\ }\href
  {https://doi.org/10.4310/ATMP.1997.v1.n2.a8} {\bibfield  {journal} {\bibinfo
  {journal} {Adv. Theor. Math. Phys.}\ }\textbf {\bibinfo {volume} {1}},\
  \bibinfo {pages} {388} (\bibinfo {year} {1998})},\ \Eprint
  {https://arxiv.org/abs/gr-qc/9711031} {arXiv:gr-qc/9711031} \BibitemShut
  {NoStop}%
\bibitem [{\citenamefont {Brunnemann}\ and\ \citenamefont
  {Rideout}(2008{\natexlab{a}})}]{Brunnemann1_2008}%
  \BibitemOpen
  \bibfield  {author} {\bibinfo {author} {\bibfnamefont {J.}~\bibnamefont
  {Brunnemann}}\ and\ \bibinfo {author} {\bibfnamefont {D.}~\bibnamefont
  {Rideout}},\ }\bibfield  {title} {\bibinfo {title} {{Properties of the volume
  operator in loop quantum gravity: I. Results}},\ }\href
  {https://doi.org/10.1088/0264-9381/25/6/065001} {\bibfield  {journal}
  {\bibinfo  {journal} {Class. Quant. Grav.}\ }\textbf {\bibinfo {volume}
  {25}},\ \bibinfo {pages} {065001} (\bibinfo {year} {2008}{\natexlab{a}})},\
  \Eprint {https://arxiv.org/abs/0706.0469} {arXiv:0706.0469 [gr-qc]}
  \BibitemShut {NoStop}%
\bibitem [{\citenamefont {Brunnemann}\ and\ \citenamefont
  {Rideout}(2008{\natexlab{b}})}]{Brunnemann2_2008}%
  \BibitemOpen
  \bibfield  {author} {\bibinfo {author} {\bibfnamefont {J.}~\bibnamefont
  {Brunnemann}}\ and\ \bibinfo {author} {\bibfnamefont {D.}~\bibnamefont
  {Rideout}},\ }\bibfield  {title} {\bibinfo {title} {{Properties of the volume
  operator in loop quantum gravity: II. Detailed presentation}},\ }\href
  {https://doi.org/10.1088/0264-9381/25/6/065002} {\bibfield  {journal}
  {\bibinfo  {journal} {Class. Quant. Grav.}\ }\textbf {\bibinfo {volume}
  {25}},\ \bibinfo {pages} {065002} (\bibinfo {year} {2008}{\natexlab{b}})},\
  \Eprint {https://arxiv.org/abs/0706.0382} {arXiv:0706.0382 [gr-qc]}
  \BibitemShut {NoStop}%
\end{thebibliography}%
\let\addcontentsline\oldaddcontentsline

\end{document}